\documentclass[twocolumn, floatfix, superscriptaddress, prb, longbibliography, 10pt]{revtex4-1}
\usepackage{extsizes}
\usepackage[paperwidth=210mm,paperheight=297mm,centering,hmargin=2cm,vmargin=2.5cm]{geometry}
\usepackage[utf8]{inputenc}

\usepackage[version=3]{mhchem} %
\usepackage[dvipsnames]{xcolor}
\usepackage{physics}
\usepackage[normalem]{ulem}
\newcommand{\todorev}[1]{ }

\usepackage{graphicx}
\usepackage{amsmath}
\usepackage{amssymb}
\usepackage{comment}
\usepackage{hyperref}
\usepackage{ulem}
\usepackage{bm}
\usepackage[capitalize]{cleveref}
\usepackage[symbols,nogroupskip,sort=none]{glossaries-extra}
\usepackage{soul}

\setcitestyle{sort&compress,numbers,super}
\usepackage{listings}
\usepackage{physics}
\usepackage{xparse}

\newcommand{\mbf}[1]{\ensuremath{\mathbf{#1}}}

\NewDocumentCommand{\rep}{s d<| d|>}{%
\IfBooleanTF{#1}{
   \IfValueTF{#2}{
       \IfValueTF{#3}{\braket{#2}{#3}}{\bra{#2}}
       }{
       \IfValueTF{#3}{\ket{#3}}{}
       }
   }{
   \IfValueTF{#2}{
       \IfValueTF{#3}{\braket*{#2}{#3}}{\bra*{#2}}
       }{
       \IfValueTF{#3}{\ket*{#3}}{}
       }
   }
}

\NewDocumentCommand{\rbra}{sm}{\IfBooleanTF{#1}{\rep<#2|}{\rep*<#2|}}
\NewDocumentCommand{\rket}{sm}{\IfBooleanTF{#1}{\rep|#2>}{\rep*|#2>}}
\NewDocumentCommand{\rbraket}{smom}{
    \IfBooleanTF{#1}{
        \IfNoValueTF{#3}{\rep*<#2||#4>}{\matrixel{#2}{#3}{#4}}
    }{
        \IfNoValueTF{#3}{\rep<#2||#4>}{\matrixel*{#2}{#3}{#4}}
    }
}
\NewDocumentCommand{\cg}{m m m}{\rep<#1; #2||#3>}

\NewDocumentCommand{\field}{o m e{_} e{^} o e{_} e{^}}{
\IfValueTF{#5}{\overline{
  #2\IfValueT{#3}{_#3}\IfValueT{#4}{^{\otimes #4}} %
  \otimes 
  #5\IfValueT{#6}{_#6}\IfValueT{#7}{^{\otimes #7}} %
  \IfValueT{#1}{;#1}
}}{
  \IfValueTF{#4}{\overline{
     #2\IfValueT{#3}{_#3}\IfValueT{#4}{^{\otimes #4}}
     \IfValueT{#1}{;#1}
  }}
  {#2\IfValueT{#3}{_#3}}
}
}

\NewDocumentCommand{\frho}{o e{_} e{^}}{
\field[#1]{\rho}_{#2}^{#3}
}
\NewDocumentCommand{\fdelta}{o e{_} e{^}}{
\field[#1]{\delta}_{#2}^{#3}
}

\newcommand{\e}{a}  %

\newcommand{\br}{\mbf{r}}
\newcommand{\bx}{\mbf{x}}
\newcommand{\bxhat}{\hat{\mbf{x}}}
\newcommand{\brhat}{\hat{\mbf{r}}}

\NewDocumentCommand{\ex}{e_}{
\IfValueTF{#1}{\e_{#1}\bx_{#1}}{\e\bx}
}  %

\NewDocumentCommand{\lm}{e_}{
\IfValueTF{#1}{l_{#1}m_{#1}}{lm}
}
\NewDocumentCommand{\nlm}{e_}{
\IfValueTF{#1}{n_{#1}\lm_{#1}}{n\lm}
}
\NewDocumentCommand{\enlm}{e_}{
\IfValueTF{#1}{\e_{#1}\nlm_{#1}}{\e\nlm}
}
\NewDocumentCommand{\en}{e_}{
\IfValueTF{#1}{\e_{#1}n_{#1}}{\e n}
}
\NewDocumentCommand{\nlk}{e_}{
\IfValueTF{#1}{n_{#1}l_{#1}k_{#1}}{nlk}
}
\NewDocumentCommand{\enlk}{e_}{
\IfValueTF{#1}{\e_{#1}\nlk_{#1}}{\e\nlk}
}
\NewDocumentCommand{\enl}{e_}{
\IfValueTF{#1}{\en_{#1}l_#1}{\en l}
}
\NewDocumentCommand{\nl}{e_}{
\IfValueTF{#1}{n_{#1}l_#1}{n l}
}

\NewDocumentCommand{\nnl}{s}{
\IfBooleanTF{#1}{n_1 n_2 l}{n_1; n_2; l}
}
\NewDocumentCommand{\ennl}{s}{
\IfBooleanTF{#1}{\en_1 \en_2 l}{\en_1; \en_2; l}
}

\NewDocumentCommand{\gslm}{s}{
\IfBooleanTF{#1}{\sigma\lambda\mu}{\sigma;\lambda\mu}
}
\NewDocumentCommand{\glm}{}{\lambda\mu}
\usepackage{bm,upgreek}
\newcommand{\bh}{\mbf{h}}

\newcommand{\bu}{\mbf{u}}
\newcommand{\bk}{\mbf{k}}
\newcommand{\bP}{\mbf{P}}
\newcommand{\bq}{\mbf{q}}

\newcommand{\krn}[0]{\operatorname{k}}
\newcommand{\dst}[0]{\operatorname{d}}

\newcommand{\Q}[0]{Q}
\newcommand{\q}[0]{q}
\newcommand{\y}[0]{y}
\newcommand{\yt}[0]{\tilde{y}}
\newcommand{\by}[0]{\mbf{\y}}

\newcommand{\feat}{\upxi}
\newcommand{\bfeat}[0]{\ensuremath{\bm{\upxi}}}
\newcommand{\bfeatc}[0]{\ensuremath{\bm{\upchi}}}
\newcommand{\bFeat}{\bm{\Xi}}

\newcommand{\D}[2][]{\ensuremath{\mathop{}\!\mathrm{d}^{#1}{#2}\,}}
\newcommand{\I}{\ensuremath{\mathrm{i}}}

\newcommand{\That}{\hat{t}}
\newcommand{\ihat}{\hat{i}}
\newcommand{\Rhat}{\hat{R}}
\newcommand{\Rthree}{{\mathbb{R}^3}}
\newcommand{\Othree}{{O(3)}}
\newcommand{\SOthree}{{SO(3)}}
\newcommand{\Ylm}{Y^m_l}

\newcommand{\rrhoislm}[1]{\frho[\gslm]_i^{#1}}

\newcommand{\lmax}[0]{{l_\text{max}} }
\newcommand{\nmax}[0]{{n_\text{max}} }
\newcommand{\fcut}[0]{{f_\text{cut}} }
\newcommand{\rcut}[0]{{r_\text{cut}} }
\newcommand{\sigmaa}[0]{{\sigma_{\e}} }
\newcommand{\nneigh}[0]{n_\text{neigh}}

\newcommand{\qmax}[0]{\q_\text{max}}

\newcommand{\vertequiv}{\rotatebox{90}{$\;\equiv\;$}}

\begin{document}

\title{Physics-inspired structural representations for molecules and materials}

\author{Felix Musil}
\affiliation{Laboratory of Computational Science and Modeling, IMX, \'Ecole Polytechnique F\'ed\'erale de Lausanne, 1015 Lausanne, Switzerland}
\affiliation{National Centre for Computational Design and Discovery of Novel Materials (MARVEL), \'Ecole Polytechnique F\'ed\'erale de Lausanne, Lausanne, Switzerland}

\author{Andrea Grisafi}
\affiliation{Laboratory of Computational Science and Modeling, IMX, \'Ecole Polytechnique F\'ed\'erale de Lausanne, 1015 Lausanne, Switzerland}

\author{Albert P. Bart\'ok}
\affiliation{Department of Physics and Warwick Centre for Predictive Modelling, School of Engineering, University of Warwick, Coventry CV4 7AL, United Kingdom}

\author{Christoph Ortner}
\affiliation{University of British Columbia, University of British Columbia,  Vancouver, BC, Canada V6T 1Z2}

\author{G\'abor Cs\'anyi}
\affiliation{Engineering Laboratory, University of Cambridge, Trumpington Street, Cambridge CB2 1PZ, United Kingdom}

\author{Michele Ceriotti}
\email{michele.ceriotti@epfl.ch}
\affiliation{Laboratory of Computational Science and Modeling, IMX, \'Ecole Polytechnique F\'ed\'erale de Lausanne, 1015 Lausanne, Switzerland}
\affiliation{National Centre for Computational Design and Discovery of Novel Materials (MARVEL), \'Ecole Polytechnique F\'ed\'erale de Lausanne, Lausanne, Switzerland}

\date{\today}
\raggedbottom

\begin{abstract}
The first step in the construction of a regression model or a data-driven analysis, aiming to predict or elucidate the relationship between the atomic scale structure of matter and its properties, involves transforming the Cartesian coordinates of the atoms into a suitable \emph{representation}. 
The development of atomic-scale representations has played, and continues to play, a central role in the success of machine-learning methods for chemistry and materials science. 
This review summarizes the current understanding of the nature and characteristics of the most commonly used structural and chemical descriptions of atomistic structures, highlighting the deep underlying connections between different frameworks, and the ideas that lead to computationally efficient and universally applicable models. 
It emphasizes the link between properties, structures, their physical chemistry and their mathematical description, provides examples of recent applications to a diverse set of chemical and materials science problems, and outlines the open questions and the most promising research directions in the field.
\end{abstract}
\maketitle

\tableofcontents
\todorev{

\begin{itemize}
\item Z-matrix figure
\item Decide what to do with universal SOAP (GC)
\item librascal benchmarks?
\item harmonize notation with other chemrev. $F->y$ perhaps? Kernel? feature vectors.
\end{itemize}

1.) References section must have a heading (References, Literature Cited, Bibliography, etc. -Works Cited, however, is not permitted).

2.) References must adhere to ACS journal format: include the list of authors, article titles, journal abbreviation, publishing year, volume and full-page range; for example:
(1.) Hammes-Schiffer, S.; Soudackov, A. V. Proton-Coupled Electron Transfer in Solution, Proteins, and Electrochemistry. J. Phys. Chem. B 2008, 112, 14108-14123
- Abbreviate all journal names in references.
-All titles in references should have the same format -Title Case (preferred) or non-title case; mixed references formatting is not allowed.
- Include full page range in journal references.
- Include a volume number in all journal references
- References with more than 10 authors should list the first 10 authors followed by “et al.”

3.) A TOC graphic is required and make sure that it is no bigger than 5cm by 5cm.
}

\section{Introduction}

The last decade has seen a tremendous increase in the use of data-driven approaches for the modeling of molecules and materials. 
Atomistic simulation has been a particularly fertile field of use; applications range from the analysis of large databases of materials properties,\cite{isay+15cm} to the design of molecules with the desired behavior for a given application.\cite{sanc-aspu18science}
Machine learning techniques have been applied to devise coarse-grained descriptions of complex molecular systems,\cite{das+06pnas,ceri+11pnas,spiw-kral11jcp,rohr+13arpc,Kanekal2019, Jackson2019, Wang2019} to build accurate and comparatively inexpensive interatomic potentials,\cite{behl-parr07prl,braa-bowm09irpc,bart+10prl,soss+12prb,behl16jcp,deri-csan17prb,drag+18prm,chen+20nature,deri+21nature} and more generally to predict, or rationalize, the relationship between a specific atomic configuration and the properties that can be computed by electronic-structure calculations\cite{rupp+12prl,smit+17cs,schu+18jcp,paru+18ncomm,gris+19acscs,wilk+19pnas,schu+19nc,kali+21acr}.

All of these applications to atomic-scale systems share the need to map an atomic configuration $A$ -- identified by the positions and chemical identity of its $N$ atoms $\left\{\br_i,\e_i\right\}$, and possibly by the basis vectors of the periodic repeat unit $\bh$ -- into a more suitable representation.
This mapping associates $A$ with a point in a feature space, which is then used to construct a machine-learning model to regress (fit) a structure-property relation, to cluster (group together) configurations that share similar structural patterns, or to further map the conformational landscape of a data set onto a low-dimensional visualization.

The terms \emph{descriptor} or \emph{fingerprint} are used, usually interchangeably, in chemical and materials informatics to indicate heuristically-determined properties that are easier to compute than the quantities one ultimately wants to predict, but correlate strongly with them, facilitating the construction of transferable and accurate models.\cite{ouya+18prm}
Examples of descriptors include the fractional composition of a compound, the electronegativity of its atoms, a low-level-of-theory determination of the HOMO-LUMO gap of a molecule.
In this review we focus on a more systematic class of mappings that use exclusively atomic composition and geometry as inputs, and aim to characterize precisely the instantaneous arrangement of the atoms, for which we use the term \emph{representation}. We will be especially interested in those representations that apply geometric and algebraic manipulations to the Cartesian coordinates, to transform them in a way that fulfills physically-informed requirements:  smoothness and symmetry with respect to isometries. 
Commonly used representations include atom-centered symmetry functions\cite{behl+07jcp,behl-parr07prl}, Coulomb matrices\cite{rupp+12prl}, and the smooth overlap of atomic positions (SOAP)\cite{bart+13prb}.
It is important to note that representations can be expressed using different mathematical entities. In the most straightforward realisation, the space of features takes the form of a vector space, in which each configuration is associated with a finite-dimensional vector whose entries are explicitly computed by the mapping procedure.
Depending on the application, however, it may be simpler or more natural to describe the relationship between pairs of configurations.
Such relationship can be expressed in terms of a kernel function $\krn(A,A')$ (e.g. the scalar product between feature vectors), or in terms of a distance between configurations $\dst(A,A')$ (e.g. the Euclidean distance between associated features). 
As we will see, distance or kernel-based formulations implicitly define a feature space, that in most cases can be expressed (at least approximately) in terms of a vector of features, and so can be seen as equivalent to a representation of individual structures,  even in cases in which the distance or the kernel are not explicitly computed from a pair of feature vectors.%

While one can trace the origins of different representations to specific subfields of computational chemistry and materials science, the fact that representations should describe precisely the nature and positions of each atom means that they often are not specialized to a given application, but can be used with little modification for any atomistic system, from gas-phase molecules to bulk solids\cite{bart+17sa,huo-rupp17arxiv,chen+19cm}.
This generality, however, does not mean that representations are completely abstract or disconnected from physical and chemical concepts.
Over the past few years, it has become clear that representations that reflect more closely some fundamental principles -- such as locality, the multi-scale nature of interactions, the similarities in the behavior of elements from the same group in the periodic table -- usually yield models that are more robust, transferable and data-efficient.
The link between a representation and the physical concepts it incorporates is usually mediated by the strategy one uses to fit the desired structure-property relations: %
it is often possible to show an explicit relationship between linear regression models built on the representation of a structure and well-known empirical forms of interatomic potentials (such as body-ordered, or multipole expansions), and more complex, non-linear machine-learning schemes built on the same features improve the flexibility in describing structure-property relations, albeit at the price of a less transparent interpretation of their behavior.%

Given the central role of structural representations in the application of data-driven methods to atomistic modeling, it is perhaps not surprising that considerable effort is being dedicated to understanding and improving their properties.
These efforts follow several directions.
First, the efficient, scalable, and parallel implementation of the construction of a given set of features is essential to ensure computational efficiency.
Second, reduction in the number of features that is used to describe the system reduces the computational effort, and often improves the robustness of the model: feature selection aims at identifying the most expressive, yet concise, description of the system at hand.
Third, it is often desirable to fine-tune a representation so that it facilitates training a model on a small number of reference structures, by incorporating more explicitly the available prior knowledge.

This review aims to summarize recent work on the construction of efficient and mathematically sound representations of atomic and molecular structures, with a particular focus on the use for the regression of atomic-scale properties.
It is part of a special issue that covers the many facets of the application of machine learning to chemical simulations, and the interested reader may find, among others, discussions of machine learning models based on Gaussian process regression, using some of the descriptors we discuss here\cite{chemrev2021deringer}, of the construction of potentials for molecules\cite{chemrev2021gdml,chemrev2021smallmolecules} and materials\cite{chemrev2021behler},  the description of excited states\cite{chemrev2021excited}, and of unsupervised machine learning schemes\cite{chemrev2021unsupervised}.
Rather than focusing on a historical overview, we intend to provide a snapshot of the current insights on what makes a good representation, supporting our considerations with recent publications, and providing a perspective of the most promising research directions in the field.

\setlength{\glsdescwidth}{0.8\linewidth}
\setglossarysection{section}
\section{List of symbols}
\renewcommand{\glossarysection}[2][]{}
\glsxtrnewsymbol[description={An atomic structure}]{A}{\ensuremath{A}}
\glsxtrnewsymbol[description={An environment centered on the $i$-th atom of the structure $A$}]{Ai}{\ensuremath{A_i}}
\glsxtrnewsymbol[description={Position of the $i$-th atom}]{bri}{\ensuremath{\br_i}}
\glsxtrnewsymbol[description={Vector separating the $i$-th atom and its $j$-th neighbor, $\br_j-\br_i$}]{brji}{\ensuremath{\br_{ji}}}
\glsxtrnewsymbol[description={Generic continuous index enumerating the components of an atomic representation}]{S}{\ensuremath{\Q}}
\glsxtrnewsymbol[description={Generic discrete index enumerating the components of an atomic representation}]{x}{\ensuremath{\q}}
\glsxtrnewsymbol[description={A representation of a structure $A$ indexed by an unspecified label or set of labels $\Q$ }]{repSA}{\ensuremath{\rep<\Q||A>}}
\glsxtrnewsymbol[description={Feature vector (with elements indexed by $\q$)  associated with an atom-centered environment, $\feat_\q(A_i)=\rep<\q||A_i>$}]{bfeatAi}{\ensuremath{\bfeat(A_i)}}
\glsxtrnewsymbol[description={Feature matrix combining the features associated with multiple structures/environments} ]{bFeat}{\ensuremath{\bFeat}}
\glsxtrnewsymbol[description={A column in a feature matrix, where $({\bfeatc}_{\q})_i =\feat_\q(A_i)$} ]{bfeatc}{\ensuremath{\bfeatc_\q}}

\glsxtrnewsymbol[description={An atom-centered property, or its systematic approximation in terms of an atom-centered representation $\rep|A_i>$ }]{yAi}{\ensuremath{\y(A_i)}}
\glsxtrnewsymbol[description={A non-linear model that approximates \gls{yAi} using the feature vector \gls{bfeatAi} }]{yt}{\ensuremath{\yt(\bfeat)}}
\glsxtrnewsymbol[description={A (non-)linear kernel computed between two structures or environments, represented by the corresponding feature vectors $\bfeat(A)$ }]{Kt}{\ensuremath{\krn(A,A')}}
\glsxtrnewsymbol[description={A distance computed between two structures or environments }]{D}{\ensuremath{\dst(A,A')}}
\glsxtrnewsymbol[description={Structure representation  based on a smooth atom density }]{rrho}{\ensuremath{\rep|\rho>}}
\glsxtrnewsymbol[description={Representation of an environment centered on atom $i$, that can be obtained by symmetrizing \gls{rrho} over translations  }]{rrhoi}{\ensuremath{\rep|\rho_i>}}
\glsxtrnewsymbol[description={Symmetrized $\nu$-point correlation of the atomic density built on the atom-centered representation \gls{rrhoi} }]{rrhoinu}{\ensuremath{\rep|\frho_i^\nu>}}
\glsxtrnewsymbol[description={Dirac-$\delta$ limit of the smooth atom density \gls{rrho}. Analogous symmetrized versions are indicated as $\rep|\fdelta_i>$ and $\rep|\fdelta_i^\nu>$ }]{rdelta}{\ensuremath{\rep|\delta>}}
\glsxtrnewsymbol[description={Atom-density field representation, suitable to describe long-range correlations}]{rv}{\ensuremath{\rep|V>}}

\printunsrtglossary[type=symbols,style=long]

\section{Representations for materials and molecules}
\label{sec:history-wishes}

Even though this review has no intention of providing an exhaustive historical account of the development of descriptors for atomic structures, it is worth providing a brief overview.
A ``data-driven'' philosophy emerged early in the field of chemical and molecular science, where the combinatorial extent of the space of possible molecules,\cite{blum-reym09jacs} and the possibility of accessing this space with comparatively simple synthetic strategies,  encouraged the development of quantitative structure/property relationships (QSPR) techniques, attempting to map\cite{kare00book} descriptors of molecular structure -- based on cheminformatics fingerprints,\cite{wein88jcim, Todeschini2010}chemical-intuition driven descriptors\cite{will+20jcim}, molecular graphs,\cite{schn-fech05nrdd} or indicators obtained from quantum chemical calculations\cite{kare+96cr} -- to the behavior of a selected compound, usually focusing on properties of direct applicative interest\cite{Gaulton2012, Sterling2015, Kim2016} such as solubility, toxicity,\cite{Wu2018} or pharmacological activity.\cite{obre+07jcim,Mysinger2012}

\begin{figure}[tbp]
    \centering
    \includegraphics[width=1.0\linewidth]{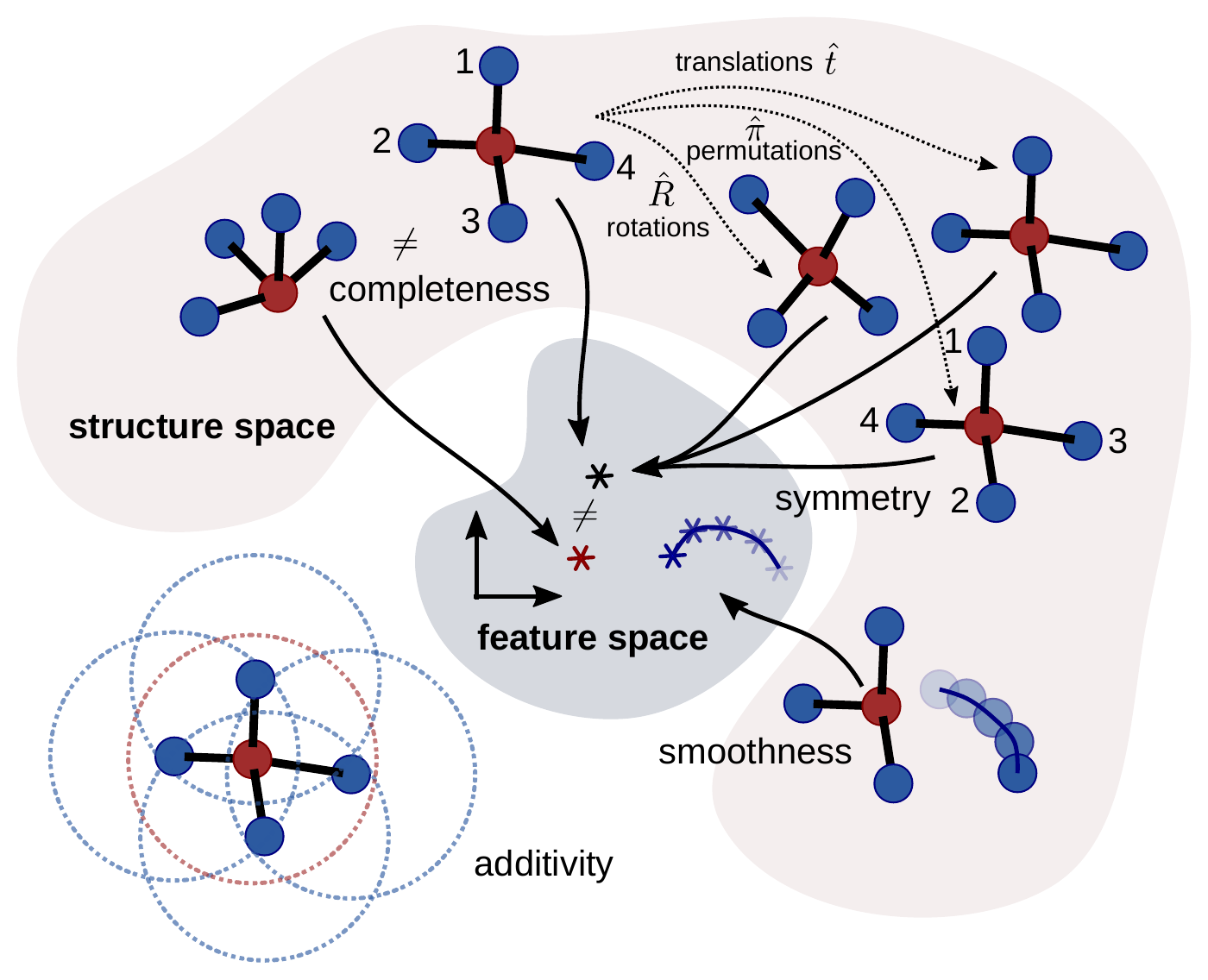}
    \caption{A schematic overview of the requirements for an effective structural representation. The mapping between structures and feature space should obey fundamental physical symmetries (equivalent structures should be mapped to the same features); should be complete (inequivalent structures should be mapped to distinct features); should be smooth (continuous deformations of a structure should map to a smooth deformation of the associated features). Furthermore, whenever dealing with datasets that are not homogeneous in molecular size, the representation should be additive: a structure should be decomposed in a sum of local environments (usually atom-centered), ensuring transferability and extensivity of predictions.
    }
    \label{fig:rep-wishes}
\end{figure}

\begin{figure*}[tbp]
\centering
\includegraphics[width=1.0\linewidth]{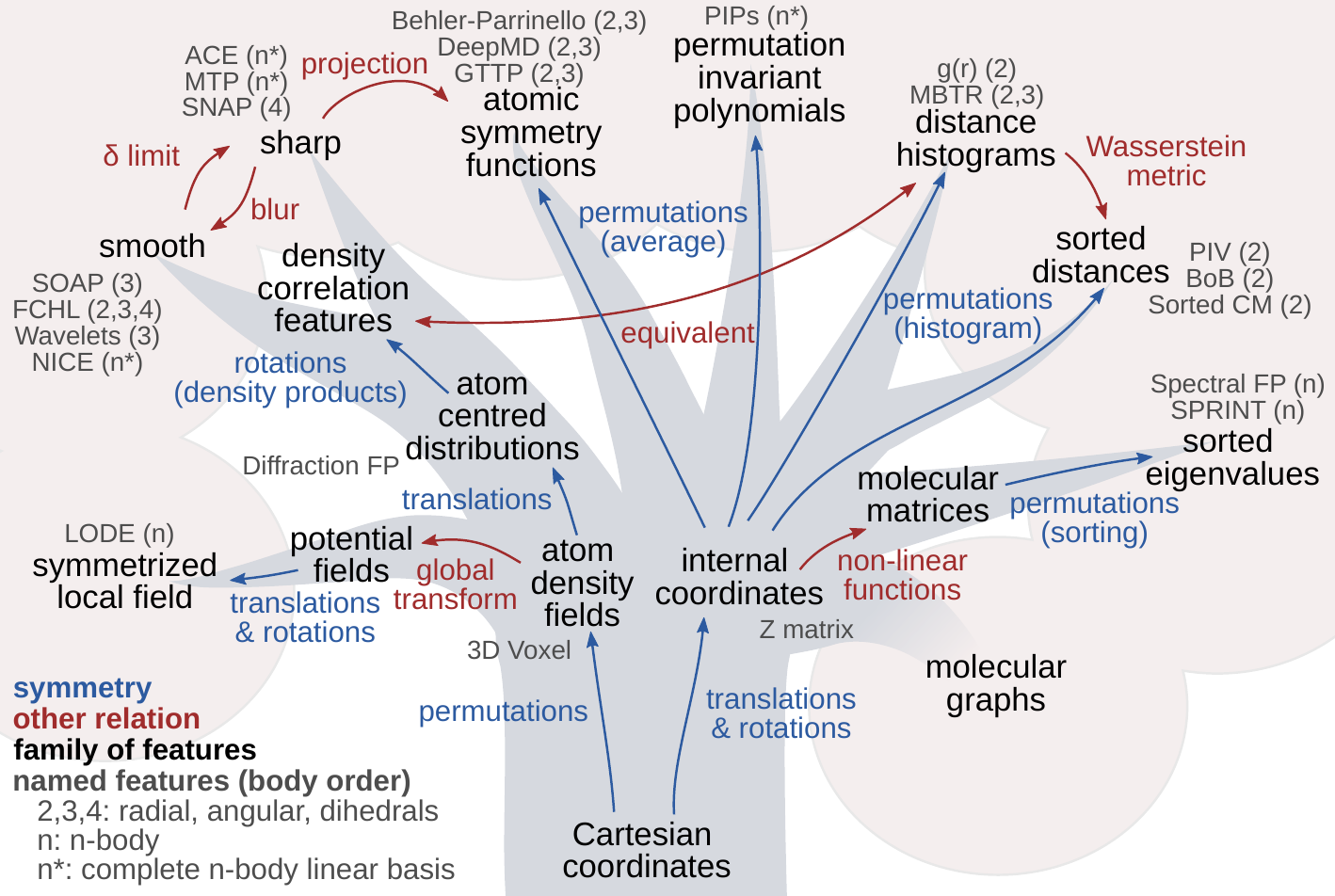}
\caption{A phylogenetic tree of structural representations for materials and molecules. Arrows indicate the relationship between different groups of features. Lists of names, in gray, indicate the most common implementations for each class.  Classes that appear as ``leaves'' of the tree are fully symmetric.
\label{fig:tree}
}
\end{figure*}

This approach should be contrasted with that of ``bottom-up'' predictions, that aim to use models of the interactions between the atomic constituents of a material to simulate the behavior of the system on an atomic time and length scale.
Starting from the early days of molecular simulations\cite{jone24prsa,Alder1970, rahm-stil71jcp,car-parr85prl} the objective was to predict the energy, the forces, or any other observable of interest, for a specific molecular configuration, and use them to search for (meta-)stable configurations, or to simulate the evolution of the system by molecular dynamics\cite{alle-tild90book,fren-smit02book}.
In the absence of reliable reference values for the properties of specific atomic configurations, interatomic potentials (also called empirical force fields) were built using physically-inspired functional forms, combining harmonic terms to describe chemical bonds with Coulomb and $1/r^6$ terms to describe electrostatics and dispersion. Their (few) parameters were determined by matching the values of experimental observables, such as cohesive energies, lattice vectors and elastic constants.
The continuous increase in computational power, and the availability of electronic-structure techniques with a better cost-accuracy ratio\cite{parr-yang94book,burk12jcp,boot+13nature} has made it possible to compute extremely accurate energies and properties of specific configurations.
This has opened the way to \emph{ab initio} simulations of materials\cite{car-parr85prl}, but also provided a viable alternative to empirical functional forms for the construction of interatomic potentials.  Starting from the simplest compounds\cite{part-schw97jcp}, and then gradually increasing in complexity\cite{huan+05jcp}, molecular potential energy surfaces fitted by interpolating between a comparatively small number of \emph{ab initio} reference calculations provided the first practical applications of this idea. The possibility of combining very accurate calculations of the electronic structure of atomic systems with sampling of the statistics and dynamics of the nuclei on the electronic potential energy surface has allowed  theoretical predictions that do not only agree with experimental results\cite{part-schw97jcp} -- they can predict experiments\cite{russ-mano96cpl} two decades before measurements become precise enough to verify the theoretical values.\cite{kim+15science} 

Even though the ultimate goal of QSPR models and machine-learned potentials is the same -- predicting scientifically and/or technologically relevant properties of molecules and materials -- the approaches they follow to achieve this goal are quite different,
which is reflected in the way an atomic structure is translated into an input for a machine-learning model.
Cheminformatics descriptors, or fingerprints, are built \textit{ad hoc}, incorporating both descriptors of molecular structure and composition, and easy-to-estimate molecular properties.
They usually rely on a considerable amount of prior knowledge, are often system and problem specific, and are meant to label a compound rather than a specific configuration of its atoms. This is a logical consequence of the fact that QSPR aims for an end-to-end description of a thermodynamic property, which is not an attribute of an individual configuration, but of a thermodynamic state of matter.
In the case of bottom-up modeling, instead,  one aims first at building a very accurate surrogate model that is capable of reproducing precisely and inexpensively the outcome of quantum calculations for a specific configuration of the atoms. The end goal of predicting thermodynamic properties is achieved by coupling these prediction with statistical sampling methods\cite{alle-tild90book,fren-smit02book,tuck08book} aimed at computing averages over the appropriate classical (or quantum\cite{feyn-hibb65book,mark-ceri18nrc}) distribution of atomic configurations.
As a consequence, the representations used as inputs of these surrogate quantum models are usually rather generic, constructed based exclusively on atomic coordinates and chemical species. They aim to establish a precise mapping between a specific structure and the associated atomic-scale quantities, and for this reason have also proven very useful to \emph{analyze} atomistic configurations\cite{sade+13jcp,de+16pccp,bern+19acie}, an application we discuss in detail in Section~\ref{sec:maps}.
Even though we focus our discussion on this latter class of features, it is worth mentioning the recent, and rather successful, attempts to use descriptors that incorporate information from electronic-structure calculations, that we briefly summarize in Section~\ref{sub:electronic-structure-descriptors}.

In the rest of this section, we discuss the properties that are desirable for a representation used in atomistic machine learning, which are graphically summarized in Figure~\ref{fig:rep-wishes}.
The mapping between structures and features should be consistent with basic symmetries -- i.e. reflect the fact that the properties associated with a structure do not change when the reference system or the labelling of identical atoms are modified; be smooth, so that models built on the features inherit a regular behavior with changing atomic coordinates; be complete, so that fundamentally distinct configurations are never mapped to the same set of features.
Furthermore, many machine-learning tasks benefit greatly from being based on \emph{local} features, which describe atoms or groups of atoms. Even though this is a less stringent requirement and, as we discuss below, global descriptors have been used very successfully, representations based on local environments are usually associated with higher transferability, reflecting a ``divide and conquer'' approach to materials modeling\cite{yang91prl,gall-parr92prl}.
Finally, less fundamental but not less important requirements are the numerical stability and computational efficiency of the structure-representation mapping, which we discuss in Section~\ref{sec:efficiency}.

\subsection{Symmetry}

The Cartesian coordinates of the atoms encode all the information that is needed to reconstruct the geometry of a structure.
Yet, it is obvious that they cannot be used directly as the input of a regression model.
The fact that the Cartesian description of a molecule depends on its absolute position and orientation in space, and the order by which atoms are listed, means that configurations that are completely equivalent can be represented by many different Cartesian values, which makes any regression, classification or clustering scheme inefficient and potentially misleading.
Over the years, many different approaches have been proposed by which translations, rotations, inversion and atom permutation symmetries can be enforced, which is reflected in the variety of alternative frameworks to achieve an effective representation to be used of the input of an atomistic machine-learning scheme.
In fact, symmetry is such a central principle underpinning these efforts that it can be used to construct a  ``phylogenetic tree'' of representations, organized according to the strategy that is used to incorporate symmetry in their construction, as shown in Figure~\ref{fig:tree}.

The need to remove the trivial symmetries, namely the dependency of the Cartesian coordinates on the origin and orientation of the reference system, has been recognized very early in the field of chemical and materials modeling.
Different sets of internal coordinates\cite{pula+79jacs} (bonds, angles, torsions) have been proposed, based on chemical intuition, as invariant descriptors of molecular geometry, and most of the molecular forcefields that have been so effective in the modeling of biological systems\cite{mayo+90jpc,vano+09jcc,halg96jcc,damm+97jcc} rely on internal coordinates to define bonded interactions.
A collection of internal coordinates that is sufficient to fully characterize the geometry of a structure, often referred-to as the Z-matrix, is a paradigmatic example of this class of representations. Even though the efficiency of this approach has often been questioned\cite{bake-hehr91jcc,bake-chan96jcc}, particularly because there is no unique way to define the Z-matrix, internal coordinates are still ubiquitous, and are effective whenever the system being studied has a well-defined, persistent bonding pattern (see Ref.~\citenum{Harrison2018} for a recent review).
In these cases, internal coordinates can be seen as the initial step in the construction of discretized molecular representations, such as a molecular graph. Even though very widely used in chemical machine learning\cite{duve+15nips,sanc-aspu18science}, these graph based schemes are not meant to describe the exact arrangement of the atoms, but just their bonding pattern, and so fall outside the scope of this review.

The limitations of an internal-coordinates description become most apparent when one wants to model a chemically-active system, as the bonding patterns can change during the course of a simulation, and therefore the invariance to atom index permutations becomes crucial to achieve a consistent model. The Empirical Valence Bond (EVB) method\cite{kame-wars11wcms} has been used to simulate bond-breaking events, but the generality of the EVB approach is limited as the possible assignments need be pre-determined.
This led to the development of representations that are intrinsically independent on the ordering of the atoms, such as permutation-invariant polynomials (PIPs)\cite{brow+04jcp,braa-bowm09irpc,bowm+10jpcl,xie-bowm10jctc,jian-guo13jcp} which are obtained by summing functions of the internal coordinates over all possible orderings.
In their original implementation, the exponentially increasing cost of evaluating these sums limited their applicability to molecules with a small number of degrees of freedom.
It is worth mentioning that the problem of fitting molecular potential energy surfaces, particularly for applications to gas-phase physical chemistry, has led to approaches that anticipate several of the ideas that have become central to modern machine-learning techniques: the need to symmetrize appropriately atomic structures,\cite{coll-pars93jcp} the systematic fitting to databases of configurations computed with high levels of quantum chemistry,\cite{part-schw97jcp} and even the use of ``neural network potentials'' \cite{blan+95jcp,gass+98jpca} are just a few examples of the pioneering contributions from this field. 

In the condensed phase, a similar pioneering role was played by the construction of systematic expansions of the potential energy of alloys\cite{sanc+84pa}, and of bond order potentials based on the moments of the density of states \cite{pett89prl,hors+96prb2,ozak+00prb}. 
Both anticipate the use of an atom-centered description of the energy, the role of symmetry, and the notion of building a systematic expansion of the target property in terms of a convergent hierarchy of terms of increasing complexity. 
The first successful attempt of explicitly bringing machine-learning ideas to the construction of interatomic potentials for condensed-phase materials can be attributed to Behler and Parrinello, who in Ref.~\citenum{behl-parr07prl} introduced the concept of atom-centered symmetry functions (ACSF), which rely on a local expansion of the energy and on the construction of a symmetric description of atomic environments.
Similarly to PIPs, ACSF are translationally and rotationally invariant because they are functions of angles and distances, and permutationally invariant because they are summed over all possible atomic pairs and triplets within an atomic environment.
The computational cost of ACSF is kept under control by restricting the range of interactions (which we discuss further in subsection~\ref{sub:locality}) and the body order of the correlations considered. Despite these restrictions, ACSF models have been shown to achieve comparable accuracy to that reached by PIPs\cite{nguy+18jcp}. Indeed, the recently proposed \emph{atomic PIPs}\cite{vand+20mlst} use the same polynomial basis as global PIPs, but avoid the unfavorable scaling with increasing molecule size by combining locality (via a distance cutoff) and a truncation of the order of the expansion.

Internal coordinates are also the fundamental building block of molecular matrix representations, which are based on functions of the interatomic distances within a structure.
Coulomb matrices, which list the formal electrostatic interactions $q_i q_j / r_{ji}$ between each atomic pair in a structure, have been extensively explored in early applications of the machine learning of molecular properties\cite{rupp+12prl}, with the main limitation being connected to the lack of permutation invariance\cite{mous12prl}, which has also been tackled by approximate symmetrization, summing over a manageable number of randomized orderings of the atoms\cite{mont+12nips,hans+13jctc}.
We discuss alternative approaches to symmetrizing Coulomb matrices, as well as other representations based on molecular matrices, in Subsection~\ref{sub:smoothness}.

The phylogenetic tree in Fig.~\ref{fig:tree} shows that a large number of existing representation take a different strategy to achieve symmetrization: rather than using internal coordinates that are inherently invariant to rotations and translations, they first -- implicitly or explicitly -- describe the system as an atom density $\sum_i g(\bx-\br_i)$, obtained by summing over localized functions centered on the positions $\br_i$ of all atoms in the system.  Such a density is naturally invariant to permutations, and only at a later stage one proceeds to symmetrize it over translations and rotations.
We discuss in great detail this second approach in Section~\ref{sec:symmetry-fields}. It suffices to say, at this point, that even if the construction of symmetrized density representations is conceptually very different from those based on internal coordinates, there are many direct and indirect links between the two branches, sketched in Figure~\ref{fig:tree}, which we will discuss when reviewing specific classes of representations.

\subsection{Smoothness} \label{sub:smoothness}

The overwhelming majority of atomic-scale properties are continuous, smooth functions of the atomic coordinates. Function regularity is crucial for creating efficient ML models, and is therefore one of the requirements for a good structural representation.
Features  constructed from a symmetrized atom density %
are naturally smooth functions of atomic coordinates, and it is usually not a problem to maintain this regular behavior upon symmetrization over translations and rotations. The level of smoothness can be adjusted by smearing the atomic density, or by expanding it on a smooth basis (effectively a Fourier smoothing), as we discuss more extensively in Section~\ref{sec:symmetry-fields}.
Internal coordinates are also usually smooth, but the process of manipulating them to achieve a permutation invariant representation can affect the smoothness of the mapping. %

One way to obtain permutation invariance without incurring the exponential scaling of the cost associated with enumerating all possible permutations of atomic indices involves sorting the entries in a distance or Coulomb matrix\cite{mont+12nips,rupp+15jpcl}, an approach that has also been used with permutation invariant vectors (PIV)\cite{pipo+17prl}, ``bag of bonds'' features (BoB)\cite{hans+15jpcl}.
Similar descriptors based on sorted distances have been also used to identify recurring structures in structure optimization algorithms\cite{vilh-hamm12prl,vilh-hamm14jcp}, and more recently generalized to lexicographically-sorted lists of $k$-neighbors distances\cite{chen+18jctc}.
Computing the eigenvalues of (functions of) interatomic distances, which underlies the SPRINT method\cite{piet-andr11prl} as well the overlap matrix eigenvalue fingerprints\cite{sade+13jcp, zhu+16jcp}, also effectively achieves permutation invariance by similar means, since the vector of eigenvalues is taken to be sorted in ascending or descending order.  The earliest implementation of the DeepMD scheme\cite{wang+18cpc} also relied on sorting a local distance matrix.
However, the sorting operation introduces derivative discontinuities in the mapping between Cartesian coordinates and features, because the order of the distance vector changes as atoms are displaced in the structure.

\begin{figure}
    \centering
    \includegraphics[width=0.9\linewidth]{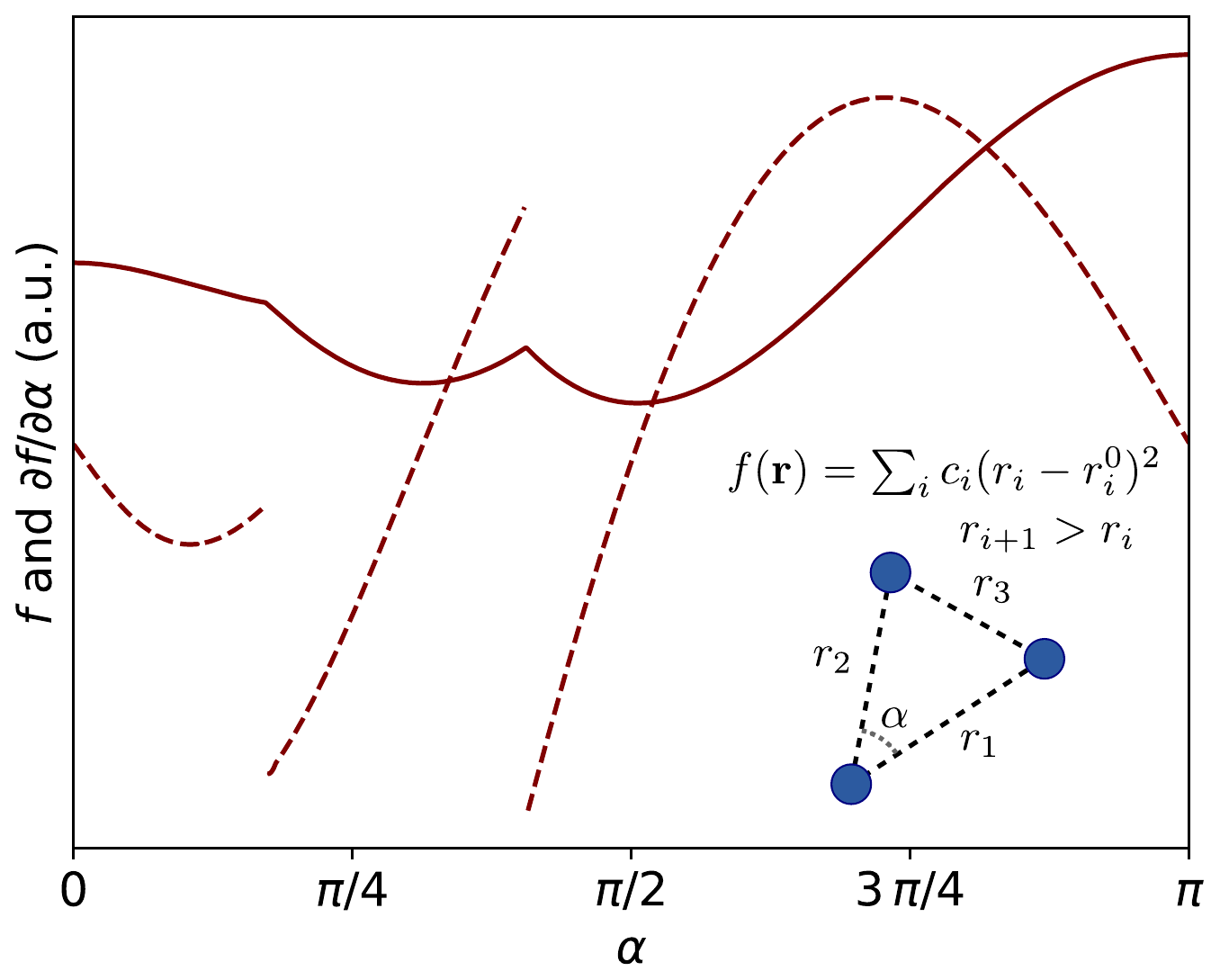}
    \caption{Toy model demonstrating a non-smooth property (solid line) and its discontinuous derivative (dashed line) that are defined as functions of the ordered list of interatomic distances for a three-atom cluster.
    }
    \label{fig:3B_discont}
\end{figure}

Figure~\ref{fig:3B_discont} illustrates the discontinuity of the derivatives of a function that is built from an ordered list of features.
Consider a system of 3 atoms that is uniquely defined by the 3 interatomic distances $r_i$, where the index $i$ denotes the position of the interatomic distance $r_i$ in the ordered list of distances. We define a smooth function of the sorted distances, $f = \sum_i c_i \left(r_i-r_i^0 \right)^2$ parameterized by $\mathbf{c}$ and $\mathbf{r}^0$.
The function $f$ is indeed invariant to the permutations of the atom order in the trimer, but at the price of introducing kinks in $f$ and discontinuities in its derivative when the distance ordering changes. Fitting any smooth function of the trimer geometry by optimizing the parameters $\mathbf{c}$ and $\mathbf{r}^0$ would necessarily lead to poor approximation accuracy.

The lack of regularity has implications for the accuracy and stability of machine-learning models built on such features, as has been shown recently by using a Wasserstein metric to compare Coulomb matrices in a permutation-invariant manner\cite{cayl+20mlst}.
In this context it is worth noting the remarkable connection linking the Euclidean distance between vectors of sorted distances and the Wasserstein distance between radial distribution functions (Section III.F in Ref.~\citenum{will+19jcp}), which builds a formal bridge between conceptually unrelated families of atomic-scale representations.

\subsection{Locality and additivity} \label{sub:locality}

The overwhelming majority of empirical interatomic potentials are expressed as an additive combination of local terms, or of long-range pairwise contributions.
Early models built to fit molecular potential energy surfaces were built explicitly as a function of the coordinates of all atoms in the system.\cite{part-schw97jcp,isch-coll94jcp,ho-rabi96jcp}
Besides the issues of computational cost, this approach is problematic, as it hinders the application of the potential to a molecule with a different number of atoms, or chemical composition.
The work of Behler and Parrinello\cite{behl-parr07prl} did not only have the merit of emphasizing the importance of symmetries in atomistic machine learning, but it also applied to ML interatomic potentials an additive expansion of the molecular energy $E(A)$, writing it as a sum of atom-centered contributions, $E(A)\approx \sum_{i\in A} E(A_i)$.

The notion of an additive decomposition of properties, which is implicit in the functional forms of most interatomic potentials, has far reaching consequences in terms of the data efficiency of the model, as discussed in Subsection~\ref{sub:feature-optimization}.
Combined with the requirement that the atomic contributions only depend on the position of atoms within a finite range of distances, which is needed for the method to be computationally practical and is supported by fundamental physical principles\cite{prod-kohn05pnas}, the additivity assumption breaks down the problem of predicting the properties of a complex structure into simpler, short-range problems.
An additive decomposition is also the most straightforward way to ensure extensivity of predictions\cite{jung+20csc}, i.e. that the prediction of a property for two copies of a molecule at infinite distance from each other is equal to twice the prediction for a single molecule.

It is not by chance that also in the field of molecular machine learning, for which many of the early representations aimed at a \emph{global} description of a molecule\cite{rupp+12prl,fabe+15ijqc,huo-rupp17arxiv,stuk+19jcp}, most of the recent approaches have moved to additive, atom-centered representations\cite{fabe+18jcp,chri+20jcp}, that yield more accurate and transferable models, at least for extensive properties\cite{rama-vonl15cijc}.
Oftentimes it is possible, and relatively straightforward, to modify a global representation to describe an atom-centered environment\cite{barr-kato13prb,sade+13jcp,vand+20mlst}, or to combine atom-centered representations to build a global description\cite{de+16pccp}, e.g. by summing or averaging the values of all the atom-centered features that are present in the structure, as we discuss in Section~\ref{sub:measuring-similarity}.
In fact, one could regard the list of atom-centered features for all the atoms in a structure as an \emph{equivariant} global representation of the structure --  one in which the entries in the feature vector transform according to the permutation of the atomic indices. This notion underlies for instance the concept of self-attention\cite{zhen19jcim,shin2019self}, which has been very fruitfully applied in the construction of neural networks and models for cheminformatics.
The connection between symmetry, locality, additivity, and the nature of the structure-property relation that one wants to model is essential to the construction of effective and transferable machine-learning models.

\subsection{Completeness}

The requirements of symmetry, smoothness and locality can be seen as geared towards reducing the complexity of the structural representation, eliminating redundant structures, reducing the resolution to the intrinsic length scale over which the target property exhibits substantial variations, and breaking down complicated compounds into simple fragments.
This simplification should not, however, come at the expense of the completeness of the representation, meaning that the mapping between Cartesian and feature spaces should keep inequivalent structures distinct.
For example, it has been known for some time that a histogram of interatomic distances (discarding the identity of the connected atoms) is insufficient to fully characterize a structure composed of more than three atoms\cite{bout-kemp04aam,bart+13prb,huan-vonl16jcp}.
More recently, counterexamples have emerged showing that atom-centered correlations -- at least those of low order -- are also insufficient to preserve the injectivity of the structure-feature mapping (see Ref.~\citenum{pozd+20prl} and Section~\ref{sub:geo-completeness} for a more thorough discussion).

Besides completeness in terms of the geometric structure-feature mapping, one should also consider whether \emph{for a chosen regression scheme} the feature-property mapping can be converged to arbitrary accuracy. More complex, non-linear models can often provide good results even when using a representation that involves excessive smoothing, or an highly truncated version of a family of features. The interplay between model and features is discussed in more detail in Section~\ref{sec:models}, and the (largely open) problem of completeness in Section~\ref{sec:incompleteness}.

\section{Symmetrized atomic field representations}\label{sec:symmetry-fields}

As discussed in the previous Section, a multitude of representations have been introduced over the past decade, attempting to incorporate basic principles of symmetry and locality at the very core of atomistic machine learning. The differences between them are much less fundamental than it appears at a first glance, and in fact several works have recently pointed at the existence of a unified framework, in which an explicit formal connection can be established between the vast majority of representations.\cite{will+18pccp, will+19jcp,drau19prb,Bachmayr2019}
In this Section we summarize the construction of a class of features, that we refer to as ``symmetrized atomic field representations'', emphasizing the role played by symmetry and locality, as well as hint to the connection between this class of features and a linear mapping between structure and properties, which is discussed in more detail in Section~\ref{sec:models}.

\subsection{Dirac notation for atomic representations}
\label{sub:bra-ket}

We formalize a notation, that extends the one introduced in Refs.~\citenum{will+18pccp,will+19jcp} and used in Ref.~\citenum{langer2020arxiv} to compare different kinds of local and global representations, which expresses the feature vectors associated with the representation of a structure in a way that mimics Dirac notation in quantum mechanics.
At the most basic level, this notation can be seen as a way to indicate  expressively the nature of the representation used, and to tidily enumerate the components of the associated feature vector.
Much like in the quantum case, the real value of the formalism is that it emphasizes the basis-set independence of the class of representations we concentrate on, and that it provides visual cues that help recognizing at a glance the linear operations that occur in the construction and manipulation of the feature vectors, and of the models built on them.\footnote{To facilitate the use of this notation in \LaTeX documents, we provide a set of macros at \url{https://github.com/cosmo-epfl/cosmo-tools/tree/master/tex/dirac-rep}}
We will use this notation consistently throughout this review as a neutral medium to express general results that reflect concepts shared by many of the most widespread representation, but occasionally make a link to the different notations that have become established to describe specific frameworks. 

\begin{figure}[tbp]
    \centering
    \includegraphics{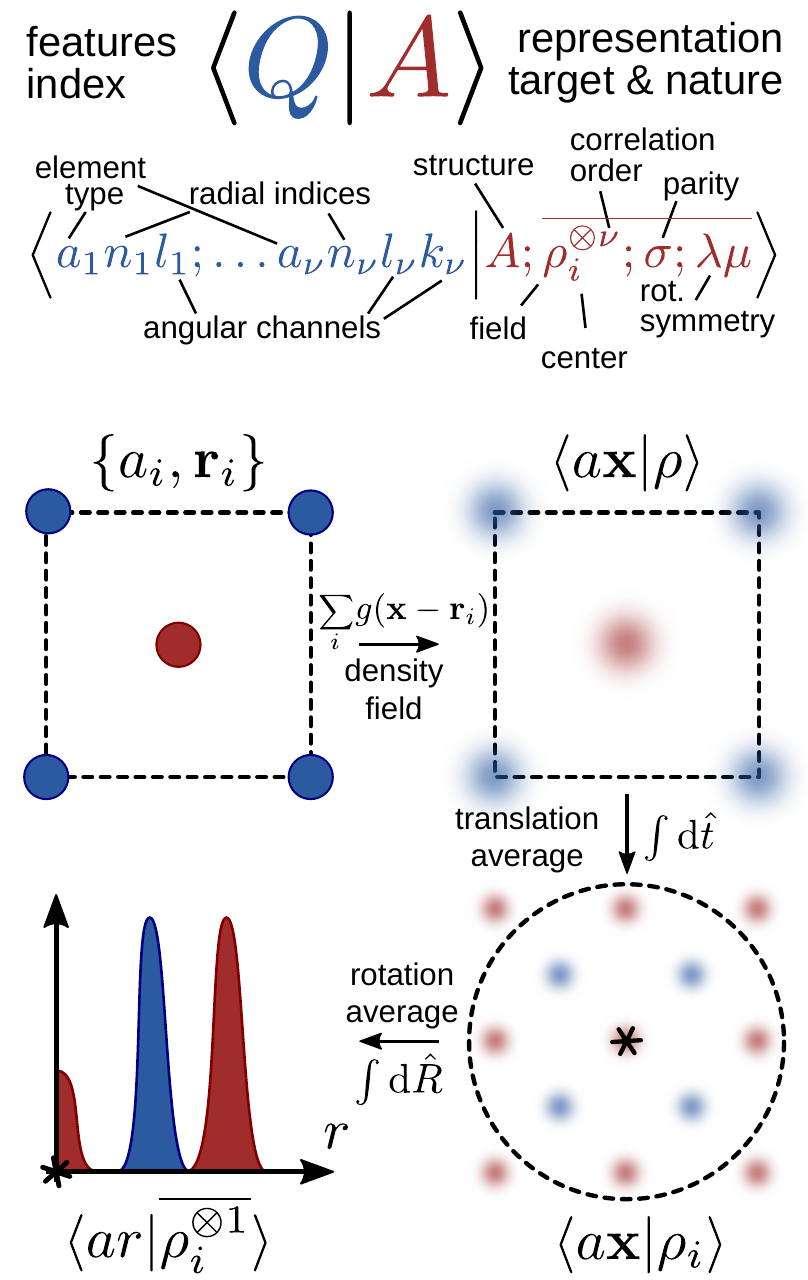}
    \caption{Top: overview of the notation we use to indicate the features that represent an atomistic structure; bottom: summary of the steps in a symmetrized field construction.
    }
    \label{fig:fields}
\end{figure}

\paragraph*{Representations in bra-ket notation}

We use a ket $\rep|A>$ to indicate an abstract feature vector associated with a structure $A$, and -- when necessary -- complement the indication of the structure with one or more symbols and indices (e.g. $\rep|A;\alpha>$) that describe the nature of the representation. 
These indices might specify the portion of the structure the representation refers to, its symmetry properties, or serve as a reminder of the way the representation was constructed. 
When we need to explicitly enumerate the elements of the feature vector, we use one or more indices in the bra, leading to expressions of the form $\rep<\Q||A>$. In this review, we use $\Q$ to indicate a generic continuous index, and $\q$ to indicate a discrete feature index.

Both the ket and the bra indices can (and will) be used with some looseness, to emphasize the most relevant elements of a representation while keeping the notation slim. 
For instance, as shown in Fig.~\ref{fig:fields}, one can indicate explicitly multiple bra indices when their meaning in the definition of a representation is important, separating with a semicolon groups of indices that are conceptually related, or condense them in a compound index when the substructure is irrelevant. 
Occasionally, e.g. when juxtaposing different choices of basis functions, one may also include qualifiers in the bra, e.g. $\rep<n; \text{GTO}|$ to indicate that Gaussian type orbitals are used as a basis.  
Moreover, when discussing the construction of a representation, the reference structure is not important, and so one may drop the structure index from the notation and write $\rep|\alpha>$ instead of $\rep|A; \alpha>$.
Conversely, when the representation of choice is well-established -- e.g. when writing expressions that describe the regression scheme after having discussed the choice of representation -- one may omit the specifics of the representation and write simply $\rep|A>$.

The indices and the qualifiers that are associated with the structure index (typically in the ket) describe the essential nature of the representation and will be reflected in the architecture of a model built on it. 
The indices in the bra, instead, simply enumerate features that are of homogeneous nature, are usually manipulated together in the construction of the model, and can be transformed, contracted or sub-selected in a way that does not change the fundamental properties of the representation.
In many cases, it is possible to describe the construction of a representation as a combination of kets, without indicating explicitly the use of a particular basis.

This notation can be applied in a way that yields usage patterns that are very similar to those that are common in quantum mechanics,  e.g. bra and ket can be interchanged using the convention $\rep<A||\Q> = \rep<\Q||A>^\star$. 
However, much as in the case of the formalism we take inspiration from, a rigorous characterization of the mathematical relations between bras and kets is problematic\cite{gier00rpp}.
It is better to see this notation as a form of symbolic calculus that facilitates memorizing and applying correctly recurring operations and transformations. 
Let us give a few examples, which also provide a reference of how the notation will be applied in this review. 

\paragraph*{Change of basis.}
A change in the basis that is used to practically compute a representation can be written as a linear transformation,
\begin{equation}
\rep<T||A> = \int \D{\Q} \rep<T||\Q> \rep<\Q||A>,
\end{equation}
where $\rep<T||\Q>$ indicates the coefficients that enact the change of basis. This kind of manipulations will be used in Section~\ref{sub:lm-rep} to convert between a real-space description of the atom-centred density and one based on radial functions and spherical harmonics.\footnote{Note that, much as it is the case in quantum chemistry, non-orthogonal bases introduce some ambiguity in the bra-ket notation, because the coefficients in the expansion of a function in the basis differ from the scalar product between the basis and the function -- the two being related by the overlap matrix of the basis. 
The notation should be treated with some care when translating it into a practical implementation if the basis used is not orthonormal.}
All of the expressions discussed here as integrals over a continuous index can be formulated as sums over (finitely or infinitely) countable, discrete indices
\begin{equation}
\int \D{\Q} \rep|\Q>\rep<\Q| \sim \sum_\q \rep|\q>\rep<\q|.
\end{equation}

\begin{figure*}[tbp]
    \centering
    \includegraphics[width=0.9\linewidth]{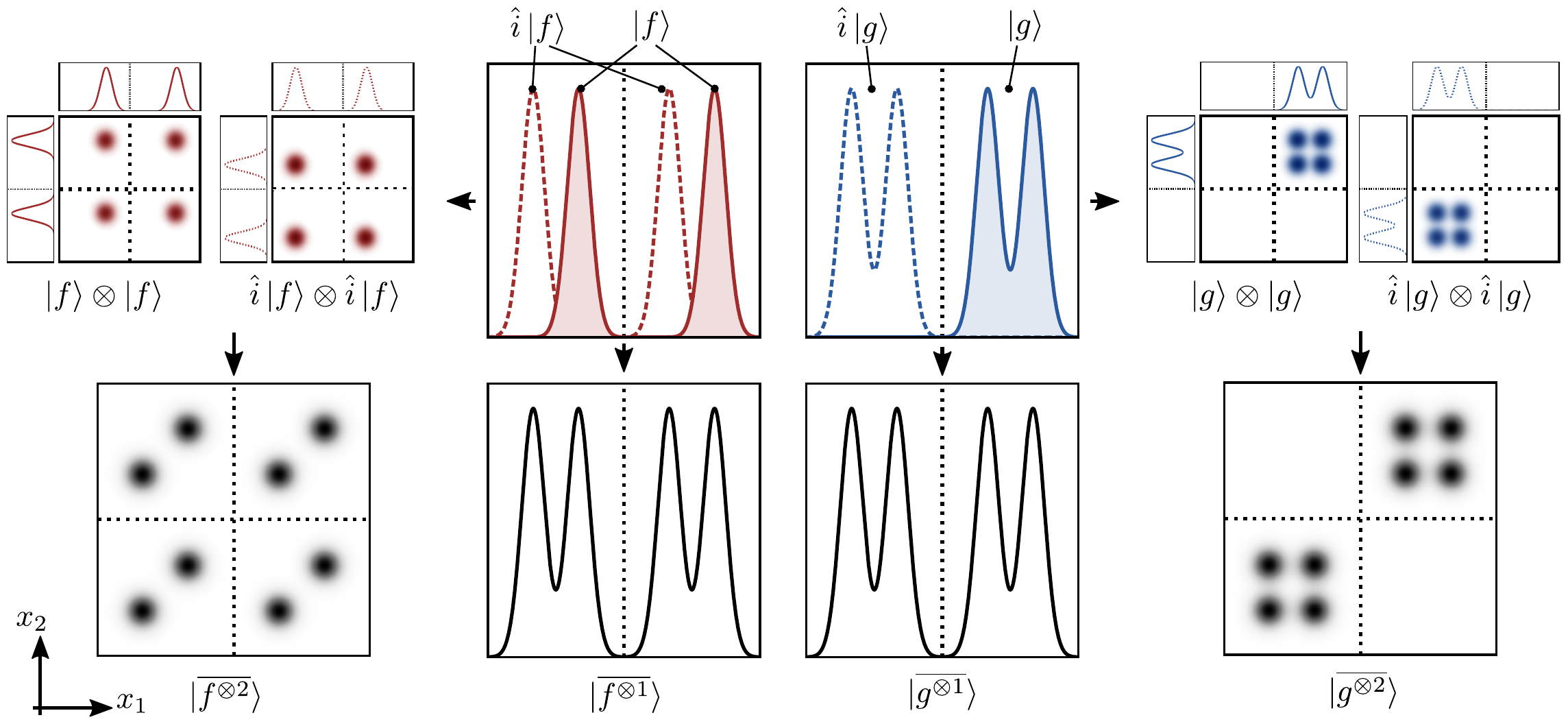}
\caption{To obtain features that are invariant to inversion with respect to the vertical dotted line, Haar integration over the symmetry group in this case just corresponds to summing over two symmetry related images. Starting from two distinct functions $\rep|f>$ (left panels, red) and $\rep|g>$ (right panels, blue), the functions (full lines) and their mirror transformation (dotted lines) are summed to obtain invariant features (bottom row).
Direct symmetrization is depicted in the central panels, yielding $\rep|\field{f}^1>$, while the external panels visualize the construction of tensor-product features, their symmetrization and summation, yielding $\rep|\field{f}^2>$. }
    \label{fig:inversion_invariance}
\end{figure*}

\paragraph*{Scalar product and kernels.}
The scalar product between the features of two structures $A$ and $A'$ can be written using a complete basis indexed by $Q$ as
\begin{equation}
\rep<A||A'> =  \int \D{\Q}
\rep<A||\Q> \rep<\Q||A'>,
\end{equation}
where one recognizes an expression that is reminiscent of a completeness relation $\int \D{\Q} \rep|\Q>\rep<\Q| = 1 $.
This definition only holds for a complete, orthogonal basis and might entail an approximation when computed with a finite basis. 
The notation $\rep<A||A'>$ can  also be used to refer to a kernel  $\krn(A,A')$ that expresses the similarity between two configurations; this is obvious when considering a linear kernel, but can also be used for non-linear kernels, keeping in mind that it might not be possible to write explicitly the features that correspond to the Hilbert space that reproduces the kernel.\cite{aron50tams} 

\paragraph*{Linear models.}
The bra-ket notation implicitly assumes linearity in the transformation between different choices of basis, and in the modeling of target properties. 
Even though the features can be used as an input of an arbitrarily complex nonlinear regression scheme (see Section~\ref{sub:non-linear-model}), we will often investigate their behavior in the context of linear models, because they reveal more transparently how a given representation reflects structure-property relations.
When using a representation $\rep|A; \alpha>$ to describe structures, a linear model for a property $\y(A)$  can be written as
\begin{equation}
 \y(A) \equiv \rep<\y||A> \approx \int \D{\Q} \rep<\y; \alpha||\Q> \rep<\Q||A; \alpha>, \label{eq:dirac-lr}
\end{equation}
where $\rep<\y; \alpha||\Q>$ indicates the regression weights for a model based on $\rep|A; \alpha>$. 
Leaving aside (important) issues related to regularization, this expression emphasizes that one can transform simultaneously the weights and the features to a different basis, and the predicted value is unchanged.
The expression $\rep<\y||A>$ can also be seen as a hint of the fact that a collection of properties could be used as descriptors for a structure $A$, although this is an approach we only discuss briefly in this review.

\paragraph*{Tensor product.}
A pattern we use frequently in what follows, and that mimics a construction used in quantum mechanics, is the combination of multiple kets to build a tensor-product space, e.g.
\begin{equation}
\rep|(A; \alpha) \otimes (A'; \alpha')> = \rep|A; \alpha> \otimes \rep|A'; \alpha'>. \label{eq:dirac-tensorproduct}
\end{equation}
The construction of a tensor-product representation is well-defined even without indicating explicitly the basis used to describe either side of Eq.~\eqref{eq:dirac-tensorproduct}, and it is often possible to use either an explicit Cartesian product of the bases on the right-hand side, or a combined basis
\begin{equation}
\rep<\Q_1; \Q_2||A\otimes A> \equiv \rep<\Q_1||A> \rep<\Q_2||A> \rightarrow \rep<T||A\otimes A>, \label{eq:dirac-tensorproduct-bras}
\end{equation}
using only $\rep|A>$ as a special case of Eq.~\eqref{eq:dirac-tensorproduct} in which  $A\equiv A'$, and $\alpha\equiv \alpha'$ can be omitted.

\paragraph*{Operators and symmetry averages.}
Finally, we can consider the action of an ``operator'' on a ket, that is to be interpreted as a linear map that transforms the atomic structure.
Taking for instance the operator $\ihat$ associated with inversion symmetry, $\ihat\rep|A>$ indicates the representation associated with structure $A$ after the coordinates of all atoms have been reflected relative to the origin.
Much as in quantum mechanics, the operator can also be applied to the bra, where it corresponds to a transformation of the basis. In terms of symmetry operations, this corresponds to the active or passive transformations, acting on the structure or on the reference frame.
By summing over the operators associated with a  symmetry group, an operation which is also referred to as Haar integration\cite{nachbin1976haar}, one can build symmetrized representations that are covariant under the actions of the elements of the group, e.g. for the $C_i$ point group, 
\begin{equation}
\rep|\ev{A\otimes A}_{C_i}; \sigma> = \rep|A>\otimes \rep|A> + \sigma ( \ihat\rep|A>\otimes \ihat\rep|A> ).
\end{equation}
The index $\sigma$ takes the value $-1$ for representations that change sign under inversion, and $+1$ for invariant features; in the invariant case, $\sigma$ may be omitted. 
When the resulting symmetric representation is used often, and the symmetry group is clear from the context, we indicate the averaging with an overline and omit the explicit indication of the group it has been symmetrized over, e.g., $\rep|\ev*{A\otimes A}_{C_i}>\rightarrow\rep|\field{A}[A]>\rightarrow\rep|\field{A}^2>$.
Figure~\ref{fig:inversion_invariance} illustrates the notation and the Haar integration in one dimension. Two distinct functions, $f$ and $g$ are plotted using their usual real-space features, $f(x) \equiv \rep<x || f >$ and $g(x) \equiv \rep< x || g>$. Applying inversion yields $\rep<x|\ihat\rep|f> = f(-x)$. An inversion-invariant feature can be created by symmetrizing:
$\rep|\field{f}^1>  = \rep|f> + \hat{i} \rep|f>$, but our choice of $f$ and $g$ leads to a degenerate description, as $\rep|\field{f}^1>=\rep|\field{g}^1>$. A second order feature may be obtained by generating the tensor product of the functions, e.g. $\rep|f>  \otimes \rep|f> $, which in real space results in $\rep<x_1;x_2 || f \otimes f > \equiv f(x_1) f(x_2)$.
Symmetrizing this tensor product yields features $\rep|\field{g}^2>$ and $\rep|\field{f}^2>$ that  are also inversion-invariant, but are still able to distinguish between the two functions.

\paragraph*{An example: SOAP in bra-ket notation}

To give a concrete example of the use of this formalism, let us compare the functional notation used in Refs.~\citenum{de+16pccp,bart+13prb} to indicate the components of a SOAP feature vector with the corresponding bra-ket notation. 
The reader who is unfamiliar with the SOAP construction will find  the remainder of this Section, and in particular Section~\ref{sub:lm-rep}, to give a very detailed account of this family of features, and might better skip this brief overview, that assumes knowledge of the derivation from Ref.~\citenum{bart+13prb}. 
The SOAP power spectrum describes the two-point correlations between the atom density centered around the $i$-th atom of structure $A$, expanded in terms of atomic species (labeled by the indices $\e_{1,2}$), radial basis functions (labeled by $n_{1,2}$) and angular momentum channels (labeled by $l$). 
The density expansion coefficients can be written as
\begin{equation}
\begin{split}
\rep<\enlm||A; \rho_i> =& \int \D{\bx} \rep<n||x> \rep<lm||\bxhat> \rep<\ex||A; \frho_i> \\
&\hfill \vertequiv \hfill \\
c^{i,\e}_{\nlm} = &
\int\D{\bx} R_n(x)^\star \Ylm(\bxhat)^\star \rho_i^\e(\bx).
\end{split}
\end{equation}
In this expression, $\rep<\ex||A; \frho_i>\equiv \rho^{i,\e}(\bx)$ indicates the atom-centred density, $\rep<x||n>\equiv R_n(x)$ an orthonormal set of radial functions, and $\bra{\bxhat}\ket{lm}\equiv \Ylm(\bxhat)$ the spherical harmonics. 

The SOAP features for the environment $A_i$ can be written as
\begin{multline}
\rep<a_1 n_1; a_2 n_2; l||A; \frho_i^2>  \propto \hfill \quad\\ \sum_m 
\rep<A; \rho_i||\en_2 l m >\rep<\en_1 l m||A; \rho_i>
\\
\hfill \vertequiv \hfill \\ 
p^{i,a_1a_2}_{n_1 n_2 l} \propto  \sum_m c^{i,\e_1}_{n_1 lm} (c^{i,\e_2}_{n_2 lm})^\star.
\label{eq:dirac-soap}
\end{multline}
In the functional notation, one relies on the convention that $c$ corresponds to the density expansion coefficients and $p$ to the power spectrum, while the Dirac notation uses the more expressive symbols $\rho_i$ to indicate the $i$-centered atom density, and $\frho_i^2$ as a reminder that SOAP features can be derived as a symmetry-averaged 2-point correlation of $\rep|\rho_i>$.
This expanded notation is indicative of the place of the SOAP powerspectrum in the hierarchy of density-correlation features, and is useful to distinguish between different kinds of features (radial correlations, power spectrum, bispectrum \ldots). When it is clear that one is only using one type of representation, the compact (and generic) form $\rep|A_i>$ can be used instead.
When it comes to the indices labelling different features, the functional notation mixes the indices ($\e$) associated with the chemical species of the neighbors and the index $i$ of the central atom, separating them from those associated with the radial channel ($n$). This reflects how SOAP was originally introduced to describe single-element systems. In the Dirac notation, on the other hand, the $(\en_1)$ and $(\en_2)$ indices are grouped together to indicate that they are conceptually linked in the construction as a tensor product of two densities, and the index indicating the identity of the central atom is associated with the ket.

\subsection{Global field representations}

The starting point for the construction of a symmetry-adapted field representation is a field that describes the structure in terms of the distribution of its atoms -- or, more generally, of points that are associated with the building blocks of the material, as one would have in a coarse-grained model. In the simplest possible case, one would take localized functions $g$ centered on each atomic position $\br_i$ and define
\begin{equation}
\rep<\bx||A;\frho>
\equiv \sum_{i\in A} \rep<\bx||\br_i; g>, \label{eq:a-rho-simple}
\end{equation}
where $\rep<\bx||\br_i; g>\equiv g(\bx-\br_i)$ is a localized function (e.g. a Gaussian) centered on the $i$-th atom, and the $\rho$ in the ket indicates the kind of field used to describe the structure. As we discuss in more detail in Section~\ref{sub:lm-rep}, the atomic density functions can be either finite-width Gaussians, which leads to representations akin to SOAP features\cite{bart+13prb}, or Dirac $\delta$ distributions, which recovers representations similar to the current implementation of moment tensor potentials\cite{shap16mms} or the atomic cluster expansion\cite{drau19prb}. 
To indicate the $g\rightarrow\delta$ limit, we use the notation $\rep|\rho>\rightarrow\rep|\delta>$.
Atoms, or more generally, ``point particles'' such as those one could associate to a coarse grained description of a molecular system, can be further characterized by internal attributes, that could be discrete (e.g. the chemical nature of an atom, or a molecule, which we indicate as $\e_i$) or continuous (e.g. an atomic or molecular dipole $\bu_i$)
\begin{equation}
\rep<\e\bu\bx||A; \field{\rho u}> \equiv \sum_{i\in A} \delta_{\e \e_i} \rep<\bu||\bu_i; g> \rep<\bx||\br_i; g>.\label{eq:a-rho-messy}
\end{equation}
In this form, Eq.~\eqref{eq:a-rho-messy} can be seen as an abstraction of the many real-space ``voxel'' representations of materials,\cite{kaji+17sr,Noh2019} that are used often in the context of generative models and reinforcement learning\cite{chri+20jcp2}.
The ket $\rep|A; \rho>$ defined by expressions like~\eqref{eq:a-rho-simple} or~\eqref{eq:a-rho-messy} could be equally well expressed in a different basis, e.g. expanded in plane waves
\begin{equation}
\rep<\bk||A; \rho> =\frac{1}{(2\pi)^{3/2}} \int\D{\bx} e^{-\I \bk\cdot\bx} \rep<\bx||A; \rho> =\sum_{i\in A} \rep<\bk || \br_i; g>, \label{eq:a-rho-k}
\end{equation}
which also shows how the change of basis can be applied directly to the atom-centred density contributions.
Eqs.~\eqref{eq:a-rho-simple} and~\eqref{eq:a-rho-k} contain the same amount of information, and can be seen as special cases of a formal definition of the representation for the structure $A$ as a sum of atomic representations,
\begin{equation}
\rep|A; \rho> =\sum_{i\in A} \rep| \br_i; g>. \label{eq:a-rho-formal}
\end{equation}
Even though the choice of a basis can be very important to simplify  analytical derivations or practical implementation, representations can be regarded as abstract objects that can be defined independently of the basis set, much as it is the case for the wavefunction in quantum mechanics.

\subsection{Translational invariance and atom-centered features}\label{sub:translational-sym}
One way to make $\rep<\bx||\rho>$ translationally invariant is to sum over the continuous translation group, $\int\D{\That}\rep<\bx|\That\rep|\rho>$. Summing directly over the atom density eliminates all structural information, because $\int\D{\That}\rep<\bx|\That\rep|\br_i; g>= \int \D{\mbf{t}} g(\mbf{t}-\br_i)=1$. Information loss is a usual issue with Haar integration, as exemplified in Figure~\ref{fig:inversion_invariance}.
One can avoid or reduce it by summing over \emph{tensor products} of the atom density field.
Considering the case in which atoms are described only by their position and chemical identity, integrating over translations $\That$ yields a two-point density correlation function
\begin{multline}
\rep<\ex_1; \ex_2||\ev*{\rho\otimes \rho}_{\Rthree}> \equiv \rep<\ex_1; \ex_2||\frho^2> \\
=
\int \D{\That}
\rep<\ex_1|\That\rep|\frho>
\rep<\ex_2|\That\rep|\frho>\\
= \sum_{ij}  \delta_{\e_1\e_j} \delta_{\e_2\e_i}
\int \D{\That} \rep<\bx_1 - \mbf{t}||\br_j; g> \rep<\bx_2 - \mbf{t}||\br_i ; g>
\\
\propto \sum_{ij} \delta_{\e_1\e_j} \delta_{\e_2\e_i}
\rep<(\bx_1-\bx_2)|| (\br_j-\br_i); \tilde{g}>
\label{eq:rho2t}
\end{multline}
where $\tilde{g}$ indicates the cross-correlation of two of the localized density functions. In the case of a Gaussian density, $\tilde{g}$ is simply a Gaussian with twice the variance, and outside this section we will use just $g$ to indicate the atomic density both in $\rep|\rho>$ and $\rep|\rho_i>$.
As a remindet that the representation has been obtained by averaging over translations the tensor product of two density fields, we use the superscript notation $\frho^2$, and we separate with a semicolon groups of feature indices that are associated with each factor in the tensor product, as discussed in Section~\ref{sub:bra-ket}.
Note that the representation in Eq.~\eqref{eq:rho2t} has a large null space, as it depends only on $\bx_1-\bx_2$.
One could then re-define it by labelling features using a single position vector, or transform it in a plane wave basis:
\begin{multline}
\rep<\e_1; \e_2; \bk||\frho^2> =
\int \D{\bx} e^{-\I\bk\cdot\bx}
\rep<\e_1 \mbf{0}; \e_2 \bx||\frho^2>\\
= \rep<\e_1 \bk||\rho>^\star  \rep<\e_2 \bk||\rho>
\end{multline} 
where the second equality is a consequence of the convolution theorem.
One sees that the translationally-symmetrized density is essentially equivalent to the diffraction pattern of the atomic structure $I(\mathbf{k})$, that has been already used as a descriptor to classify crystalline configurations.\cite{zile+18nc}

This construction can be taken as an inspiration to introduce an atom-centered representation
\begin{equation}
\rep<\ex||A;\frho_i> =
\sum_{j\in A} \delta_{\e\e_j} \rep<\bx ||\br_{ji};\tilde{g}>, \label{eq:rhoi-ket}
\end{equation}
where $\br_{ji}=\br_j -\br_i$. The fact that $\rep|A;\frho_i>$ is atom centered (and hence translationally invariant) is hinted at by the subscript notation $\frho_i$, and so in what follows we only use this subscript to distinguish it from its non-symmetrized counterpart~\eqref{eq:a-rho-simple} and simultaneously to indicate the central atom index.
When expressing a representation centered around atom $i$ without emphasis on its precise nature, we will use the notation $\rep|A_i>$. %

Writing the symmetrized two-point density correlation in terms of Eq.~\eqref{eq:rhoi-ket} clarifies how an atom-centered representation is a natural consequence of the translational symmetrization:
\begin{equation}
\rep<\ex_1; \ex_2||A;\frho^2> = \sum_{i \in A} \delta_{\e_2\e_i} \rep<\e_1 (\bx_1-\bx_2)||A;\frho_i>.
\label{eq:additive-rho2}
\end{equation}
When building a linear model, this expression implies an additive decomposition of the target property, as well as the use of separate models depending on the nature of the central atomic species:
\begin{equation}
\begin{split}
\rep<\y||A> \approx & \sum_{i\in A} \rep<\y; \e_i||A_i> \\
&= \sum_{i\in A} \sum_{\e} \int \D{\bx}
\rep<\y; \e_i||\ex>
\rep<\ex||A;\frho_i>.
\end{split}
\end{equation}
Note that in this case we assume that only the regression weights depend on the nature of the central atom, but one might as well fine-tune the atom-centred features depending on the central atom.
As discussed in Section~\ref{sub:linear-models}, this expression can be taken as the prototype of all pair potentials, and higher-order of many-body interaction can be incorporated by taking higher tensor powers before symmetrization, or in the subsequent step of rotational averaging.
Localization can be enforced by introducing a cutoff function in the definition~\eqref{eq:rhoi-ket}. This is far from being an inconsequential operation, as it introduces an error: atomic energies and properties cannot depend on neighbors farther than this limit, as one can measure in terms of the locality of the response of forces to atomic displacements of neighbors\cite{deri-csan17prb}. 
However, introducing a relatively short-range cutoff often results in more robust models, which perform better in the data-poor regime. We discuss this in more detail in Section~\ref{sub:feature-optimization}.

\subsection{Rotational invariance and body-ordered representations}
\label{sub:body-order-representations}

The atom-centered representation~\eqref{eq:rhoi-ket} is translationally invariant, but does depend on the orientation of the structure. One should then proceed to perform Haar integration over the rotation group and (possibly) over inversion.

\begin{figure}
    \centering
    \includegraphics[width=1.0\linewidth]{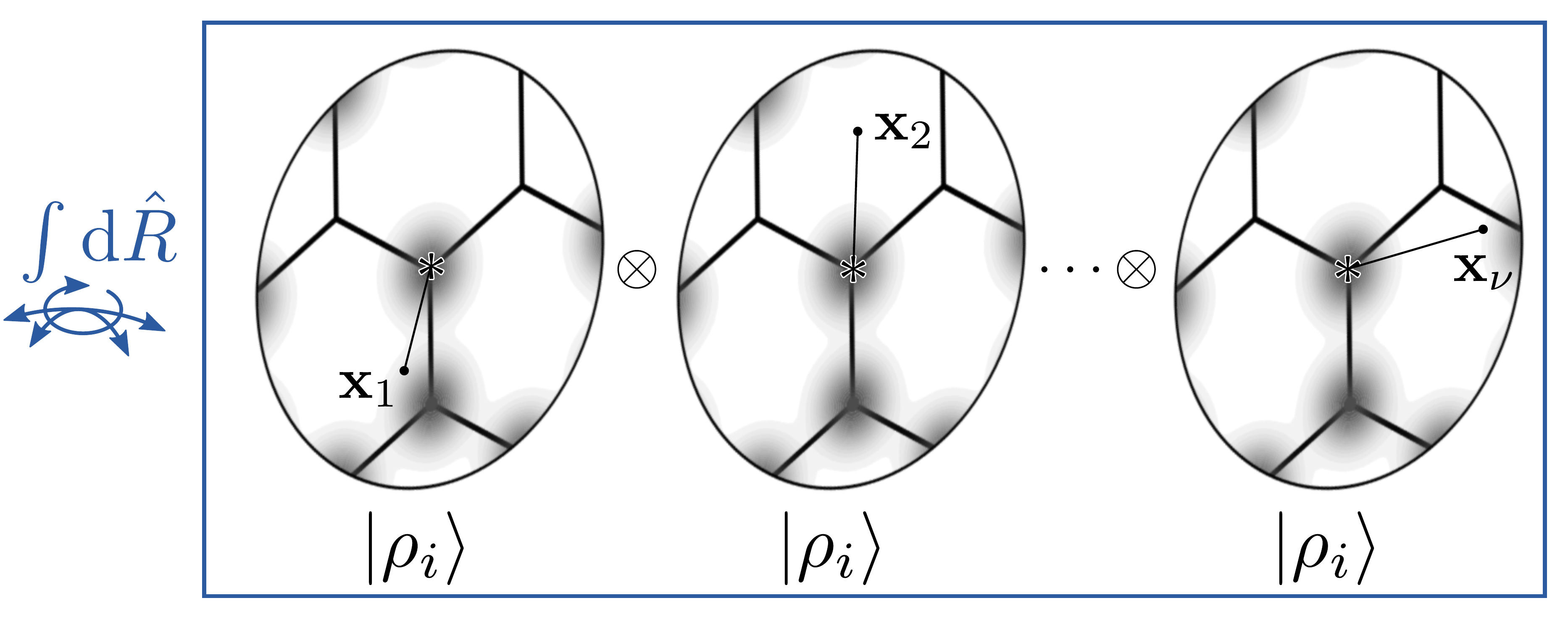}
    \caption{Graphical scheme of the construction of a $\SOthree$-symmetrized tensor product representation. Copies of the atom-centered density are evaluated at $\nu$ separate points, and the tensor product is averaged by simultaneously rotating all densities.  }
    \label{fig:rho-average}
\end{figure}

We can define the ($\nu+1$)-body order symmetrized field representation as 
\begin{multline}\label{eq:body-order}
\rep|\frho_i^{\nu}> \equiv \rep|\ev*{\underbrace{\rho_i\otimes\cdots\otimes\rho_i}_{\nu\ \text{times}} }_\Othree > \\= \sum_{k=0,1}\int_{\SOthree}\hspace{-4mm}\D{\Rhat}\,
\ihat^k\Rhat\rep|\frho_i> \otimes\ldots\otimes
\ihat^k\Rhat\rep|\frho_i>.
\end{multline}
This can be expanded on an explicit position basis 
\begin{multline}\label{eq:body-order-x}
\rep<\ex_1; \ldots \ex_{\nu}||\frho_i^{\nu}> \\
=\sum_{k=0,1}\int_{\SOthree} \hspace{-4mm}\D{\Rhat}
\rep<\ex_1|\ihat^k\Rhat\rep|\frho_i> \ldots
\rep<\ex_{\nu}|\ihat^k\Rhat\rep|\frho_i>,
\end{multline}
emphasizing that $\rep|\frho_i^\nu>$ corresponds to a symmetrized, $\nu$-point correlation of the atom density centered on the $i$-th atom (Fig. \ref{fig:rho-average}) -- a $(\nu+1)$-point correlation function, in the language used in statistical mechanics to describe the structure of liquids\cite{ande-chan72jcp,chan87book}.
Similar to the case of Eq.~\eqref{eq:rho2t}, this object has a large null space (e.g. in the $\nu=1$ case it only depends on $x_1=\left|\bx_1\right|$). As discussed in Ref.~\citenum{will+19jcp}, one can choose a more concise enumeration of the real-space correlations in terms of distances and angles, that reduces in the limit $g \to \delta$ to a sum over distances and angles between atoms. For instance, for the $\nu=2$ case one can write
\begin{multline}
\rep<\e_1 r_1; \e_2 r_2; \omega||\fdelta_i^2> \\ \propto
\sum_{jj'} \delta_{\e_1 \e_j} \delta_{\e_2 \e_{j'}}
\delta(r_1 - r_{ji}) \delta(r_2 - r_{j'i}) \delta(\omega - \brhat_{ji}\cdot\brhat_{j'i}),
\label{eq:rho2-delta}
\end{multline}
where we use $\rho\rightarrow \delta$ to indicate that the correlation function is built on the Dirac-$\delta$ limit of the atom density field.
Expressions of this kind reveal the close connection between symmetrized-field representations and atom-centered symmetry functions\cite{behl-parr07prl,behl11jcp,smit+17cs}, as well as equivalent constructions such as those used in the ANI\cite{smit+17cs} and DeepMD\cite{zhan+18prl} frameworks, and the FCHL features\cite{fabe+18jcp,chri+20jcp}.
Features that describe a chemical environment are written as a sum over tuples of neighbors of appropriate functions of their distances and angles, and can be seen as just a different choice of basis set for Eq.~\eqref{eq:rho2-delta}
\begin{multline}
\rep<\e_1\e_2 k||\fdelta_i^2> =  \int \D{r_1}\D{r_2}\D\omega \\ \times \rep<k; G^3||r_1 r_2 \omega> \rep<\e_1 r_1; \e_2 r_2; \omega||\fdelta_i^2>\\
\hfill \vertequiv \hfill \\
\sum_{jj'} \delta_{\e_1 \e_j} \delta_{\e_2 \e_{j'}} G^3_k(r_{ji}, r_{j'i}, \brhat_{ji}\cdot\brhat_{j'i}), \label{eq:rho2-g3}
\end{multline}
that demonstrates the connection between density correlations and atom-centered symmetry functions computed as a sum over groups of neighbors following the notation used in Ref.~\citenum{behl11jcp}.

Note that we choose to symmetrize the atom-centered description $\rep|\frho_i>$ -- given that this is the procedure that recovers most of the existing representations -- but one could as well proceed by averaging over tensor products of the translationally invariant representation of the full structure $\rep|\frho^2>$
\begin{multline}\label{eq:global-rot-rep}
\rep|\ev*{ \ev*{\rho\otimes\rho}_{\mathbb{R}^3} \otimes \ev*{\rho\otimes\rho}_{\mathbb{R}^3}}_\SOthree >\sim\\ \int_{\SOthree}\hspace{-4mm}\D{\Rhat}\, \sum_{ii'}
\Rhat\rep|\frho_i> \otimes
\Rhat\rep|\rho_{i'}>,
\end{multline}
as it was done for instance in Ref.~\citenum{mavr+18jpcl}.
Doing so results in the appearance of cross terms involving correlations between densities centered on different atoms, which could be used to  systematically incorporate in this framework machine-learning approaches based on convolutional, and message-passing, neural networks that combine information centered on neighboring atoms.\cite{schu+18jcp,ande+19nips}

\subsection{Density correlations in an angular momentum basis}
\label{sub:lm-rep}

More concise (and easier to evaluate) expressions for the density correlation representations can be obtained with a change of basis.
Using orthonormal radial functions $R_n(x)\equiv \bra{x}\ket{n}$ and spherical harmonics $\Ylm(\bxhat)\equiv \bra{\bxhat}\ket{lm}$ yields a discrete set of coefficients that transform as spherical harmonics
\begin{multline}
\rep<\enlm||A; \frho_i> =
\int \D{\bx} \rep<n||x> \rep<lm||\bxhat> \rep<\ex||A; \frho_i>\\
= \sum_{j\in A_i} \delta_{aa_j}
\int \D{\bx}
\rep<n||x> \rep<lm||\bxhat> \rep<x\bxhat||\br_{ji}; g>\\
=\sum_{j\in A_i} \delta_{aa_j}
\rep<nlm||\br_{ji}; g>,
\label{eq:density-nlm}
\end{multline}
where $\rep<nlm||\br_{ji}; g>$ corresponds to the expansion in radial functions and spherical harmonics of a Gaussian centered on the interatomic vector $\br_{ji}$. These expansion coefficients can be seen as functions of $\br_{ji}$, enumerated by the indices $(n,l,m)$, that can be evaluated numerically or analytically, depending on the choice of basis (see Section~\ref{sub:implementation} for a few examples).

The use of spherical harmonics $\rep|lm>$ for the angular basis is natural, and makes it easy to evaluate the rotational integral of \cref{eq:body-order} analytically, because the matrix elements $\mel{lm}{\Rhat}{l'm'}=\delta_{ll'}D^l_{m'm}(\Rhat)$ correspond to  Wigner-D matrices, an irreducible representation of $\SOthree$.
Well-known results from the theory of angular momentum,\cite{Thompson2004} such as the orthonormality and the product reduction formula for  Wigner-$D$ matrices, allow deriving explicit expressions for the symmetrized field representations of order $\nu=1,2,3$
\begin{equation}
\rep<\enlm_1||\frho_i^1> = \frac{8\pi^2}{2l_1+1}
\rep<\enlm_1||\frho_i> \delta_{l_10}\delta_{m_10}
\label{eq:nu1-basis-uncoupled}
\end{equation}
\begin{multline}
\rep<\enlm_1;\enlm_2||\frho_i^2> = \delta_{l_1l_2}\delta_{m_1m_2}\frac{8\pi^2}{2l_1+1} \\
\sum_s (-1)^{s-m1}
\rep<\enl_1 s||\frho_i> \rep<\enl_2 (-s)||\frho_i>,
\label{eq:nu2-basis-uncoupled}
\end{multline}
\begin{multline}
\rep<\enlm_1; \enlm_2; \enlm_3||\frho_i^3> = \\
\frac{8\pi^2}{2l_1+1} (-1)^{-m_1} \cg{\lm_2}{\lm_3}{l_1(-m_1)}\\
\sum_{s_1 s_2 s_3}
(-1)^{-s_1} \cg{l_2 s_2}{l_3s_3}{l_1 (-s_1)}
\rep<\enl_1 s_1||\frho_1> \\
\rep<\enl_2 s_2||\frho_2> \rep<\enl_3 s_3||\frho_3>,
\label{eq:nu3-basis-uncoupled}
\end{multline}
where $\cg{\lm_1}{\lm_2}{LM}$ is a Clebsch–Gordan coefficient.

Much as it was the case for the real-space versions of the density correlation representations, there are  several redundant indices in these expressions, resulting from the rotational averaging that leaves some of the $m_i$ as free parameters.
We can then re-label the invariant features, in a way that emphasizes the connection to existing representations, by coupling the angular basis and absorbing some of the inconsequential constant factors.
For the case $\nu=1$ one can define
\begin{equation}
\rep<\en||\frho_i^1>=\rep<\e n 0 0||\frho_i>
\label{eq:nu1-basis-coupled}
\end{equation}
which corresponds to a discretized version of a pair correlation function
\begin{multline}
\rep<\en||\frho_i^1> =
\int \D{\bx} \rep<n||x> \rep<00||\bxhat> \rep<\e (x\bxhat)||\frho_i> \\
\propto \int \D{x} x^2 \rep<n||x> \int \D{\bxhat} \rep<\e (x\bxhat)||\frho_i>\\ 
\sim  \int \D{r} r^2 R_n(r)^\star g_{\e}(r),
\end{multline}
in which we use the usual notation  $g_{\e}(r)$ to indicate the distribution of $\e$ atoms (although in this case it is restricted to an $i$-centered environment rather than averaged over an equilibrium distribution). 
For the $\nu=2$ case, \cref{eq:nu2-basis-uncoupled} can be redefined as
\begin{multline}
\rep<\ennl||\frho_i^2> = \frac{(-1)^l}{\sqrt{2l+1}} \\
\sum_m (-1)^m \rep<\en_1 lm||\frho_i> \rep<\en_2 l(-m)||\frho_i>
\label{eq:nu2-basis-coupled}
\end{multline}
This corresponds -- modulo irrelevant constants -- to the rotation invariant 3D shape descriptor\cite{Rusinkiewicz2003} and to the SOAP features, which would be written, in the notation of Refs.~\citenum{bart+13prb,de+16pccp} as
\begin{equation}
p^{i, \e_1\e_2}_{n_1 n_2 l} = \frac{1}{\sqrt{2l+1}} \sum_m
c^{i, \e_1}_{n_1lm}(c^{i,\e_2}_{n_2lm})^\star,
\end{equation}
where $c_{\nlm}^{i,\e}=\rep<\enlm||\rho_i>$ indicate the density expansion coefficients following the same notation. 
The $\nu=2$ representation can also be written on a real-space basis as $\rep<\e_1r_1; \e_2 r_2; \omega||\frho_i^2>$, emphasizing its nature as three-body density correlation function that depends on two distances $r_1$, $r_2$ and the cosine $\omega$ of the angle between the directions along which they are evaluated.
The 4-body order invariant representation becomes
\begin{multline}
\rep<\enl_1; \enl_2; \enl_3||\frho_i^3> = \frac{(-1)^{l_3}}{\sqrt{2l_3+1}}\\
\sum_{m_1 m_2 m_3} (-1)^{m_3} \cg{\lm_1}{\lm_2}{\lm_3}  \rep<\enlm_1||\frho_i> \\[-2mm]
\rep<\enlm_2||\frho_i> \rep<\enl_3 (-m_3)||\frho_i>
\label{eq:nu3-basis-coupled}
\end{multline}
corresponding  to the SOAP bispectrum\cite{bart+13prb}
\begin{multline}
b^{i,\e_1\e_2\e_3 }_{\nl_1 \nl_2 \nl_3} = \frac{1}{\sqrt{2l+1}} \sum_{m_1 m_2 m_3}  \cg{\lm_1}{\lm_2}{\lm_3} \\
c^{i,\e_1}_{\nlm_1} c^{i,\e_2}_{\nlm_2} (c^{i,\e_3}_{\nlm_3})^\star,
\end{multline}
and closely related to the bispectrum used in the spectral neighbor analysis method\cite{thom+15jcp,wood-thom18jcp}, which is essentially equivalent to a different choice of basis.
As discussed in more detail in Ref.~\citenum{niga+20jcp} and in the next sections, the relationship between the redundant expressions~(\ref{eq:nu1-basis-uncoupled}, \ref{eq:nu2-basis-uncoupled}, \ref{eq:nu3-basis-uncoupled}) that arise from the integral over rotations, and the more concise versions~(\ref{eq:nu1-basis-coupled}, \ref{eq:nu2-basis-coupled}, \ref{eq:nu3-basis-coupled}) can be seen as a transformation from the uncoupled to the coupled angular momentum basis, and starting from the $\nu=4$ additional indices $k_\nu$ must be included to account for the different ways the coupling can be realized. 
A practical implementation of these higher body order features is given by the the atomic cluster expansion (ACE), which is usually computed based on the $g\rightarrow\delta$ limit of $\rho$. The coefficients of the atom density are indicated as $\rep<nlm||\fdelta_i>\equiv A_{inlm}$ following the notation of Ref.~\citenum{drau19prb}, and 
\begin{equation}
\rep< n_1l_1k_1 \cdots n_\nu l_\nu k_\nu||\fdelta_i^{\nu}> \equiv B_{\substack{i n_1 \cdots n_\nu \\ l_1 \cdots l_\nu}}^{(\nu)}
\end{equation}
correspond to the features associated with $\nu$-order neighbor clusters. Note that each $B_{\substack{i n_1 \cdots n_\nu \\ l_1 \cdots l_\nu}}^{(\nu)}$ indicates a group of basis functions indexed by $k_1,\dots, k_\nu$.
An equivalent construction, that emphasizes the connection with angular momentum theory, is provided by the N-body iterative contraction of equivariants\cite{niga+20jcp}, that is discussed in Section~\ref{sub:equivariants}.
Through a further linear transformation (change of basis) made explicit in Refs.~\citenum{drau20prb,Bachmayr2019} the moment tensor potential (MTP) of Ref.~\citenum{shap16mms} can also be related to this construction.
The {\em philosophy} behind the density correlation features is different from that behind MTPs and ACE, in that these methods were at least originally thought of as bases for polynomial regression. While these basis functions can be equally used as symmetry-adapted features there are subtleties to be considered that we discuss in Sec.~\ref{sec:incompleteness} and in Sec.~\ref{sub:feature-selection}.
Note that even though the contracted basis $\rep<(\enlk_i)_{i=1\ldots \nu}|$ eliminates some of the redundant indices that are present in the tensor-product basis, the indices do not label a set of linearly independent features. Symmetries and selection rules --  some of which, listed in Ref.~\citenum{niga+20jcp}, can be derived from results of angular momentum theory\cite{bied-louc84book} -- restrict greatly the number of independent entries that need to be computed. However, the non-trivial interaction between the radial and angular basis component makes this list incomplete. A mixed algebraic/numerical precomputation step can further reduce the required features\cite{Bachmayr2019}.

Finally, the global SOAP-like descriptors introduced in Ref.~\citenum{mavr+18jpcl}, corresponding to Eq.~\eqref{eq:global-rot-rep}, can be readily expressed in an angular momentum basis as 
\begin{multline}
\rep<\ennl||A; \frho^2 \otimes \frho^2> = 
\frac{(-1)^l}{\sqrt{2l+1}} \\
\sum_m (-1)^m \rep<\en_1 lm||\frho^2> \rep<\en_2 l(-m)||\frho^2>,
\end{multline}
where we recall that $\rep<\enlm||A; \frho^2>=\sum_{i\in A} \rep<\enlm||\frho_i>$.

\subsection{The density trick}
\label{sub:density-trick}

A crucial point in comparing different representations is that with an appropriate discretization of the angular basis one can evaluate symmetrized high-order correlations as sum of products of the density coefficients defined in \cref{eq:density-nlm}. This ensures that the cost of computing \emph{all} coefficients of a given order $\nu$,  scales only linearly with the number of neighbors included within the cutoff around atom $i$, even though it scales exponentially with $\nu$ in terms of the number of basis functions, at least with a naive choice of basis. 
This is to be contrasted with atom-centered symmetry functions (ACSF),\cite{behl11jcp,smit+17cs,zhan+18prl} and permutation invariant polynomials (PIP),\cite{braa-bowm09irpc} in which function are evaluated over all possible tuples composed of $\nu$ neighbors of the central atom (or on all the possible tuples in a structure to yield a global descriptor). In these frameworks, the cost depends linearly on the number of basis functions, but exponentially with $\nu$ in terms of the number of neighbors. This crucial difference makes density-expansion frameworks more convenient when one wants to ramp up the value of $\nu$, and there are many neighbors. A-priori sparsification schemes, exemplified in~\eqref{eq:totaldegree}, and feature selection schemes, discussed in \cref{sub:feature-selection}, allow one to keep only the most important basis functions, and eliminate the exponential scaling  with $\nu$ altogether.

Despite this rather fundamental difference in philosophy and computational cost, the two families of representations compute entities that are essentially equivalent, which we see by writing explicitly Eq.~\eqref{eq:nu2-basis-coupled} in the $g\rightarrow\delta$ limit as a sum over neighbors $j$ and $j'$
\begin{multline}\label{eq:nu2-delta}
\rep<\ennl||\fdelta_i^2> \propto
\sum_{jj'} \frac{\delta_{\e_1\e_j} \delta_{\e_1\e_{j'}}}{\sqrt{2l+1}}
\rep<n_1||r_{ji}> \rep<n_2||r_{j'i}>
\\
\times
\sum_m (-1)^{m}
\rep<\lm||\brhat_{ji}>
\rep<l ({-m})||\brhat_{j'i}>.
\end{multline}
By using the addition formula of the spherical harmonics we get the equivalent formulation
\begin{multline}
\rep<\ennl||\fdelta_i^2> \propto \sqrt{2l+1} \\
 \sum_{jj'} \delta_{\e_1\e_j} \delta_{\e_1\e_{j'}}
\rep<n_1||r_{ji}> \rep<n_2||r_{j'i}>
\rep<l ||\brhat_{ji}\cdot \brhat_{j'i}>,
\label{eq:nu2-acsf}
\end{multline}
in which $\rep<\omega||l>\equiv P_l(\omega)$ is a Legendre polynomial of order $l$.
In Eq.~\eqref{eq:nu2-acsf}, the $\nu=2$ density correlation coefficients are computed as a function of the distances and angles between triplets of atoms including the central atom $i$.
By plugging this expression for $\rep<\ennl||\fdelta_i^2>$ into Eq.~\eqref{eq:rho2-g3}, that evaluates the value of an arbitrary atom-centered symmetry function,  %
one sees that this result is not specific to the choice of $P_l$ as angular functions: in the limit of a complete basis set, it is equally possible to compute any ACSF using a sum over neighbor tuples \emph{or} a contraction of density coefficients, drawing an explicit link between the SOAP power spectrum features,\cite{bart+13prb,de+16pccp} Behler-Parrinello symmetry functions,\cite{behl11pccp,smit+17cs} the DeepMD framework,\cite{zhan+18prl} and FCHL features\cite{fabe+18jcp}.
Similar expressions could be derived for higher-order atom-centered symmetry functions, showing the complete equivalence -- but dramatically different computational scaling with the number of neighbors -- of the two frameworks.%

\subsection{Equivariant representations and tensorial features}\label{sub:equivariants}

\begin{figure}
    \centering
    \includegraphics[width=1.0\linewidth]{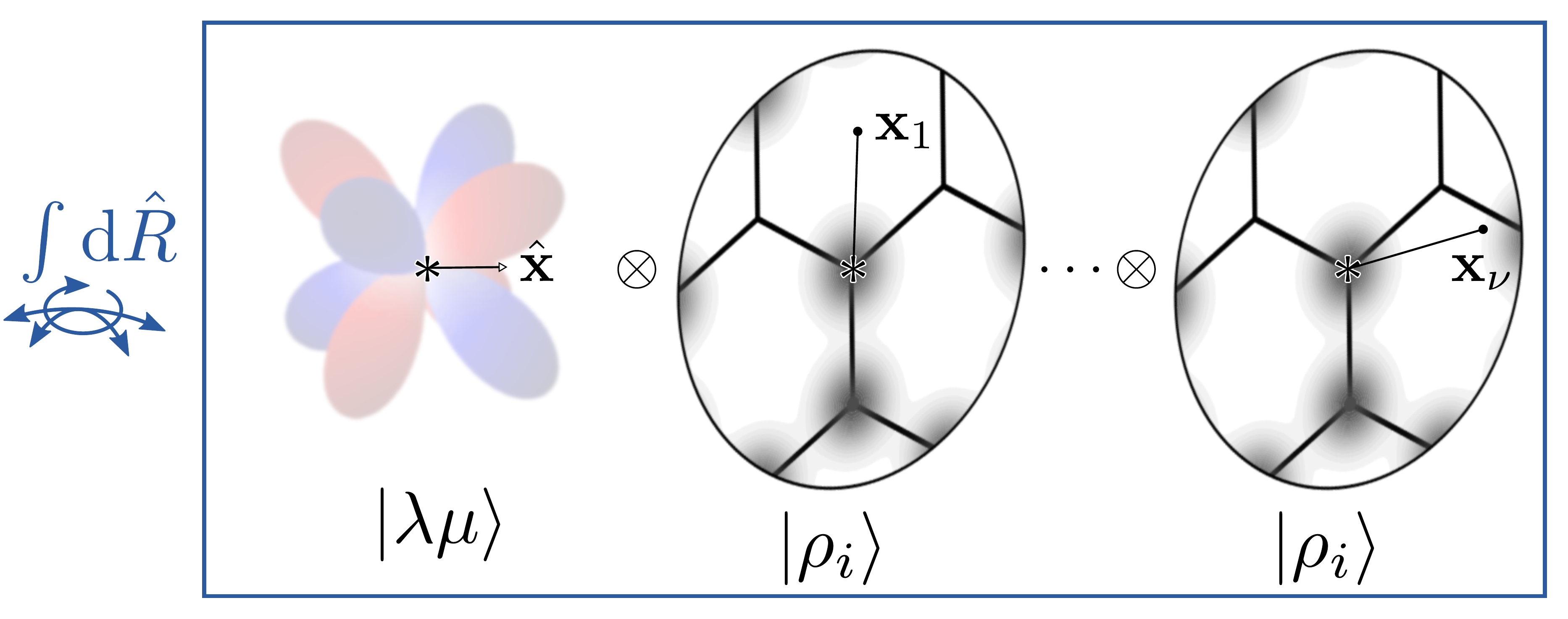}
    \caption{Graphical scheme of the construction of a $\SOthree$ equivariant tensor product representation. Copies of the atom-centered density are evaluated at $\nu$ separate points, together with a set of spherical harmonics that provide a basis to expand the components of a tensorial property. The tensor product is averaged by simultaneously rotating all densities and the $\rep|\lambda\mu>$ term.  }
    \label{fig:rho-lm-average}
\end{figure}

The previous construction is suitable to represent any rotationally-invariant atomic property. In many circumstances, however, one is interested in representing vector-valued or general tensorial quantities $\by$.
In this case, the prescribed transformations that the tensor undergoes under the symmetry operations of the $O(3)$ group (e.g. $\by(\Rhat A)=\Rhat \by(A)$) have to be incorporated into the atomic representation in the form of covariant, rather than simply invariant, features, so that the representation follows the same transformation as the target property, $ \rep|\Rhat A>=\Rhat \rep|A>$. %
Equivariance (the general concept that indicates symmetry-adapted behavior, encompassing both invariance and covariance) can be enforced by comparing environments and defining the local contribution to the target relative to a pre-defined local reference frame, which has been used to build machine-learning models of tensorial properties in molecular systems\cite{hand-pope09jctc,bere+15jctc,lian+17prb,sche+20jctc}.
A more general approach for achieving this goal consists in endowing the representation with the symmetries of spherical harmonics $\rep<\bxhat||\lambda\mu> = Y^\mu_\lambda(\bxhat)$, as well as the desired parity under the action of the inversion operator $\ihat$, which we associate with a ket $\rep|\sigma>$ such that $\ihat\rep|\sigma>=\sigma \rep|\sigma>$. The eigenvalue $\sigma$ is 1 for polar tensors, and $-1$ for pseudotensors.
Features that transform as $\rep|\sigma>\otimes\rep|\lambda\mu>$ can be achieved by  including two additional  fields\cite{will+19jcp,gris+19book} within the symmetrized tensor product of Eq.~\eqref{eq:body-order}, i.e.,
\begin{multline}\label{eq:equiv-features}
\rep|\ev*{\rho_i^{\otimes\nu}\otimes \sigma \otimes \lambda\mu}_\Othree>\equiv \rep|\frho[\gslm]_i^{\nu}>\\
=\sum_{k=0,1}\int_\SOthree\!\!\!\!\!\!\D{\Rhat}   \ihat^k\rep|\sigma>  \otimes \ihat^k\Rhat\rep|\lambda\mu> \\\otimes
\ihat^k\Rhat\rep|\frho_i>  \otimes\cdots  \otimes \ihat^k\Rhat\rep|\frho_i> \,.
\end{multline}
The operation is depicted in Fig.~\ref{fig:rho-lm-average}, showing how the $\lambda\mu$ ket corresponds to the evaluation of a set of spherical harmonics that anchors the atom-centered density to a reference frame.
The scalar and rotationally-invariant case is recovered by taking $\rep|\gslm> = \rep|1; 00>$.

This construction represents a particularly convenient framework to target the prediction of any Cartesian tensor $\by$ in terms of its irreducible spherical components,\cite{stone1975}
namely  $\y^{\sigma\lambda}_\mu$, that transform under rotation and inversion as
\begin{equation}
\begin{split}
\y^{\sigma\lambda}_\mu(\Rhat A) = & \sum_m D^\lambda_{\mu m}(\Rhat) \y^{\sigma\lambda}_m(A), \\
\y^{\sigma\lambda}_\mu(\ihat A) =& \sigma(-1)^\lambda \y^{\sigma\lambda}_m(A).
\end{split}\label{eq:tensprop}
\end{equation}
Within a linear regression model, they can be written as the combination of equivariant representations of the proper order $\lambda$ and parity $\sigma$ with a set of rotationally-invariant weights~$\rep<\Q||\by;\sigma;\lambda>$:
\begin{multline}\label{eq:tenspred}
\y^{\sigma\lambda}_\mu(A)=
\rep<\by||A; \gslm> \\\approx
\sum_i\int \D{\Q}
\rep<\by; \sigma\lambda; ||\Q>
\rep<\Q||A; \frho[\gslm]_i^{\nu}>\, .
\end{multline}
Each irreducible spherical component of $\by$ gives rise to a separate equivariant model, and the appropriate transformation rules are ensured by the fact that each equivariant feature $\rep<\Q||A; \frho[\gslm]_i^{\nu};>$ separately transforms as the spherical harmonics $\rep|lm>$ and the parity function $\rep|\sigma>$.
Much like the case of invariant symmetrized fields features, Eq.~\eqref{eq:equiv-features} can be most effectively computed by first expanding the atom-centered field on a basis of spherical harmonics, and is equivalent to an equivariant extension of the atomic cluster expansion\cite{drau20prb} or the moment tensor potentials, that are usually evaluated in the $g\rightarrow\delta$ limit.

A concrete example of these features is given by the density coefficients themselves: in fact, one can see that the $\nu=1$ equivariant reads simply
\begin{equation}
\rep<n||\frho[\gslm]_i^{1}> \equiv \rep<n\lambda \mu||\frho_i>^\star\delta_{\sigma 1}.
\end{equation}
Note how in the bra-ket notation the $(\lambda,\mu)$ indices on the two sides of this equation carry a different meaning. 
When used in the bra of the local density expansion $\rep<n\lambda \mu||\frho_i>$, they identify one of many components that are translationally invariant, but are not required to be rotationally equivariant; there is no explicit link to their behavior under rotation, and one could build a model by selecting only some of the $\mu$ values for a given $(n, \lambda)$.
When used in the  ket of an equivariant feature $\rep<n||\frho[\gslm]_i^{1}>$, they label groups of features that should be taken together, because they transform in a specific way under the symmetries of the $\Othree$ group.
By using $\rep<n||\frho[\gslm]_i^{1}>$ features in Eq.~\eqref{eq:tenspred} one obtains a model that fulfills~\eqref{eq:tensprop} (with the caveat that pseudotensors cannot be described by $\nu=1$ features) because acting on the spherical harmonics with $\Rhat$ yields a product with the associated Wigner matrix 
\begin{multline}\label{eq:rot-nu1}
\sum_{n}
\rep<\by;\lambda||n>
\rep<n||\Rhat A; \frho[\glm]_i^{1}>\, =\\
=\sum_{n}
\rep<\by;\lambda||n>
\rep<n l (-\mu)||\Rhat A;\rho_i> \\=
\sum_{n}
\rep<\by;\lambda||n>
\sum_m D^\lambda_{\mu m}(\Rhat) \rep<n l (-m)|| A;\rho_i>\\
=\sum_m D^\lambda_{\mu m}(\Rhat)
\sum_{n}
\rep<\by;\lambda||n>
\rep<n||A; \frho[\lambda m]_i^{1}>\,.
\end{multline}
The same covariant property applies to all density-correlation features, 
\begin{equation}
\rep|\Rhat A; \frho[\sigma;\lambda \mu]_i^{\nu}>= \sum_m D^\lambda_{\mu m}(\Rhat) \rep|A; \frho[\sigma;\lambda m]_i^{\nu}>.\label{eq:covariant-ket}
\end{equation}

Scalar products of these equivariant features generate matrix-valued kernels, that are suitable for symmetry-adapted Gaussian process regression -- for example  $\lambda$-SOAP kernels\cite{gris+18prl,raim+19njp}. 
Each entry in the kernel describes the coupling between the $\mu$ channels associated with the two environments,
\begin{multline}
\krn^{\sigma\lambda}_{\mu\mu'}(A_i,A'_{i'}) = \int \D{\Q} \\
\times \rep<A; \frho[\gslm]_i^{\nu}||\Q>\rep<\Q||A'; \frho[\gslm]_{{i'}}^{\nu}>.
\end{multline}
The symmetry properties of the features translate into the a kernel that transforms under rotations of the environments as 
\begin{multline}
\krn^{\sigma\lambda}_{\mu\mu'}(\Rhat A_i,\Rhat' A'_{i'}) =\\ 
\sum_{mm'}D^\lambda_{\mu m}(\Rhat) \krn^{\sigma\lambda}_{mm'}(A_i, A'_{i'}) 
D^\lambda_{\mu' m'}(\Rhat') ^\star,
\label{eq:covariant-kernel}
\end{multline}
which generalizes the covariant property for kernels introduced by Glielmo et al. for the case of Cartesian vectors.\cite{glie+17prb}

The fact that equivariant features of the form~\eqref{eq:equiv-features} follow $\Othree$ transformation rules  means that they can be combined using  established relationships in the quantum theory of angular momentum.
In particular, the coupled-basis representation used in the definition of Eqs.~(\ref{eq:nu1-basis-coupled}--\ref{eq:nu3-basis-coupled}) can be formulated for an arbitrary value of $\nu$, and in this form it is possible to express succintly\cite{niga+20jcp} a recursive formula to evaluate $\rep|\frho[\gslm]_i^\nu>$ based on lower order terms:
\begin{multline}
\rep<\ldots \nlk_\nu;\nlk||\frho[\gslm]_i^{(\nu+1)}>\propto
\delta_{s\sigma((-1)^{l+k+\lambda})}   \times \\
\sum_{m}  \cg{\lm}{k (\mu-m)}{\glm} \rep<n||\frho[\lm]_i^1> \\
\rep<\ldots; \nlk_\nu||\frho[s; k (\mu-m)]_i^{\nu}>.
\label{eq:rho-recursion}
\end{multline}
For $\nu=2$, the recursion yields the original expression for $\lambda$-SOAP equivariants\cite{gris+18prl}
\begin{multline}
\rep<\nl_1;\nl_2|| \frho[\gslm]_i^{2}> =\\\frac{\delta_{\sigma (-1)^{l_1+l_2+\lambda}}}{\sqrt{2\lambda+1}}
\sum_m \cg{l_1 m}{l_2 (\mu-m)}{\lambda \mu}\\
\times \rep<\nl_1m||\rho_i>^\star \rep<\nl_2 (\mu-m)||\rho_i>^\star .
\label{eq:soap-lambda-2}
\end{multline}
Similar recursive expressions have been independently proposed to efficiently  compute \emph{invariant} features\cite{Bachmayr2019,shap16mms}, that can be obtained by taking $\rep|\gslm> = \rep|1; 00>$ in Eq.~\eqref{eq:rho-recursion}.
The possibility of combining equivariant features using angular momentum rules is also exploited in the construction of covariant neural networks\cite{ande+19nips,thomas2018arxiv}

\begin{figure*}[tbp]
    \centering
    \includegraphics[width=0.9\linewidth]{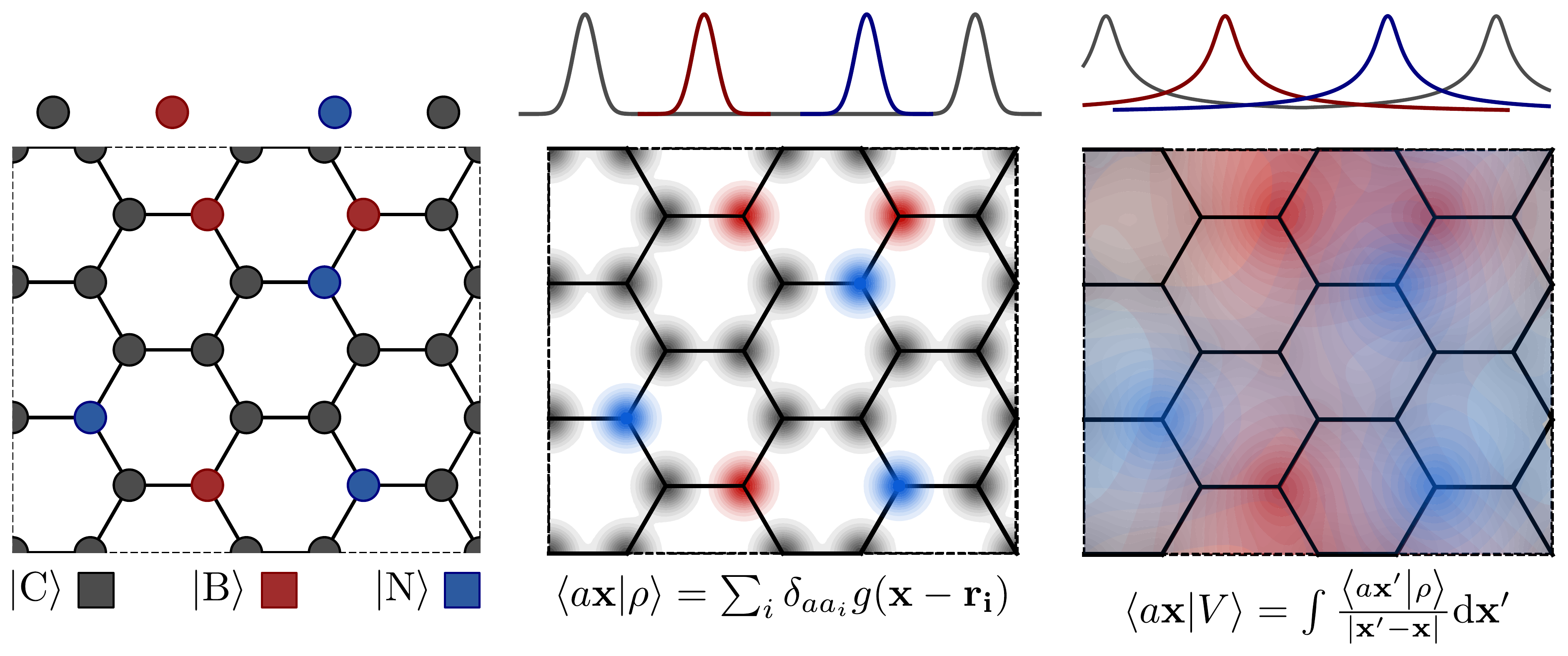}
    \caption{Relationship between Cartesian coordinates, local and long-range fields. The top row shows a 1D cartoon, and the second row a more realistic, hypothetical ``doped graphene'' system in 2D. Left: Reference structure; middle: atom-density field, divided in three elemental channels, color-coded; right: atom-density potential, color-coded. Adapted with permission from Ref.~\citenum{gris+21cs}. Copyright 2020 Royal Society of Chemistry.} 
    \label{fig:lode}
\end{figure*}

One can also build models that are imbued with the appropriate transformation properties in an indirect fashion, by learning atom-centered scalars and combining them with the atomic positions to evaluate formal (or actual) molecular multipoles. This is easily seen for the case of the dipole moment of a neutral molecule, that can be computed as
\begin{equation}
\boldsymbol{\mu}(A) = \sum_{i\in A} q(A_i) \br_i.
\label{eq:point-charge}
\end{equation}
Models of this form have been used since the early days of the construction of molecular potential and dipole moment surfaces\cite{huan+05jcp,yu-bowm19jpca}, combined with neural-network potentials to compute IR spectra in the condensed phases\cite{gast+17cs}, and more recently combined with tensorial models, to describe the interplay of atomic charges and polarization contributing to the total dipole moment\cite{veit+20jcp}.
Assigning constant formal charges $q_i$ to atoms has also been used to derive covariant kernels, in the so-called operator machine learning framework\cite{chri+19jcp}, which is also similar in spirit to the tensorial embedded atom neural network\cite{zhan+20jpcb}. The gist of the idea (although expressed in a feature rather than kernel (or NN) language) is that one can define a translationally-invariant representation that depends formally on an applied electric field, e.g.
\begin{equation}
\rep<\bx||A_i;\mbf{E}> = \sum_{j\in A}   \rep<\bx||\br_{ji}; g> (\br_{ji}\cdot \mbf{E}) q_j
\end{equation}
Deriving with respect to one of the components of \mbf{E} brings a dependency on the corresponding component of $\br_{ji}$, that upon rotational averaging (keeping in mind that $\Rhat$ acts on atomic coordinates and not on the external field) plays the same role as $\rep|\lambda\mu>$ in Eq.~\eqref{eq:equiv-features}, providing a basis of features that can be used to learn vectors covariantly. The use of local interatomic vectors to build a covariant reference system is similar to the approach adopted in Ref.~\citenum{kim+19jpcc} to define a general atomic neighborhood fingerprint, and in Ref.~\citenum{zhan+20prb} to learn the position of electronic Wannier centers. Despite the superficial similarity with the environment-dependent point-charge model of Eq.~\eqref{eq:point-charge}, this scheme more closely resembles a framework based on atomic dipoles, since its predictions can be decomposed as a sum of atom-centered equivariant terms.

\begin{figure*}[bthp]
\includegraphics[width=1.0\linewidth]{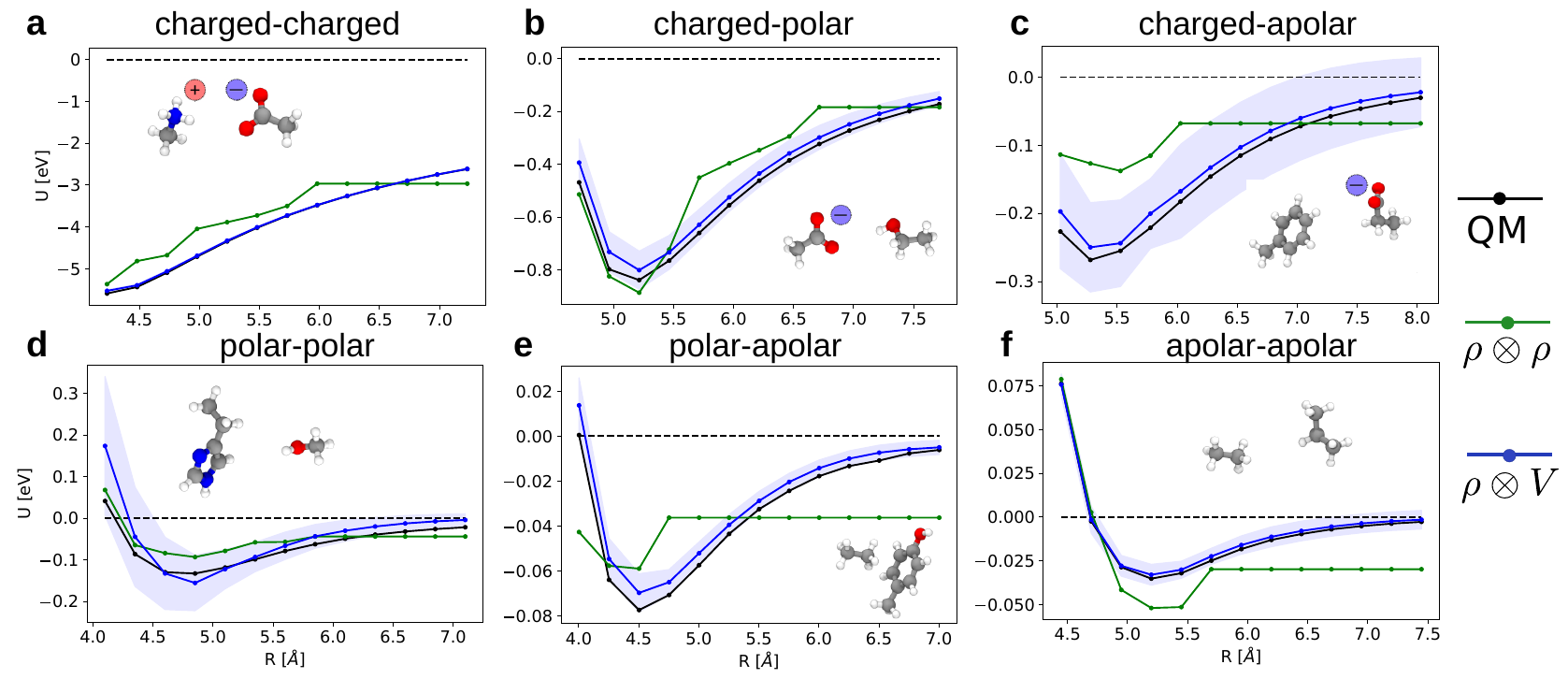}
\caption{Median-error binding curves for six different classes of intermolecular interactions, involving charged, polar and apolar molecules extracted from the BioFragment Database\cite{burn+17jcp}. (black lines) reference quantum-mechanical calculations. (green lines) predictions of a local ($\rep|\frho_i^2>$-based) model. (blue lines) predictions of a  multi-scale ($\rep|\field{\rho}_i[V]_i>$-based) model. The shaded area indicates the confidence interval for the prediction estimated from a committee model\cite{musi+19jctc}. Reproduced from Ref.~\citenum{gris+21cs}.
}
\label{fig:ms-lode}
\end{figure*}

\begin{figure}[bp]
    \centering
    \includegraphics[width=1.0\linewidth]{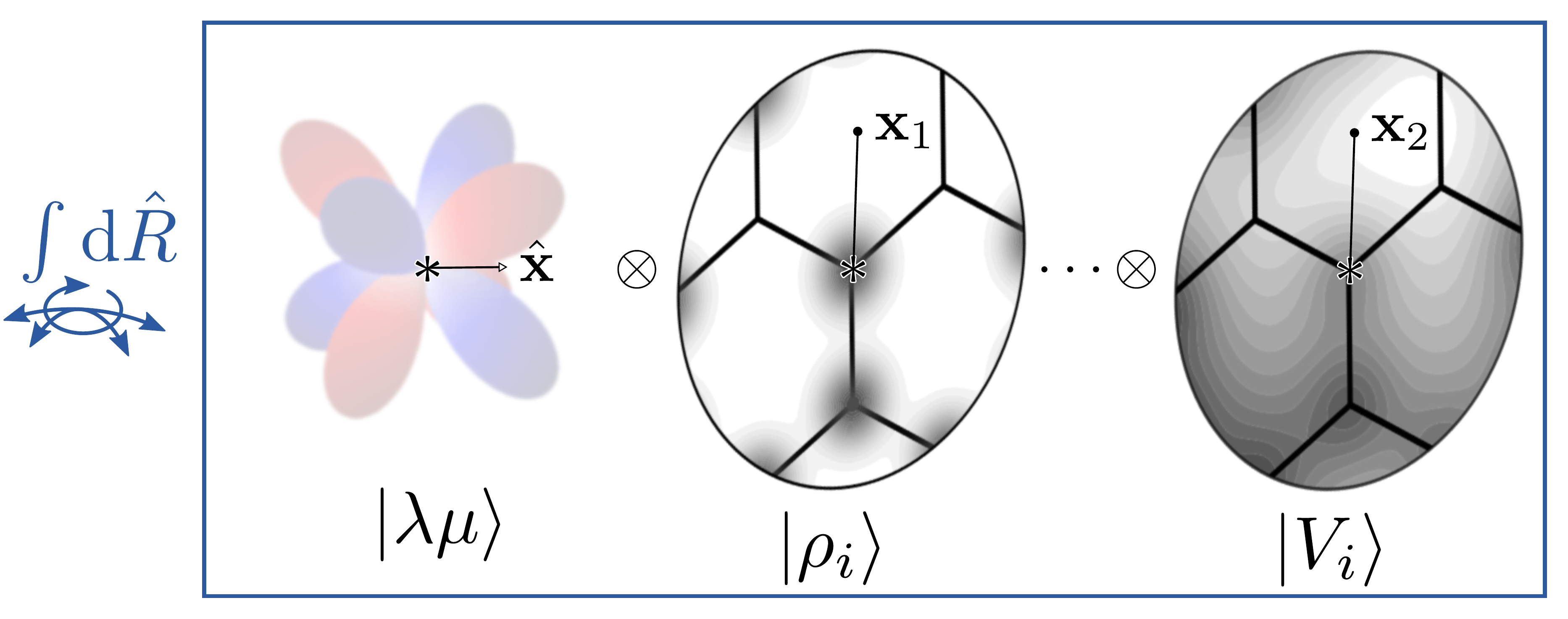}
    \caption{A multi-scale equivariant representation combining atom-centered density fields $\rep|\frho_i>$, long-range fields $\rep|V_i>$  and a set of spherical harmonics.  }
    \label{fig:rho-lm-V-average}
\end{figure}

\subsection{Long-range features}
\label{sub:long-range}

Introducing a cutoff in the definition of the local density is not only necessary to reduce the cost of evaluating the expansion coefficients, or the number of terms that have to be included to obtain a converged expansion of the density correlations. Increasing the range of the environment makes the model more complex, which often results in slower learning when limited training data is available.\cite{bart+17sa}
The problem is particularly evident when studying systems with a prominent electrostatic component\cite{gris+18prl,yue+21jcp}, but long-range physics is ubiquitous\cite{ambr+16science}, and ultimately limits the accuracy and transferability of machine-learning models\cite{wilk+19pnas,veit+20jcp,yue+21jcp}.
One pragmatic solution is to build models that explicitly incorporate a physically-motivated functional form as a baseline, which could take the form of an existing model\cite{medd+14jctc,medd+15jcp}, an electrostatic scheme based on machine-learned partial charges\cite{artr+11prb,ghas+15prb,gast+17cs} or atomic multipoles\cite{bere+15jctc,bere+18jcp}.
Alternatively, one may attempt to construct representations that are multi-scale in nature, and are therefore suitable to describe, in a data-driven manner, properties that depend on multiple length scales. This idea has been implemented by combining local representations with different cutoffs\cite{bart+17sa}, scaling atomic contributions according to distance\cite{fabe+18jcp,will+18pccp} (see also Section~\ref{sub:feature-optimization}), treating separately intra- and inter-molecular correlations\cite{veit+19jctc,metc+20jcp}, as well as by building global structural representations based on an intrinsically multi-scale wavelet scattering transform\cite{eick+17nips}.

A recently-proposed, more radical take to the problem, extends the symmetrized atomic field construction beyond the use of the atomic density as the starting point.  
In order to describe more naturally the long-range behavior that is typical of electrostatic interactions, it defines a Coulomb-like potential field based on the smoothed atomic density (Figure~\ref{fig:lode})
\begin{equation}
\rep<\ex||A; V> =\int \D{\bx'} \frac{\rep<\ex||A; \rho>}{\left|\bx - \bx'\right|}
. \label{eq:lode-v}
\end{equation}
This is a \emph{global} operation, which can however be performed efficiently by transforming the density in plane waves, using one of the many different schemes that are routinely used to model electrostatics. 
Symmetrizing $\rep|V>$ in the same way as for $\rep|\rho>$ leads to an atom-centered potential,  \begin{multline}
\rep<\ex||A; V_i> =\int \D{\bx'} \frac{\rep<\ex||A; \rho_i^{<} >}{\left|\bx - \bx'\right|} + \int \D{\bx'} \frac{\rep<\ex||A; \rho_i^{>} >}{\left|\bx - \bx'\right|}\\ \equiv
\rep<\ex||A; V_i^{<}> + \rep<\ex||A; V_i^{>}>\,, \label{eq:lode-vi}
\end{multline}
where we introduce the short-range density $\rep|\rho_i^{<}>$, restricted to the region within the cutoff, and the far-field density $\rep|\rho_i^{>}>$, restricted outside the cutoff, and the corresponding local and non-local fields $\rep|V^{<}_i>$ and $\rep|V^{>}_i>$.
Crucially, these features incorporate information on atoms outside the cutoff, yet their complexity can be kept under control by restricting the range of the spherical environment over which they are computed.
Just as for $\rep|\rho_i>$, the ket can be discretized by expanding it on an orthogonal basis of radial functions and spherical harmonics to obtain $\rep<\enlm||\field{V}_i>$.

One can then build features that are fully equivariant by averaging $\rep|V_i>$ over the symmetry operations of the $\Othree$ group, leading to $\nu$-point correlations analogous to those discussed above. Furthermore, one can combine local and long-range fields,  as in Figure~\ref{fig:rho-lm-V-average}, constructing a family of multi-scale long-distance equivariants (LODE) features\cite{gris+21cs}, that in the most general form can be written as
$\rep|\field[\gslm]{\rho}_i^\nu[V]_i^{\nu'}>$:
\begin{multline}
\label{eq:ms-lode}
\rep|\ev*{\rho_i^{\otimes \nu}\otimes V_i^{\otimes \nu'}\otimes \sigma \otimes \lambda\mu}_\Othree> = \sum_{k=0,1}\int_\SOthree\!\!\!\!\!\!\!\D{\Rhat}\,\ihat^k\rep|\sigma>  \otimes \\
\ihat^k\Rhat\rep|\lambda\mu> \otimes\underbrace{\ihat^k\Rhat\rep|\rho_i> \cdots\otimes
\ihat^k\Rhat\rep|\rho_i> }_{\nu\ \text{times}}
\otimes\underbrace{\ihat^k\Rhat\rep|V_i> \cdots\otimes
\ihat^k\Rhat\rep|V_i> }_{\nu'\ \text{times}}\,.
\end{multline}
The simplest multi-scale representation $\rep|\field{\rho}_i[V]_i>$ can be linked to physics-based models using an atom-centered multipole expansion of electrostatic interactions, as we discuss further in Section~\ref{sub:ms-model}, but are effective to learn a multitude of long-ranged interactions, from permanent electrostatics, to polarization and dispersion.
When trying to represent long-range interactions between molecular fragments, a model based on local $\rep|\frho_i^2>$ features produces a completely unphysical behavior, with the interaction reaching a plateau when the molecules are separated by more than the cutoff distance (Figure~\ref{fig:ms-lode}).
Multi-scale LODE features, instead, can describe the asymptotic tail even when using a 3\AA{} cutoff in the definition of the atom-centered environments, and are capable of representing interactions of very different chemical nature.
Using a non-local field as the starting point of the symmetrization procedure provides interesting opportunities to incorporate long-range, many-body interactions in atomistic machine learning.

\section{Representations and models}
\label{sec:models}
Even though this review focuses on the problem of representing atomic structures in terms of a vector of features, one cannot ignore the intimate connection between the choice of features and how they are used to construct models of symmetric properties, such as site energies, which are then used in the context of regression schemes.\cite{bart-csan15ijqc,behl-parr07prl,shap16mms,drau19prb,ande+19nips,schu+18jcp,gilm+17mlr,duve+15nips,Chmiela2017,gris+18prl,gris+21cs,vand+20mlst} The purpose of this section is therefore to discuss the interplay between representations and models. Given a set of symmetric features $\bra{\q}\ket{A_i}$ of an atomic environment $A_i$, we explore how to use it to represent a symmetric property $\y(A_i)$. We discuss linear approximations, %
\begin{equation}  \label{eq:models:linear}
    \y(A_i) \approx \sum_{\q} \bra{\y}\ket{\q} \bra{\q}\ket{A_i}
\end{equation}
and show that the family of features we introduced in Section~\ref{sec:symmetry-fields} lead to natural generalisations of well-established models of interactions between atoms and molecules in terms of a body-ordered expansion.
These relatively simple models put stringent requirements on the quality of the feature sets. We then go on to review how highly non-linear models may provide more flexibility in describing the relationship between a structure and its properties, and yield satisfactory results even with a rather simple, imperfect choice of features.
Here, and in the following, we always understand implicitly that equality in these approximations can only be attained in the limit of an infinite cutoff radius and suitably converged parameterisation.

\subsection{Linear models and body-order expansion}
\label{sub:linear-models}
An advantage of linear models is that they can often be connected to classical physics-inspired frameworks, and bring to light physical-chemical insights on the nature of the underlying representations
An example of this connection involves the construction of interatomic potentials in terms of a body-ordered hierarchy of atom-centered energy terms
\begin{equation}
E(A) = \sum_{i\in A} E(A_i) = \sum_\nu \sum_{i\in A}  E^{(\nu+1)}(A_i),
\end{equation}
in which each term can be written as a sum over $\nu$ neighbors of the central atom
\begin{equation}
E^{(\nu+1)}(A_i) =
\sum_{j_1 < \cdots < j_\nu} v^{(\nu+1)}(\br_{ji_1}, \ldots \br_{ji_\nu}). \label{eq:bo-sum-nu}
\end{equation}
This kind of expansion underlies the vast majority of empirical force fields, that are customarily written as a combination of pair potentials, and short-range 2, 3, and 4-body bonded terms. 

Most potentials truncate this expansion at body-order three, i.e. $\nu = 2$ -- a notable exception being the dihedral angle potentials used in force fields, that are four-body but involve selected groups of atoms rather than a sum over all possible triplets. 
This is because the cost of a naive evaluation of the sum $\sum_{j_1 < \dots < j_\nu}$ scales exponentially with the body order $\nu$, i.e. as $\mathcal{O}(N_i^\nu)$ for an environment containing $N_i$ atoms. 
More sophisticated ways of symmetrizing the body-ordered terms, such as those discussed in Refs.~\citenum{bowm+09jcp} and~\citenum{vand+20mlst}, alleviate this behavior. In the following paragraphs we demonstrate, in particular, how this exponential scaling can be overcome by using the density correlation representations discussed in Section~\ref{sec:symmetry-fields}.

\paragraph*{The three-body case.}
It is illuminating to first discuss in full detail the representation of a 3-body site potential, written traditionally in internal coordinates, in the form
\begin{equation} \label{eq:models:naive3b}
    E(A_i) = \sum_{j} v^{(2)}(r_{ji})
        + \sum_{j < j'} v^{(3)}(r_{ji}, r_{j'i}, \omega_{ijj'}),
\end{equation}
where $\omega_{ijj'} := \hat{\br}_{ji} \cdot \hat{\br}_{j'i}$.
In order to connect to the atomic density correlations we first rewrite this as
\begin{multline}
    E(A_i) = \sum_{j} \Big(v^{(2)}(r_{ji}) - \frac{1}{2} v^{(3)}(r_{ji}, r_{ji}, 0) \Big) \\
     \qquad + \frac{1}{2} \sum_{j j'} v^{(3)}(r_{ji}, r_{j'i}, \omega_{ijj'}) \\
    \label{eq:models:selfinteraction3b}
    =:
    \sum_{j} u^{(2)}(r_{ji})
    + \sum_{j j'} u^{(3)}(r_{ji}, r_{j'i}, \omega_{ijj'}),
\end{multline}
adding and subtracting a self-interaction from the 3-body term.

Approximating $u^{(2)}(r)$ in terms of a radial basis $\rep<r||n>\equiv R_n(r) $
yields
\begin{multline}
E^{(2)}(A_i)=
\sum_{j} u^{(2)}(r_{ji})\\
\equiv  \sum_j \rep<u^{(2)}||r_{ji}>\approx\sum_{j}\sum_n \rep<u^{(2)}||n>\rep<n||r_{ji}>\\
=\sum_n \rep<u^{(2)}||n> \int \D r \rep<n||r>\sum_j \delta(r-r_{ji}) \\
=\sum_{n}\rep<u^{(2)}||n>\rep<n||\fdelta_i^1> %
\end{multline}
where $\rep|\fdelta_i>$ is the $g \to \delta$ limit of the atom-centered density $\rep|\frho_i>$.
As in Eq.~\eqref{eq:dirac-lr}, the use of the Dirac notation   to express the pair potential highlights the fact that (atom-centered) properties can be seen as a type of representation, and that in this sense a linear model is nothing but an expansion in a discrete basis of $\rep<u^{(2)}||r>\equiv u^{(2)}(r)$.

For the three-body term we revisit \eqref{eq:rho2-g3}: first, we approximate $u^{(3)}$ in terms of the radial basis $\rep<r||n>\equiv R_n(r)$ and the Legendre polynomials $ \rep<\omega||l>\equiv P_l(\omega)$,
\begin{multline}
u^{(3)}(r_{ji}, r_{j'i}, \omega_{ijj'}) \\
 \approx
    \sum_{n n' l}
    \rep<{u^{(3)}}||{nn'l}>
    \rep<n||r_{ji}> \rep<n'||r_{j'i}> \rep<l||\omega_{ijj'}>.
\end{multline}
Applying Legendre's addition theorem to expand the $P_l$ in terms of spherical harmonics $\bra{\hat\br}\ket{lm}\equiv Y_l^m(\hat\br)$,
\[
\rep<l||\omega_{ijj'}>
=  \frac{4\pi}{2l+1}
\sum_{m = -l}^l (-1)^m
    \rep<lm||\brhat_{ji}>
    \rep<l(-m)||\brhat_{ji}>,
\]
absorbing the $\frac{4\pi}{2l+1}$ into the weights
$\bra{u^{(3)}}\ket{nn'l}$ and reordering the summation yields
\begin{multline}
u^{(3)}(r_{ji}, r_{j'i}, \omega_{ijj'})    =
\sum_{nn'l} \rep<u^{(3)}||{nn'l}>
    \rep<{n}||{r_{ji}}> \rep<{n'}||r_{j'i}> \\[-3mm]
\qquad \qquad \times \sum_{m = -l}^l (-1)^m
    \rep<lm||\brhat_{ji}>
    \rep<l(-m)||\brhat_{j'i}> \\
=
\sum_{nn'l}  \rep<u^{(3)}||{nn'l}>
    \sum_m (-1)^m \rep<nlm||\br_{ji}> \rep<n'l(-m)||\br_{j'i}>.
\end{multline}
Finally, we sum over all $(j,j')$ and reorder the summation to arrive at

\begin{multline}
\sum_{j j'} u^{(3)}(r_{ji}, r_{j'i}, \omega_{ijj'})
 =
\sum_{nn'l} \rep<{u^{(3)}}||{nn'l}>
\\
 \sum_m (-1)^m
    \sum_j \rep<nlm || \br_{ji}; \delta>  \sum_{j'}
    \rep<n'l(-m)|| \br_{j'i}; \delta>
\\
\label{eq:models:magic_scaling}
=\sum_{nn'l} \rep<{u^{(3)}}||{nn'l}> \sum_m (-1)^m \rep<nlm||\fdelta_i> {\rep<nl(-m)||\fdelta_i>}
\\
=\sum_{nn'l} \rep<u^{(3)}||nn'l>
     \rep<nn'l||\fdelta_i^{2}>.
\end{multline}

In summary, we have written an arbitrary 3-body site potential in terms of 1- and 2-correlations of the atomic density,
\begin{equation}\label{eq:models:result_3body}
    \begin{split}
    E(A_i)
    &= \sum_n \rep<{u^{(2)}}||{n}> \rep<{n}||\fdelta_i^1> \\
& \qquad +
   \sum_{nn'l} \rep<{u^{(3)}}||{nn'l}> \rep<{nn'l}||\fdelta_i^{2}>
\end{split}
\end{equation}
Aside from connecting classical body-ordered interatomic potentials and $\nu$-correlations of the atomic density this formulation has significant advantages in terms of computational complexity which we discuss below after generalising the argument to arbitrary body-order.

\paragraph*{General ($\nu+1$)-body order potentials.}
The systematic expansion to arbitrary body orders has been applied to the description of alloys in terms of a cluster expansion, a procedure that was very early shown to provide a complete description of the problem\cite{sanc+84pa}, to the rationalization of fragment-based electronic structure methods\cite{rich-herb12jcp}, and to the construction of last-generation potentials for water and aqueous systems\cite{medd+14jctc}.

We adopt the generalisation of  \eqref{eq:models:selfinteraction3b} that includes self-interaction,
\begin{equation}
E^{(\nu+1)}(A_i) =
\sum_{j_1,  \ldots, j_\nu} u^{(\nu+1)}(\br_{j_1i} \ldots \br_{j_\nu i}), \label{eq:bo-sum-nu-self}
\end{equation}
which can be obtained from the more natural formulation \eqref{eq:bo-sum-nu} by incorporating the self-interaction terms into the $\nu$-body-order energy similarly to Eq.~\eqref{eq:models:selfinteraction3b}.%

To connect \eqref{eq:bo-sum-nu-self} to the density correlations we represent the rotationally invariant $(\nu+1)$-body function $u^{(\nu+1)}$ as
\begin{equation}
\begin{split}
& u^{(\nu+1)}(\br_{j_1 i}, \dots, \br_{j_\nu i})   \\
&=
\int_{\Othree}\!\!\!\!\!\D\Rhat\int \D{\Q} \rep<u^{(\nu+1)}||\Q>  \rep<\Q|\Rhat\rep|\br_{j_1 i}, \dots, \br_{j_\nu i}>,
    \end{split}
\end{equation}
where we use $\Q$ as a shorthand for $(\bx_1;\ldots \bx_\nu)$, so that
$\rep<\Q||\br_{j_1 i}, \dots, \br_{j_\nu i}>\equiv \prod_{k=1}^\nu \delta(\bx_k - \br_{j_k i}) $. %
The rotation can be made to act on the atomic positions or on the basis, depending on convenience.
The $(\nu+1)$-order site energy is obtained by summing over clusters of neighbors
\begin{multline} \label{eq:Enu1beforedensitytrick}
E^{(\nu+1)}(A_i)\\\approx \!\! \sum_{j_1, \dots, j_\nu}\!\! \int_{\Othree}\!\!\!\!\!\!\! \D{\Rhat} \! \int \!\D{\Q }\!
    \rep<u^{(\nu+1)}||\Q> \!\rep<\Q|\Rhat\rep| \br_{j_1 i} \cdots \br_{j_\nu i}> \\
=\!\!\int\! \D{\Q} \!\rep<u^{(\nu+1)}||\Q> \int_{\Othree}\!\!\!\!\!\!\! \D{\Rhat}\!\!\! \sum_{j_1, \dots, j_\nu}
  \!\!\!\!   \rep<\Q|\Rhat\rep| \br_{j_1i} \cdots \br_{j_\nu i}>.
\end{multline}
The symmetrized sum can be reordered to show that it corresponds to the $\nu$-point density correlation
\begin{multline}
\int_{\Othree}\!\!\!\!\! \D{\Rhat} \sum_{j_1, \dots, j_\nu}
     \rep<\bx_1; \ldots \bx_\nu|\Rhat\rep| \br_{j_1 i} \cdots \br_{j_\nu i}> \\
=\int_{\Othree}\!\!\!\!\! \D{\Rhat}
\sum_{j_1\ldots j_\nu} \prod_k \delta(\Rhat \bx_k-\br_{j_ki})\\
=\int_{\Othree}\!\!\!\!\! \D{\Rhat}
\prod_k \sum_{j_k} \delta(\Rhat \bx_k- \br_{j_ki})
\\
=\!\!\int_{\Othree}\!\!\!\!\! \D{\Rhat}
\prod_k \rep<\Rhat\bx_k||\delta_i>=\rep<\bx_1; \ldots \bx_{\nu}||\fdelta_i^{\nu}>,
\end{multline}
which is precisely Eq.~\eqref{eq:body-order-x} written in the $g\rightarrow\delta$ limit.
Thus we have explicitly represented $E^{(\nu+1)}$ in terms of the symmetry-adapted density correlations. We emphasize again that this calculation {\em required} the inclusion of the self-interactions as the starting point \eqref{eq:bo-sum-nu-self} -- even though, if one wishes so, they can be removed from the final result\cite{jinn+20jcp}.

\paragraph*{Linear completeness.}
For a practical implementation we can choose a
finite, discrete basis, approximating $E^{(\nu+1)}$ as
\begin{equation}
    \label{eq:Enu_correlations}
    E^{(\nu+1)}(A_i)
    \approx \sum_{\q}
    \rep<u^{(\nu+1)}||\q> \rep<\q||\fdelta_i^\nu>.
\end{equation}
Any complete implementation of $\nu$-order density correlation features\cite{shap16mms,drau19prb,Bachmayr2019,niga+20jcp} provides a basis to expand $u^{(\nu+1)}$ and approximate the $(\nu+1)$-order term, that contributes to the body-ordered expansion of $E(A)$.
The foregoing discussion shows that these bases are complete in the following sense. An (infinite) collection of symmetrized features $\{\bra{q}\ket{A_i}\}_{q\in \bq_{\rm total}}$ is a {\em complete linear basis} if there exists a sequence of finite subsets $\bq \subset \bq_{\rm total}$ such that
\begin{equation}
    \y(A_i) \approx \y_{\bq}(A_i) := \sum_{\q \in \bq} \bra{\y}\ket{\q} \bra{\q}\ket{A_i},
\end{equation}
i.e. $\y_{\bq}$ approximates $\y$ to within arbitrary accuracy in the limit as the number of features tends to infinity. We stress here that the weights $\rep<\y||\q>$ depend on the entire choice of feature set $\bq$ and not just the single index $\q$.
Therefore the density correlation features provide a universal, complete linear basis to approximate body-ordered potentials and, more generally, body-ordered expansions of properties that can be meaningfully written as a sum of atom-centered contributions.

For the specific choice
\begin{equation}
    \rep<\q| = \otimes_{\alpha = 1}^\nu \rep< \nlm_\alpha| 
\end{equation}
Eq.~\eqref{eq:Enu_correlations} is the ACE model\cite{drau19prb,Bachmayr2019}. Note that the symmetrized correlations $\rep<\q||\fdelta_i^\nu>$ can be efficiently and conveniently evaluated as already hinted at in Section~\ref{sub:lm-rep}.
Since MTPs provide an alternative basis set for the same space, they are complete as well, and in the same sense. We also emphasize that a rigorous proof of completeness of MTPs was already given by \citet{shap16mms}, and the essence of the idea can be traced back to the cluster expansion theory of alloys\cite{sanc+84pa}.
The ``density trick'', i.e., expanding in terms of the density correlations, ensures linear scaling in terms of the number of neighbors $N_i$ rather than the $\binom{N_i}{\nu}$ scaling of the naive representation \eqref{eq:bo-sum-nu}, which enables modeling very high body-orders. A recursive evaluation of the $\nu$-correlations implemented by the MTP and ACE bases, or by the NICE formalism, avoids an unfavorable scaling of the evaluation of the high-order terms (see Section~\ref{sub:implementation} for a summary of these techniques).

\subsection{Density smearing.}\label{sub:smearing}
The real-space view of the density correlation features may be more intuitive when considering finite smearing of the atomic contributions to $\rep|\rho_i>$, that gives rise to a smooth function that can be seen as a proxy for the electronic density, and is reminiscent of the atoms-in-molecules\cite{bade94book} description of the electronic structure of a molecule or a condensed-phase system as a collection of atom-centered densities.
In the literature using SOAP features, the width of the atom-centrered Gaussians has been often indicated as a hyperparameter with an important influence on the robustness\cite{deri+18prl} and accuracy\cite{caro19prb,kondati2020arxiv} of the resulting machine-learning models. 
Since we derived the link between density correlations and body-ordered potentials, and in particular the proof of the completeness of the linear expansion, only in the limit of a sharp density we now discuss whether a similar formal guarantee holds for a general $\rep|\rho_i>$, admitting in particular smearing of the atomic contributions. With tensor-product bases, all statements derived for higher correlation orders can eventually be reduced to a one-dimensional description, that is sufficient to reveal the essential features of the problem. Note that the following discussion provides only {\em theoretical guarantees}; we explain below that excessive smearing creates severe numerical ill-conditioning which must be carefully considered in practical implementations.

We begin by noting that the expansion of a smeared density in a basis $\rep<x||n>$ is identical to the expansion of a $\delta$-like density in the corresponding smeared (a.k.a. {\it mollified}) basis  $\rep<x||n; g>\equiv \int \D{x'} \rep<n||x'> g(x-x')$: 
\begin{multline}
\rep<n||\rho> = \int \D{x} \rep<n||x> \sum_i g(x-x_i) \\= 
 \int \D{x} \sum_i \delta(x-x_i) \int \D{x'} \rep<n||x'> g(x-x') \\= 
 \int \D{x} \sum_i \delta(x-x_i) \rep<n; g||x> = \rep<n; g||\delta>.
\end{multline}
With this observation in hand showing that $\rep<x||n; g>$ inherits completeness from $\rep<x||n>$ is sufficient to ensure that all our results apply also to smeared densities.

We first consider the case of standard monomials. Any continuous function $f(x)$ can be expanded to within arbitrary accuracy into polynomials $x^n$ :
\begin{equation}\label{eq:fn-poly}
f(x) \approx f_\nmax(x) = \sum_{n = 0}^\nmax c_n x^n \underset{\nmax\rightarrow\infty}{\rightarrow}f(x).
\end{equation}
We want to check whether we can also represent $f$ in terms of smeared polynomials,
\begin{equation}
    p_n^g(x) =  g \ast x^n = \int (t -x)^n e^{-t^2/2\sigma^2}/\sqrt{2\sigma^2\pi} \,\D{t}.
\end{equation}
For the particular choice of Gaussian smearing we can evaluate this expression explicitly and obtain
\begin{equation}
p_n^g(x) = x^n + \text{lower order terms},
\end{equation}
i.e., $p_n^g$ is in fact still a polynomial  with leading-order term $x^n$ and this means it forms a basis. In particular we can now again represent $f_\nmax(x)$ {\em exactly} as
\begin{equation}
f_\nmax(x) = \sum_{n=0}^\nmax c_n' p_n^g(x)
\end{equation}
And in the limit $\nmax \to \infty$ we recover $f$.

In the more general case,  suppose that we have an arbitrary complete basis $\rep<x||j>$. Then we can approximate $x^n \approx \sum_j b_{nj} \rep<x||j>$. The smearing operator $g\ast \cdot$ is bounded, which allows us to write
\begin{multline}
p_n^g(x) = g \ast x^n \approx \sum_j b_{nj} \int \D{x'}g(x-x')  \rep<x'||j> \\ =\sum_j b_{nj} \rep<x||j; g> .\quad\quad
\end{multline}
Given that $p_n^g$ are dense, it follows that also the smeared basis functions $\rep<x||j; g>  \equiv g  \ast \rep<x||j>$  are dense.
From these arguments it is reasonable to  conclude that the smeared density correlations also form a complete linear basis. 

As already mentioned above, this is a purely theoretical statement, and there is an important caveat: The inverse of the smearing operator is unbounded, which implies that the coefficients of the expansion of $f$ in terms of the smoothed polynomial basis necessarily blow up when the size of the basis is increased, even if $f$ has a stable expansion in a polynomial basis.
Therefore, in practice, the smoothing of the density, the truncation of the basis, and the regularisation of the regression, must be carefully coordinated and adapted to the natural scale of the variations of the target function $f$, i.e. to its ``natural'' smoothness.
Failure to do so may result in a representation that has insufficient resolution to describe the response of the target property to structural deformations, or vice versa to one that contains redundant information and is prone to overfitting. 
\subsection{Long-range features and potential tails} \label{sub:ms-model}
A similar formal correspondence with well-established functional forms of physical interactions can be derived when using (scalar) multiscale LODE features~\eqref{eq:ms-lode} within an additive, linear learning model, using as target the electrostatic energy $U(A)$,
\begin{equation}\label{eq:epred}
U(A) =\sum_{i\in A} U(A_i) =\sum_{i\in A} \int\D{\Q} \rep<U||\Q> \rep<\Q||A;\field{\rho}_i[V]_i>.
\end{equation}
The fact that the representation is linear both in the density and in the potential fields allows one to derive rigorous asymptotic relationships for the interaction between two distant portions of the system, that resemble the electrostatic interactions between the multipoles of a localized charge density distribution and any other charge that is located arbitrarily far away.\cite{gris+21cs}
Focusing only on the long-range contribution $U^>$ to $U(A_i)$, that is associated with the part of $\rep|A; V_i>$ generated by the far-field density, $\rep|A; V_i^{>} >$, one can write
\begin{multline}
\label{eq:ulr-multipoles}
U^>(A_i) = 
\sum_{l=0}^{\lmax}\int\! \D{r_1}\!\D{r_2}\!\rep<U||r_1 r_2 l> \rep<r_1 r_2 l ||\overline{\rho_i^<\otimes V_i^>}>\\
=\sum_{l=0}^{\lmax} \sum_{m=-l}^{+l}  \int_{\rcut}^{\infty} \!\!\!\D{r}  \frac{1}{r^{l+1}}\rep<\lm||M_i^<(U)> \rep<\rho^>_i||r \lm>\,.
\end{multline}
In this expression, in which the reader can recognize the similarity with the multipole expansion of the electrostatic potential,\cite{stone1975} $\rep|\rho^{>}_i>$ indicates the atom density outside the cutoff, which is not computed explicitly but is encoded in the expansion of the local atomic potential~\eqref{eq:lode-v}.
The coefficients $\rep< \lm||M_i^<(U)>$ can be written as a combination of the regression weights $\rep<r_1r_2l||U>$ and the local density coefficients $\rep<rlm||\rho_i^{<}>$, and can be interpreted as adaptive multipole coefficients that depend in a general manner on the atomic distribution within the environment.

\begin{figure}[tbp]
    \centering
    \includegraphics[width=0.9\linewidth]{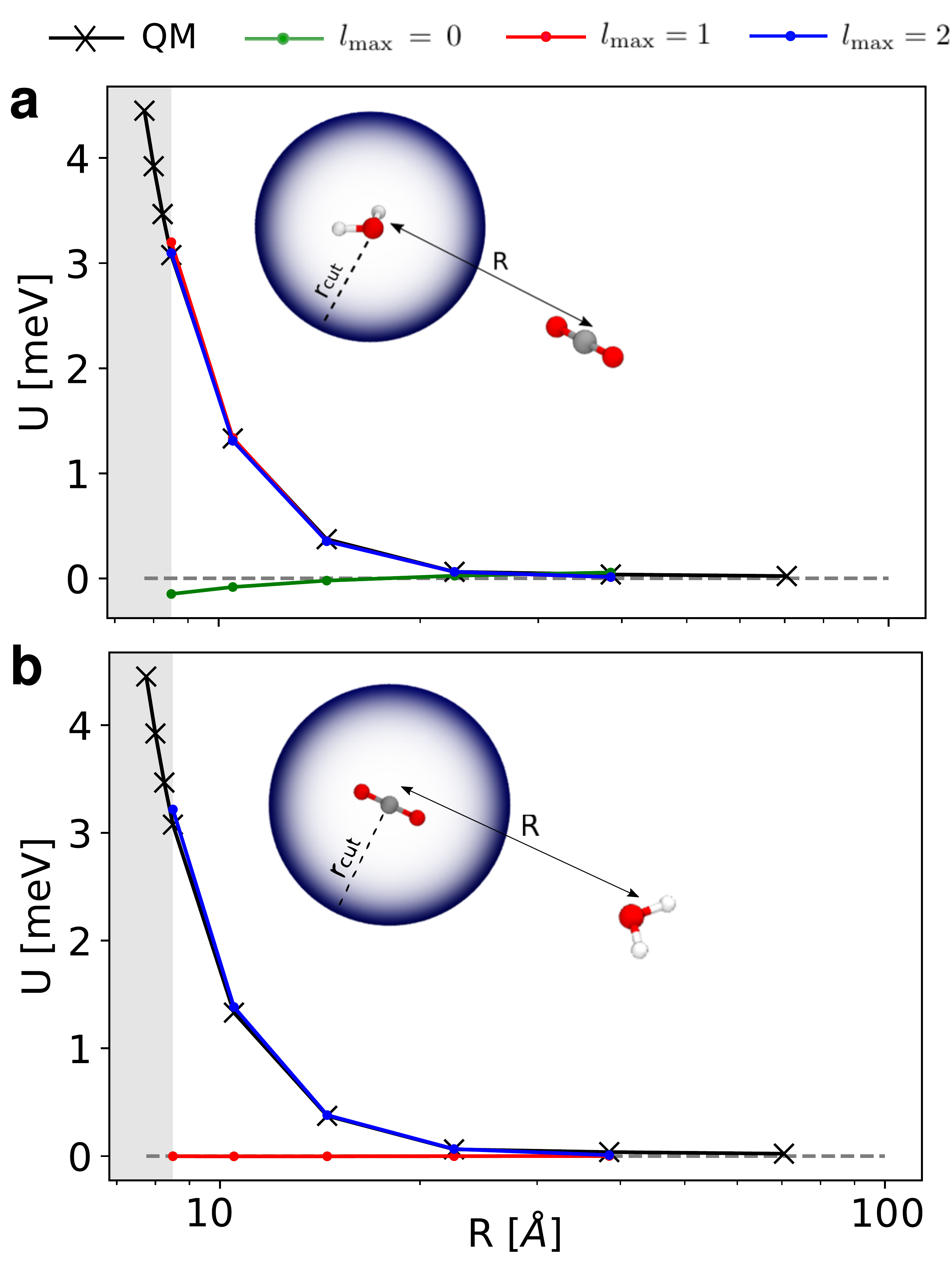}
    \caption{Extrapolated asymptotic interaction profiles for a given configuration of H$_2$O and CO$_2$ at different angular cutoff values $\lmax$. Top and bottom panels show the results of the asymptotic extrapolation when centring the representation (a) on the oxygen atom of H$_2$O  and (b) on the carbon atom of CO$_2$. Adapted with permission from Ref.~\citenum{gris+21cs}. Copyright 2020 Royal Society of Chemistry.}
    \label{fig:multipoles}
\end{figure}

Given that the atomic densities and potentials are not the physical charge density and electrostatic potential of the system, it is the role of the regression procedure to modulate the multipoles so as to reproduce the reference data for the electrostatic energy.
In Fig.~\ref{fig:multipoles} we report an example where this is demonstrated by extrapolating the long-range interaction between a pair of rigid H$_2$O and CO$_2$ molecules, upon training the multiscale LODE model on the long-range, yet not asymptotic, interaction profiles associated with 33 different reciprocal orientations of the two molecules.
The figure compares the asymptotic extrapolation performance upon centering the representation on different atoms, as well as by truncating the angular expansion at different $l_\text{max}$. It is apparent that the angular cutoff chosen reflects the number of multipoles introduced in the expansion of Eq.~\eqref{eq:ulr-multipoles} and thus determines sharp crossovers of the prediction accuracy across critical $l_\text{max}$ values. For instance, a model that uses only features centered on the oxygen atom of \ce{H2O} improves dramatically its performance when $l_\text{max}$ is increased from zero to one. A model using the carbon atom of \ce{CO2} as the only environment shows a similar, sharp improvement in accuracy when going from $l_\text{max}=1$ to $l_\text{max}=2$.
This is consistent with the primarily dipolar nature of the electrostatic field generated by a water molecule, and  with the quadrupolar nature of the center-symmetric carbon dioxide.
Even though this example showcases the link between a linear model based on $\rep|\field{\rho}_i[V]_i>$ and multipole electrostatics, the representation is sufficiently flexible to describe also other kinds of interactions, as demonstrated in Figure~\ref{fig:ms-lode}.

\subsection{Non-linear models}
\label{sub:non-linear-model}

Historically, linear representations used basis sets in internal coordinates (typically interatomic distances or simple transformations of them) that exploded in size with body order, see e.g. Refs.~\citenum{coll-pars93jcp,bowm+09jcp}, and with exponential scaling in their computational cost of prediction due to the need to sum over all $\nu$-clusters in a configuration or atomic environment.
Moreover, it is clear that high body orders would be needed to obtain the desired accuracy, especially for models of materials.
About a decade ago, {\em nonlinear} fits using low body order ($\nu=2$) descriptors appeared, with the surprising result that a few hundred degrees of freedom were enough to get good potentials\cite{behl-parr07prl,bart+10prl}.
Contrary to linear modeling where the symmetry-adapted features $\rep<\q||A_i>$ are used as a basis, in the context of non-linear regression they are best thought of as a coordinate transformation. 
In a linear setting the choice of a basis, and the details of the implementation, are a matter of computational performance but can be converged to a well-defined, basis-set independent limit. 
When taken as the input of a non-linear model, instead, the entries of the feature vector must always be precisely defined, because there is no complete basis set limit in which the models become equivalent.
To emphasize that many of the formal manipulations that are possible in a linear context take on a different meaning when features are used for a non-linear model, we abandon the Dirac notation and indicate as $\bfeat(A_i)$ the feature vector that describes the atom-centred environment $A_i$, whose components are
$\feat_{\q}(A_i)=\rep<\q||A_i>$.
If $\y(A_i)$ is a symmetric property such as a site energy, we aim to construct approximations of the general form
\begin{equation}\label{eq:models-nonlinear}
    \y(A_i) \approx \yt\big(\bfeat(A_i)\big).
\end{equation}

The two most commonly used models for $\yt$ are artificial neural networks\cite{behl11pccp,gast+17cs,schu+18jcp,smit+17cs,Mills2019,gilm+17mlr,Kondor2018,wang+18cpc,chemrev2021behler,chemrev2021smallmolecules} (ANN) and kernel ridge regression\cite{bart+17sa,fabe+18jcp,bere+18jcp,caro19prb,glie+18prb,Chmiela2017,paru+18ncomm,Botu2015,gris+19acscs,chemrev2021gdml} (KRR) models.
In KRR models,\cite{Saunders1998} one builds a kernel matrix $\mbf{K}$ with elements
\begin{equation}
        K_{ij} = \krn(\bfeat(A_i) , \bfeat(A_j) ),
\end{equation}
which provides a similarity measure between the environments $A_i$ and $A_j$, measured in terms of the similarity between the corresponding feature vectors $\bfeat(A_i)$ and $\bfeat(A_j)$.
Useful kernel functions, $\krn$,  are nonlinear, e.g. polynomials, Gaussians, etc.\cite{rasm05book}.
The kernel inherits the symmetry of the feature vectors, and therefore a model for a symmetry-invariant property $\y(A_i)$ can be obtained as
\begin{equation}
    \yt(A_i) = \sum_{j\in M} b_j \krn(\bfeat(A_i) , \bfeat(M_j) ),
\end{equation}
where, in the simplest setting, the $M_j$ are scattered interpolation points, but more generally are simply a collection of ``centers'' which induce a basis $\{\krn(\cdot, \bfeat(M_j))\}_j$ in the symmetrized feature space.
The weights $b_j$ are then obtained by a linear regression.
Kernel models have two main advantages over ``naive'' linear regression using the same features. (1) They introduce implicitly a non-linear mapping between the inputs and a ``reproducing kernel Hilbert space'' $\rep|A_i>\rightarrow \rep|A_i; \krn>$, which has a larger (often infinite) dimensionality, allowing for a more flexible approximation of $\y(A_i)$. (2) Given that the basis is centered on the training points, it is adapted to the geometry of the data set in feature space.
For example, if the centers $\rep|A_i; \krn>$ in feature space fall on (or close to) a low-dimensional manifold then the KRR model naturally exploits this. For a comprehensive discussion of the use of kernel methods in atomistic modeling, see Ref.~\citenum{chemrev2021deringer}.
In the context of body ordered features discussed above, the non-linearity in the kernel effectively increases the body order of the features used in the regression model, but in a rather special way: only those high body order terms are present that can be obtained as functions of low body order features.
See \cref{sec:incompleteness} on completeness for a more detailed discussion.

\begin{figure}
    \centering
\includegraphics[width=1.0\linewidth]{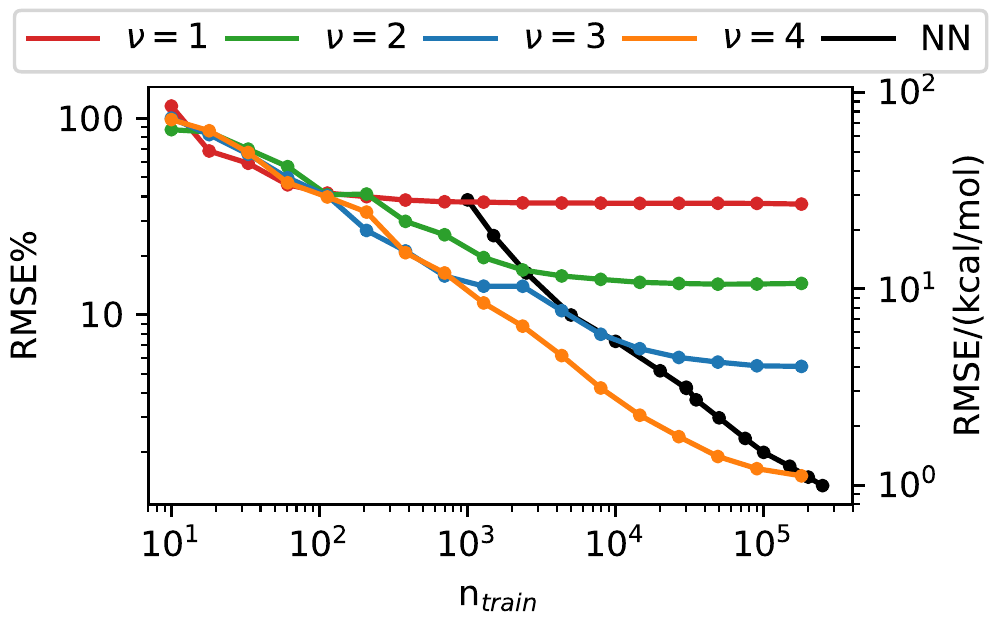}
\caption{Learning curves for the formation energy of CH$_4$ structures using linear models based on NICE features truncated to increasing body order $\nu$ ($\rcut= 6$\AA{}, $\nmax= 10$, and $\lmax= 10$, up to 3200 invariants retained at each body order) and an ANN model using NICE features up to $\nu= 4$. Errors are expressed both in absolute terms and as a percentage of the standard deviation of the dataset. The models are trained using features centered on both C and H. 
Reproduced with permission from Ref.~\citenum{niga+20jcp}. Copyright 2020 American Institute of Physics.
}
    \label{fig:nice-methane}
\end{figure}

While nonlinear models are by their very nature more flexible in representing complex high-dimensional features, linear models come with different advantages.
As we have shown in \cref{sub:linear-models} and \cref{sub:ms-model}, they tend to be more easily ``interpretable'', e.g. in terms of a body-ordered expansion of the target properties, or in terms of  physically-motivated asymptotic forms of the interactions.
But are nonlinear models necessary to achieve high accuracy?
This notion is challenged by  the SNAP,\cite{thom+15jcp,wood-thom18jcp}, the MTP,\cite{shap16mms}, the ACE\cite{drau19prb} and the NICE\cite{niga+20jcp} representations: the ``density trick''  and its generalisations to higher body-orders, replacing polynomials with correlations of the atom-centered density, circumvents both the explicit symmetrization as well as the summation of all $\nu$-clusters of traditional body ordered expansions.
Particularly when using density correlations above $\nu=2$, it is critical to fully exploit the computational cost gains offered by permutation symmetric properties.
Even if one were to initially specify a model in terms of the ``natural'' body-order expansion~\eqref{eq:bo-sum-nu},
one should convert it for computationally efficient evaluation to one of the many representation built in terms of $\nu$-correlations.
By employing the recursive evaluations introduced in Refs.~\citenum{shap16mms,Bachmayr2019,niga+20jcp}, this transformation makes it possible to truncate at very high body-orders without significant penalty in computational cost, as discussed in more detail in~\cref{sub:implementation}.
As an illustration of how a linear fit based on high-quality density-correlation representations can compete with non-linear models we show in \cref{fig:nice-methane} the learning curves resulting from the regression of the atomization energy for a very large and geometrically diverse database of \ce{CH4} configurations (generated by randomly displacing the H atoms around the central carbon, in a sphere with a radius of 3.5\AA{}).
The plot reflects a tradeoff between model complexity and the availability of training data. Saturation of the learning curves indicates that the model does not have sufficient flexibility to describe fully the underlying structure-property relations.\cite{huan-vonl16jcp,bart+17sa}
Thus, linear models based on NICE features incorporating higher and higher body order are capable of describing the structure-property relations to a higher degree of accuracy, which is apparent in the delayed saturation of the learning curve.
One sees that a $\nu=4$ model starts saturating around $n_\text{train}=10^5$, even though the system is composed of 5 atoms, and so the body-ordered expansion should be fully converged.
This is because a linear model requires a \emph{complete} basis, while here we select only a few 1000s invariants at each body order.
A NN model can be designed to be more flexible and  beat this saturation, at the expense, however, of performance in the small data set limit - which, in a more chemically and structurally diverse regression exercise, usually translates to poorer transferability.

\section{Alternative notions of completeness}
\label{sec:incompleteness}
Suppose we are given a finite collection of symmetry adapted features $\bfeat(A) = \left\{\rep<\q||A>\right\}_\q$ which we wish to use as a {\em descriptor} for atomic structures or environments, for example symmetrized correlations of the density as described in the foregoing sections.
In Section~\ref{sec:models} we discussed two classes of models built from such equivariant features: linear models, 
\begin{equation}
    A \mapsto \sum_q \rep<y||q>\rep<\q||A>,
\end{equation}
for which the representation $\bfeat(A)$ plays the role of a basis to expand the target property; and nonlinear models, 
\begin{equation}
    A \mapsto \yt(\bfeat(A)),
\end{equation}
where the representation plays the role of a coordinate transformation generating a finite-dimensional feature vector used as the argument of a non-linear function $\yt$.
In order to guarantee systematic convergence of these models to an arbitrary target, in suitable limits, we require that the employed set of features is {\em complete}. We already hinted in Section~\ref{sub:non-linear-model}  that these two scenarios lead to different requirements on the notion of completeness. In this section we provide a more in-depth discussion of the completeness issue in the nonlinear setting, and point out open problems.

Recall from Section~\ref{sub:non-linear-model} that for linear models the correct notion of completeness is the well-known and well-understood concept of a complete (linear) basis from linear algebra. In the context of a nonlinear model $\yt(\bfeat(A))$ it is instructive to think of $\yt$ as a universal approximator in feature space (e.g., an ANN, GP, etc). We then ask the question whether (in a suitable limit) the model can represent an arbitrary symmetric property $\y(A)$, i.e., whether 
\begin{equation}
    \y(A) = \yt(\bfeat(A)),
\end{equation}
is achievable. This is the case if and only if the mapping $A \mapsto \bfeat(A)$ is {\em injective}: this means that any two atomic configurations that are {\em not} related by symmetry are mapped to different descriptors. In particular knowledge of $\bfeat$ would then enable us in principle to reconstruct the configuration $A$. When this is the case, we say that the descriptor $\bfeat$ is {\em geometrically complete}.

\subsection{A pedagogical example}
\label{sub:field-completeness}
The ideal goal would be to have complete {\em finite} feature sets, that allow to approximate any symmetric function of the coordinates to arbitrary accuracy.  As an elementary introduction to how such a construction might be achieved in principle, we consider a collection of $N$ particles in 1D, $\{x_i\}_{i = 1}^N$. As a concrete example, one can take two particles with positions $(x_1, x_2)$.
In the absence of an angular component, we only need to consider the projection of the density $\rho(x) = \sum_i \delta(x - x_i)$ onto the monomial basis $x^n$:
\begin{equation} \label{eq:n-rho-1d}
\rep<n||\rho> = \sum_{i = 1}^N x_i^n, \quad n \in \mathbb{N}
\end{equation}
For example, if $N=2$, $\rep<1||\rho>=x_1+x_2$, $\rep<2||\rho>=x_1^2+x_2^2$, etc.
In this simple setting, one sees easily how the $\nu$-point density correlations form a basis of symmetric polynomials
\begin{equation} \label{eq:nu-rho-1d}
    \rep<n_1 \cdots n_\nu||\rho^{\otimes \nu}> =
    \sum_{i_1 \ldots i_\nu} x_{i_1}^{n_1} \cdots x_{i_\nu}^{n_\nu}
    = \prod_{k = 1}^\nu \rep<n_k||\rho>
\end{equation}
which is complete (in the sense of a linear basis) because it contains all possible symmetrized monomials. In analogy to what we did in Section~\ref{sub:linear-models}, we use the ``self-interaction'' formulation in which the sum extends over all the tuples of particle indices.
For the case of two particles, linear combinations of $\rep<n_1 n_2||\rho^{\otimes 2}> = x_1^{n_1+n_2}+ x_1^{n_1} x_2^{n_2} + x_2^{n_1} x_1^{n_2} + x_2^{n_1+n_2} $ are sufficient to write any symmetric polynomial of the particle positions.

Thus, if we allow for \emph{algebraic} operations on the $\rep<n||\rho>$, it is clear that the $\nu=1$ coefficients provide a sufficient basis, because the elements of the linear basis~\eqref{eq:nu-rho-1d} can be obtained as a product, e.g. $\rep<n_1 n_2||\rho^{\otimes 2}> = \rep<n_1||\rho> \rep<n_2||\rho>$.
In fact, well-established results from the theory of symmetric polynomials\cite{macd15book} allow making an even stronger statement. The first $N$ power sum polynomials $\left(\rep<n||\rho>\right)_{n=1}^N$ provide an algebraically-complete basis to write any symmetric polynomial function of the coordinates of $N$ particles. For instance, for $N=2$ we can express the $n=3$ term as a polynomial of $\rep<1||\rho>$ and $\rep<2||\rho>$
\begin{multline}
\rep<3||\rho> = x_1^3+x_2^3 =
\frac{3}{2}(x_1+x_2)(x_1^2+x_2^2)-\frac{1}{2} (x_1+x_2)^3
\\=\frac{3}{2}\rep<1||\rho>\rep<2||\rho>-\frac{1}{2}\rep<1||\rho>^3
\end{multline}
This result implies, in general, that the mapping
\begin{equation}
    \left\{ x_i \right\}_{i = 1}^N \mapsto \bfeat=\left\{\rep<n||\rho>\right\}_{n=1}^N
\end{equation}
is injective: knowledge of the first $N$ features $\rep<n||\rho>$ allows us to uniquely reconstruct the configuration (but not the index of the atoms). 
That is, this {\em minimal feature set} $\bfeat(A)$ is indeed
 {\em geometrically complete}.
It is not too difficult to construct similar complete and finite feature sets for finitely many particles in two and three dimensions as long as only permutational symmetry is considered. However, incorporating also rotational symmetry into the equivalence of particle configurations makes this much more challenging as we discuss next.

\subsection{Geometric completeness of density correlations}\label{sub:geo-completeness}
In general, for three-dimensional atom configurations it is clear that taking {\em all} $\nu$-correlations provides a complete set of features (after all, they are even complete in the sense of forming a complete linear basis), however, as we explained at the beginning of Sec.~\ref{sec:incompleteness}, this is not a practically useful property when considering nonlinear regression schemes.
As we explain next, it remains an open problem how to construct a minimal complete feature set in this general setting.

It is clear just based on dimensionality arguments that a descriptor that has fewer than $3N-6$ components (the number of elements in the Cartesian position vectors, subtracting the degrees of freedom associated to translations and rotations) cannot be complete for $N$ particles. On the other hand, the descriptors based on $\nu$-point correlations have a number of components that scales with $N^\nu$. But having more than the necessary minimum number of components does not ensure that a descriptor is complete.

\begin{figure}[btp]
    \newlength{\pica}\settowidth{\pica}{$\boldsymbol{\pi-\alpha}$}
    \newlength{\gnat}\settowidth{\gnat}{4'}
    \newlength{\gnatt}\settowidth{\gnatt}{$\pi-\alpha$}
    \centering
    \includegraphics[width=3cm]{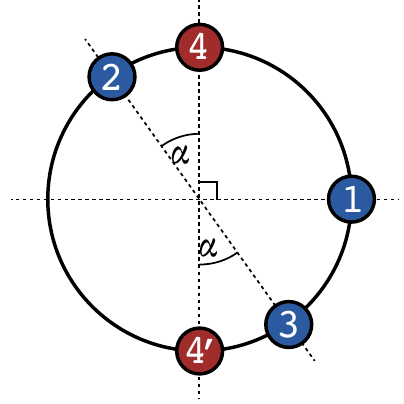}
    \vbox{
    \hbox{\begin{tabular}{l|cccc}
     &\hbox to\gnatt{\hfil1\hfil}&2&3&4\\\hline
    1&0&$\frac{\pi}{2}+\alpha$ & $\frac{\pi}{2}-\alpha$ & $\frac{\pi}{2}$\\
    2&&0&$\pi$&\colorbox{lightgray}{\makebox[\pica]{$\boldsymbol{\alpha}$}}\\
    3&&&0&\colorbox{lightgray}{$\boldsymbol{\pi-\alpha}$}\\
    \hbox to\gnat{4}&&&&0
    \end{tabular}}
    \medskip
    \hbox{\begin{tabular}{l|cccc}
     &\hbox to\gnatt{\hfil1\hfil}&2&3&4'\\\hline
    1&0&$\frac{\pi}{2}+\alpha$ & $\frac{\pi}{2}-\alpha$ & $\frac{\pi}{2}$\\
    2&&0&$\pi$&\colorbox{lightgray}{$\boldsymbol{\pi-\alpha}$}\\
    3&&&0&\colorbox{lightgray}{\makebox[\pica]{$\boldsymbol{\alpha}$}}\\
    4'&&&&0
    \end{tabular}}
    }
    \caption{A pair of environments that are not distinguished by two-correlations (sets of distances from the origin and central angles), formed from the blue atoms (1-3) and either one of 4 or 4'. The angle $\alpha$ is arbitrary. The tables on the right show the angles (or equivalently, distances) between the numbered particles in each configuration. The two environments are not related by symmetry, but the sets of distances are identical, only a pair are swapped, that are highlighted with the gray background.
    }
    \label{fig:degeneracy}
\end{figure}

Although it was appreciated for a long time that symmetrized two-correlations for entire structures are not complete, i.e. knowing the set of distances between points is not enough to reconstruct the point set\cite{bart+13prb,vonl+15ijqc, bout-kemp04aam}, it was not until recently that the connection to environment descriptors was made\cite{pozd+20prl}.
The fact that degenerate pairs of inequivalent environments mapping to the same descriptor exist for two-correlation (distance-angle) representations came as a surprise because so many ``successful'' models for potential energy surfaces have been published based on such descriptors in the past decade\cite{behl16jcp,chemrev2021deringer}.
An example of such ``degenerate pair'' is given in Figure~\ref{fig:degeneracy}. The construction involves an environment with four neighbors on the unit circle, with the two structures corresponding to the labels $(1,2,3,4)$ and $(1,2,3,4')$ being different, but having the same unordered list of distances and angles. The total number of degrees of freedom for this layout is three (because one neighbor can be fixed on the $x$ axis), and there is one degree of freedom in the construction of the degenerate pair (the angle labelled $\alpha$ in Fig.~\ref{fig:degeneracy}).
Thus, this manifold of pairs of degenerate configurations has a {\em codimension} of two, i.e. it has a dimensionality that involves two fewer degrees of freedom than the total.
A more general construction, that yields a family of 3D degenerate pairs including an arbitrary number of neighbors, is discussed in Ref.~\citenum{pozd+20prl}.
The fact that the degenerate pairs form a manifold does not mean that there is a {\em degenerate manifold}, i.e. a manifold of configurations all mapping to the same descriptor. This type of  degeneracy occurs between pairs of configurations which are typically far from one another, and so this degeneracy problem differs from that of assessing the sensitivity of a representation to small atomic displacements\cite{onat+20jcp,pars+20mlst}.

As was shown in Ref.~\citenum{pozd+20prl}, in order to break this degeneracy, the correlation order has to be increased. Three-correlations ($\rep|\frho_i^3>$, equivalent to the unordered set of central tetrahedra, and the bispectrum of the atomic density) indeed distinguish environments such as those in Fig.~\ref{fig:degeneracy}. It is however possible to build pairs of environments, composed of 7 or more neighbors, which are distinct but have the same three-correlations.
This example, also discussed in Ref.~\citenum{pozd+20prl}, raises a number of open mathematical questions: (i) is the $\nu=3$ descriptor complete for $N < 7$ neighbors, (ii) are all $\nu$-correlations degenerate for sufficiently many neighbors, (iii) what is the codimension of the manifold of degenerate configurations for $\nu > 2$?

The concept of completeness applies both to representing entire structures and to atomic environments, but the relationship between these two cases is subtle.  Given an entire structure, it can be considered to be the ``environment'' of the point at the origin, and the same symmetries apply. However, specific representations appear differently in the two views. For example, the $\nu=2$ correlations around a central atom contain information on the full set of interparticle distances between the neighbors, and so any pair of environments that are degenerate in terms of $\rep|\frho_i^2>$ is also (removing the particle at the origin) a pair of \emph{structures} with a degenerate description in terms of distances.\footnote{While the opposite is generally not true, the usual example of the pair of degenerate tetrahedra, as well as any pair of degenerate structures that can be inscribed in a sphere, correspond to a pair of environment placed at the center of such a sphere that is degenerate for $\nu=2$ correlations.}
Note that the problem of completeness for entire structures is exactly the same as the problem of reconstructing point sets\cite{bout-kemp04aam}.

One way to break the degeneracy between the representations of two entire structures involves combining information on different environments.
For instance, one can describe the entire structure using an additive combination of atom-centered features analogous to Eq.~\eqref{eq:additive-rho2}.
Following the above reasoning, a pair of environments that are degenerate in terms of the list of  distances and angles are also (removing the central atom) structures that are degenerate in terms of the list of distances.
However, these structures are not necessarily degenerate in terms of the combined list of distance and angle histograms of each local environment.
Thus, taking non-linear transformations of atom-centered features cannot resolve the environment-level degeneracies, but can provide a way to differentiate entire structures.\cite{pozd+20prl}
The construction of injective yet concise representations for environments and structures is still an open problem, whose solution may help to improve the accuracy and computational efficiency of machine-learning models.

Note that in this discussion we are implicitly taking atomic structures related by symmetry as identical, and  we focus on whether the injectivity holds for the domain of the descriptor map being the original atomic structures. The case of whether the same consideration hold for general scalar fields (e.g. those arising in the LODE construction) is a separate problem. For the case of translation symmetry (torus geometry) it is well-known that no finite correlation order suffices to reconstruct all signals\cite{Yellott1992-gh}, however {\em most} signals can be reconstructed already from the bi-spectrum ($\nu=3$). To the best of our knowledge it is an open problem whether analogous results hold for the case of rotational symmetry of 3D spherical geometry\cite{Kakarala:2012dt,kaka12jmiv}. See also \citet{Uhrin2021} for an excellent review connecting 3D signal processing and reconstruction of atomic configurations.

\subsection{Spectral representations}
\label{sub:spectral_rep}
As we explained above the set of all $(N-1)$-correlations is complete for $N$ particles, because it is equivalent to the completeness of polynomial basis sets such as MTP\cite{shap16mms}, PIP\cite{bowm+09jcp}, aPIP\cite{vand+20mlst}, ACE\cite{drau19prb,seko+19prb} and NICE\cite{niga+20jcp} (see also Section~\ref{sub:linear-models}).  Any of these bases can be expressed in terms of the $\nu$-correlations via a linear transformation, and vice-versa. Even for fixed maximum polynomial degree, these are enormous representations. Depending on how $\nu$-correlation features are chosen their number might scale as rapidly as $\binom{\qmax+\nu}{\nu}$, where $\qmax$ is the number of one-particle features.

There is a class of much lower-dimensional descriptor maps based on the eigenspectra of overlap matrices\cite{sade+13jcp,zhu+16jcp} that lifts the degeneracy for the known examples, although their actual completeness is unknown. A simplified construction of these ``spectral representations'' proceeds as follows: First, one constructs an artificial overlap matrix based on the positions of atoms within the $i$-centered environment $A_i$:
\begin{equation} \label{eq:overlap-elements}
    T_{jj'} = f_{\rm cut}(r_{ji}) f_{\rm cut}(r_{i j'}) t(r_{jj'}),
\end{equation}
where $t : \mathbb{R} \to \mathbb{R}$. Then, one computes the ordered spectrum $\{\tau_k\}_{k=1}^N$ of $T$. If $T$ is invariant (or covariant) $\{\tau_k\}_k$ is an invariant descriptor of $A_i$. Due to eigenvalue crossings, the mapping $A_i \mapsto \{\tau_k\}_k$ is non-smooth, hence one may wish to  project it on a smooth basis, e.g. polynomials,
\begin{equation} \label{eq:overlap-eva}
 \rep<n||A_i; \mbf{T}>  := \sum_k (\tau_k)^n.
\end{equation}
The spectral features (or, {\em fingerprints} as they are also called\cite{zhu+16jcp}) $\{ \rep<n||\mbf{T}>\}_{n = 1}^N$ correspond to the moments of the histogram of eigenvalues, and contain precisely the same information.

An alternative way to write $\rep<n||T>$ is
\begin{equation} \label{eq:goed:defn_Fk_v2}
    \rep<n||\mbf{T}> = \operatorname{Tr} \mbf{T}^n,
\end{equation}
which is {\em not} computationally more efficient, but highlights the close connection between $\{\rep<n||\mbf{T}>\}_n$ and the body-ordered features we discussed in previous sections.
From \eqref{eq:goed:defn_Fk_v2} we observe that
\begin{equation} \label{eq:goed:proj_nbodyhist}
\begin{aligned}
  \rep<1||\mbf{T}> &= N t(0), \\
  \rep<2||\mbf{T}> &= \sum_{j_1, j_2}  t(r_{j_1 j_2})^2 \cdot f_{\rm cut}(r_{i j_1}) f_{\rm cut}(r_{i j_2}) \\
  \rep<3||\mbf{T}> &= \sum_{j_1, j_2, j_3} t(r_{j_1j_2})t(r_{j_2j_3})  t(r_{j_3j_1}) \cdot \prod_{\alpha=1}^3 f_{\rm cut}(r_{i j_\alpha}),
\end{aligned}
\end{equation}
and so forth. That is, $\rep<n||\mbf{T}>$ contains the projection of the histogram of $n$-simplices onto a single basis function. In other words, for $n=2$, the cutoff function $f_{\rm cut}$ and the overlap function $t$ play the role of $R_n$ and $P_l$ in~\eqref{eq:nu2-acsf}.
More in general, $\rep<n||\mbf{T}>$ describes $n$-neighbors correlations, and so it could be written, in principle, as a linear combination of a complete set of $\rep|\frho_i^n>$ features. 
Thus, the $\rep<n||\mbf{T}>$ provide invariant high body-order features at relatively low computational cost, even though each scalar overlap matrix $\mbf{T}$ contains information on a single feature per body order.
If one takes $t$ to be scalar (as we have done here) then there are at most $N$ invariant features for $N$ neighbors, but $3N-6$ independent coordinates -- so that the spectral features \eqref{eq:overlap-eva} must be grossly undercomplete.
This source of incompleteness is easily lifted by simply taking multiple overlap matrices with different $t$ functions, or taking $t$ to be matrix-valued, as done in Ref.~\citenum{zhu+16jcp}.
However, even with that modification in mind, it is not at all understood whether these features are complete or can be made complete with limited modifications. For example it can be shown\cite{Bachmayr2019, niga+20jcp} that {\em most} high-body order features are actually polynomials of low body-order features, which means that they do not contain {\em genuine} high correlation information. This can be observed very easily with a seemingly trivial modification to the spectral representation construction.
Consider $N$ particles on the unit-circle at positions $\br_{ji}$, as in Fig.~\ref{fig:degeneracy}. In particular we then have only $N-1$ independent variables, which means that a scalar $t$ is {\em in principle} sufficient to identify the configuration. However, choosing
\[
    T_{jj'} := \cos \theta_{ijj'}
\]
it is straightforward to see that the two overlap matrices $T$ for the two configurations of Figure~\ref{fig:degeneracy} have eigenvalues $\{0, 0, 1, 3\}$. That is, this particular choice of spectral descriptor is unable to distinguish them nor any two configurations for different $\alpha$.

Even for a general atomic environment, $T_{jj'} = r_{ji} r_{j'i} \cos \theta_{ijj'}$ is the Gram matrix of the interatomic distance vectors, which has at most three non-zero eigenvalues -- and hence the collection $(\rep<n||\mbf{T}>)_{n = 1}^N$ contains at most three independent features even though {\em formally}, $\rep<n||\mbf{T}>$ has body-order $n$. For a configuration in which the neighbors lie on a sphere, this case can be written as an overlap matrix by choosing an appropriate, monotonically decreasing $t(r_{jj'})$, and for the general case with an appropriate (albeit contrived) choice of $\fcut$ and $t$.
The purpose of these examples is to highlight that, although spectral descriptors offer some attractive features such as their computationally cheap high body-order nature, understanding under which conditions they are {\it complete} is subtle and requires a much deeper investigation.

\subsection{Completeness: summary and open challenges}
To conclude our discussion of {\em completeness of representations} we briefly review and contrast the two key notions of completeness that we introduced and also mention a third concept that we implicitly encountered in Sec.~\ref{sub:field-completeness}. In the following, let $\bfeat(A) = \{ \rep<q||A>\}_q$ again denote a finite or infinite collection of equivariant features of a configuration or environment $A$.

{\it Complete linear basis:} This is the correct notion of completeness of $\bfeat$ for {\em linear models}, $\sum_q \rep<y||q> \rep<q||A>$, such as PIPs, aPIPs, MTP, ACE, NICE. It is now well-understood how to systematically generate such a complete linear basis in a variety of different ways. This is the strongest requirement one can make on a feature set.
    
{\it Geometric completeness:} This is the correct notion of $\bfeat$ completeness for nonlinear models, $\yt(\bfeat(A))$, i.e., it is the minimal requirement to ensure systematic convergence of such a model. Ensuring only injectivity of the mapping $A \mapsto \bfeat(A)$, means it is a much weaker requirement than being a complete linear basis. We therefore expect that complete feature vectors are generally significantly sparser, which is important for the performance of nonlinear regression schemes. At present, there is no systematic construction of minimal geometrically complete feature sets.

{\it Algebraic completeness:} We say that {\it $\bfeat$ is algebraically complete} if every element of a complete linear basis $\rep<q||A; \frho_i^\nu>$  can be written as a polynomial of the entries of $\bfeat$, $p_q(\bfeat(A_i))$.
This is precisely the concept we used to construct a geometrically complete feature set in the pedagogical example of Sec.~\ref{sub:field-completeness}. The set of invariants used to construct PIP\cite{bowm+09jcp} and aPIP\cite{vand+20mlst} potentials form a minimal algebraically complete descriptor. The concept was also proposed as part of the NICE framework\cite{niga+20jcp} as a mechanism to reduce the size of descriptor set.

In general, algebraic completeness is strictly stronger than geometric completeness and an algebraically complete feature set will be larger than a minimal geometrically complete one. It is nevertheless an interesting and useful concept: (i) it provides a stepping stone towards a theoretical understanding of geometric completeness; (ii) for the purpose of effective regression schemes it may in fact prove to be more important since it preserves polynomials, while  inverting a minimal geometrically complete descriptor is likely to introduce singularities. Indeed, reducing algebraic dependence is a common technique in the signal processing literature. \citet{Uhrin2021} reviews those techniques and modifies them for the construction of descriptors with relatively few entries, that can in principle be made complete.

\begin{figure*}
\centering
\includegraphics[width=0.31\linewidth]{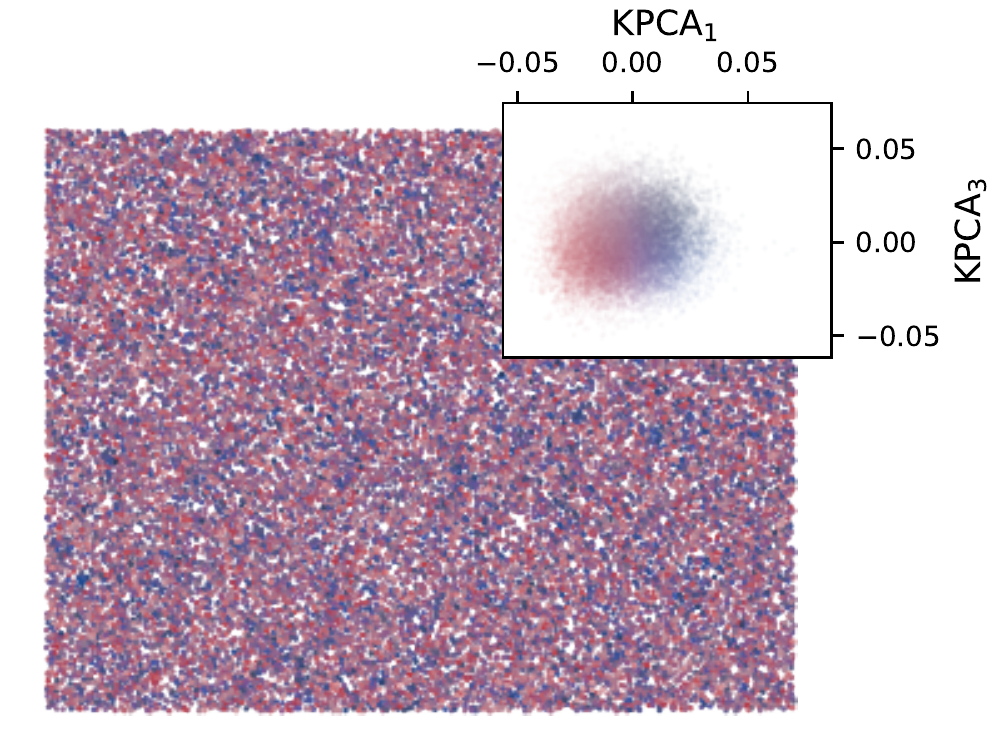}\
\includegraphics[width=0.31\linewidth]{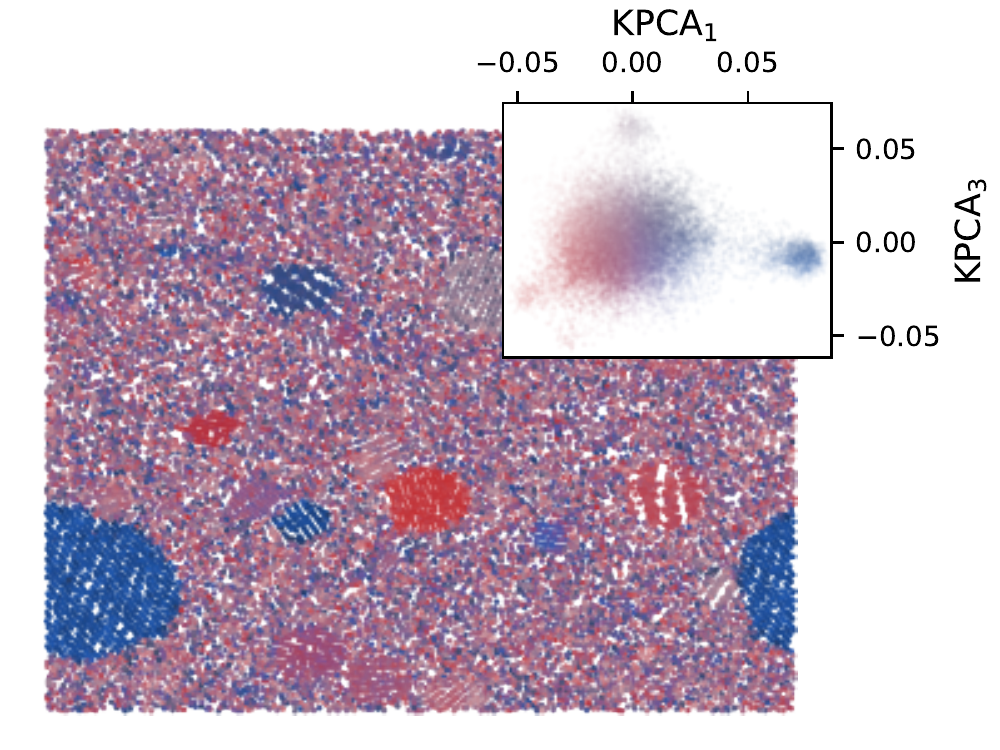}\
\includegraphics[width=0.31\linewidth]{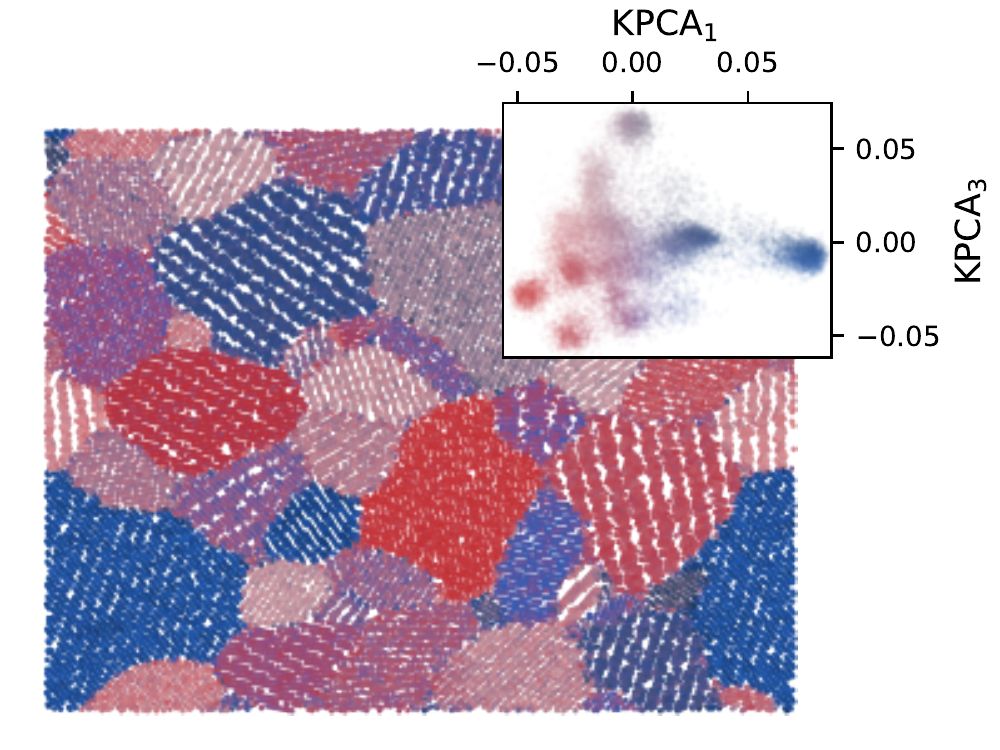}\\
\includegraphics[width=0.31\linewidth]{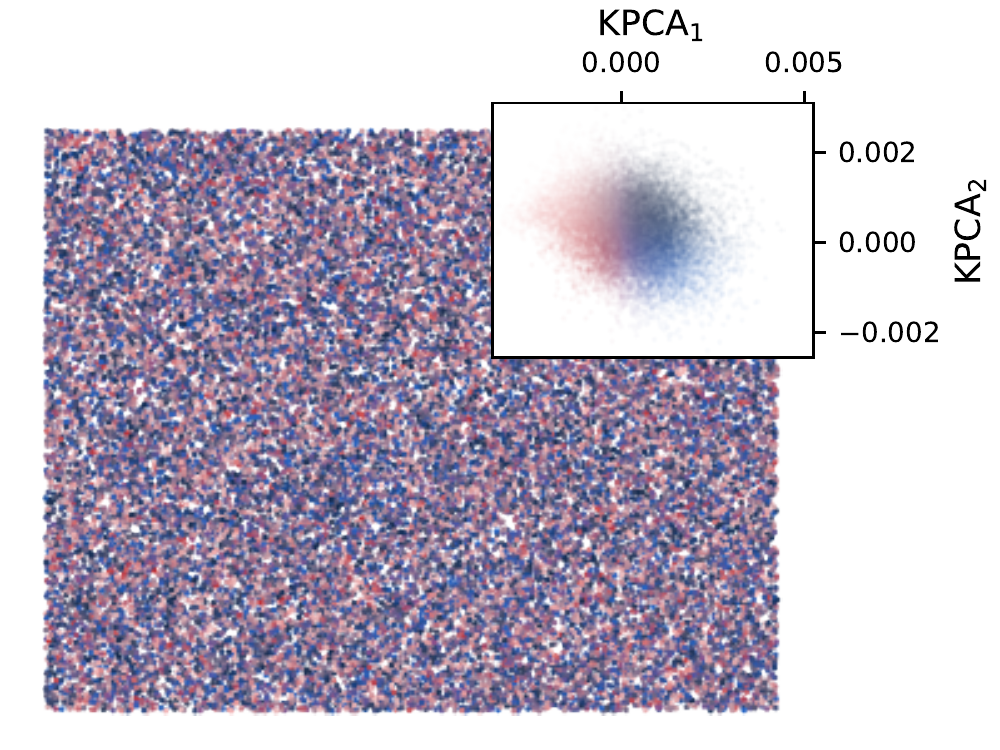}\
\includegraphics[width=0.31\linewidth]{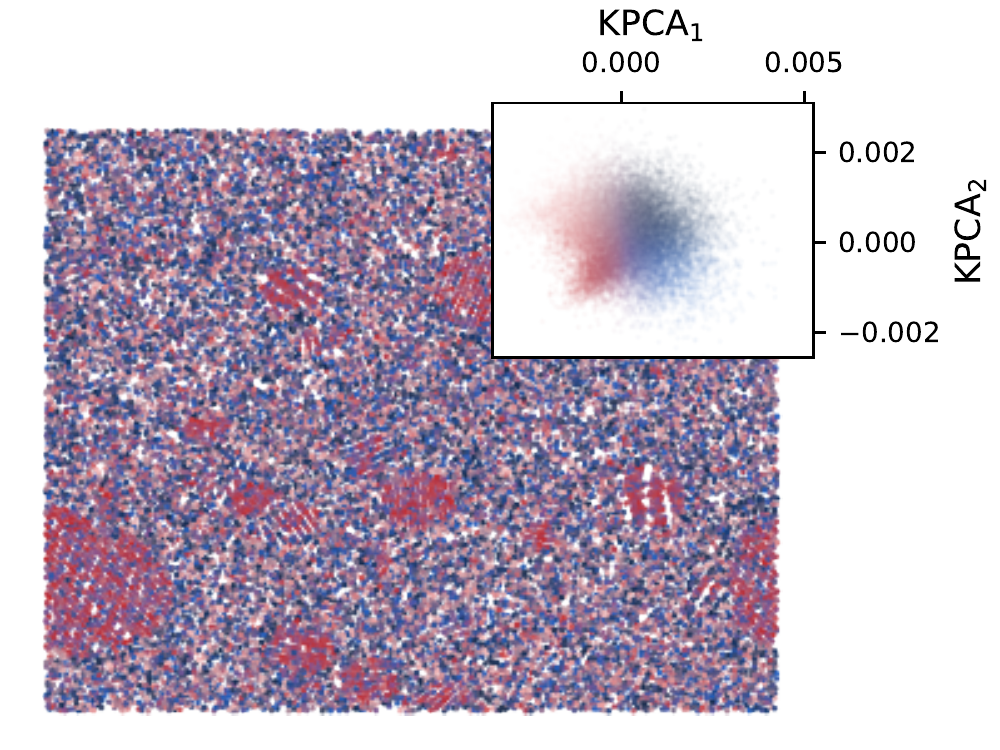}\
\includegraphics[width=0.31\linewidth]{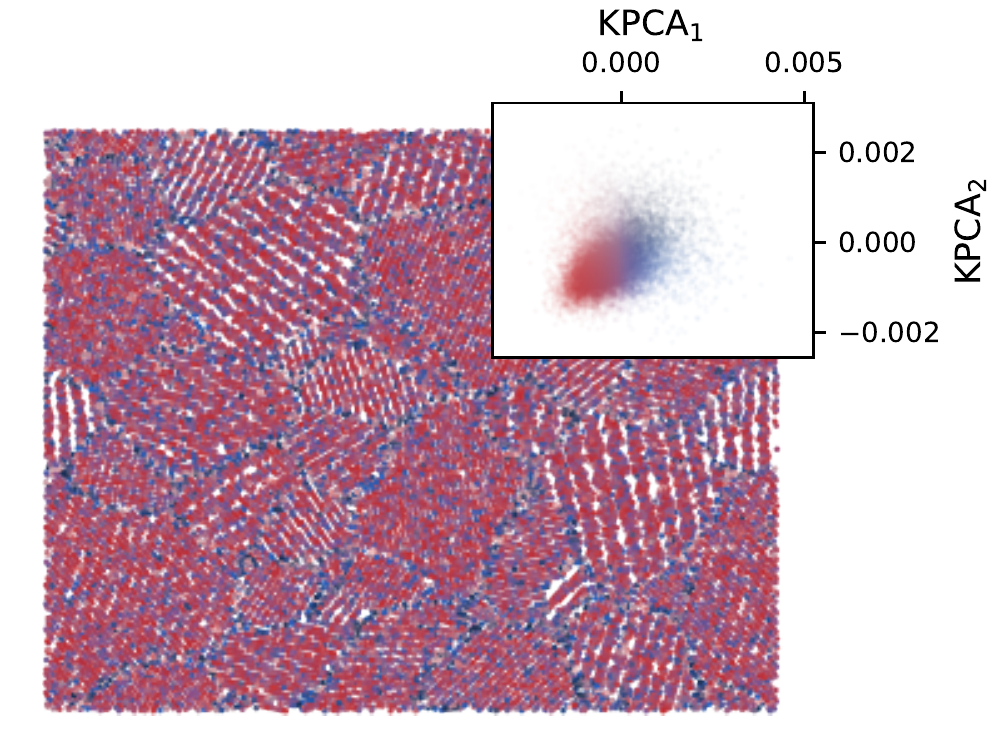}\\
\caption{Visualizing crystallization in a million-atoms simulation of undercooled iron (data from Ref.~\citenum{shib+16am}). The inset shows a KPCA map of the environments, and the atoms are color-coded following the same scheme.
Top: map and coloring based on translationally-invariant $\rep<\nlm||\rho_i>$. Bottom: map and coloring based on fully-invariant $\rep<\nnl*||\frho_i^2>$. }
\label{fig:shibuta}
\end{figure*}

\section{Representations, structures, properties and insights}
\label{sec:maps}
A mathematical representation of the structure of an atomic configuration is not only useful as the starting point of supervised-learning algorithms, aimed at predicting its energy and properties.
It can also be used, in combination with unsupervised learning schemes, to compare structures in search for repeating atomic patterns\cite{macq67proc,este+96proc,cafl06cosb,rupp+08ps,gasp-ceri14jcp,rodr-laio14science,murt-cont17wir,gasp+18jctc,piet-mart15jcp,piag-parr18pnas,kahl+18prm,chen+20arc}, to obtain low-dimensional projections that help visualize complex datasets\cite{coif+05pnas,ferg+10pnas,rohr+11jcp,ceri+11pnas,rohr+13arpc,isay+15cm,ramp+17npjcm,lemk-pete19jctc}, and more generally to describe the lie of the land in (free)energy landscapes and interpret structure-property relationships in complex systems\cite{wale03book,karp+93bc,tord-vang94jcc,helf+19jcp,chemrev2021unsupervised}.
There is a long-standing tradition of developing domain-specific descriptors to use in the automatic analysis of structural data. For instance, simulations of polypeptides have been interpreted in terms of backbone dihedral angles\cite{rama+63jmb}, discrete secondary-structure categories\cite{fris-argo96peds,kabs-sand83bp}, as well as sophisticated continuous fingerprints of secondary structure and backbone chirality\cite{piet-laio09jctc,piet+11jcc}.
Simulations of clusters and condensed-phase systems have often used more general indicators, such as Steinhardt order parameters\cite{stei+83prb}, cubic harmonics\cite{angi+10prb,cari+16pccp}, radial distribution functions (either directly\cite{vall-ogan09jcp,vall-ogan10ac} or in the form of entropy-inspired fingerprints\cite{piag-parr17jcp}), histograms of coordination numbers\cite{ceri+11pnas}, that can be seen as precursors of the atom-density correlation representations that we discuss in Section~\ref{sub:body-order-representations}.
More broadly, general-purpose descriptors that can be understood, more or less transparently, as a special case of the density-correlation features $\rep|\frho_i^\nu>$ have been developed and used in unsupervised-learning contexts as much as in the context of regression models.  A few examples include the diffraction-based fingerprints of Ziletti et al.\cite{zile+18nc}, the local order metric of Martelli et al.\cite{mart+18prb}, the spectral representations of Sadeghi et al.\cite{sade+13jcp}, the Minkowski structure metric of Mickel et al.\cite{mick+13jcp} (that closely resembles and anticipates the construction of the moment tensor potentials), and the use of SOAP features to analyze materials and molecules\cite{de+16pccp, de+16jci, bern+19acie}.

\begin{figure*}
    \centering
    \includegraphics[width=0.8\linewidth]{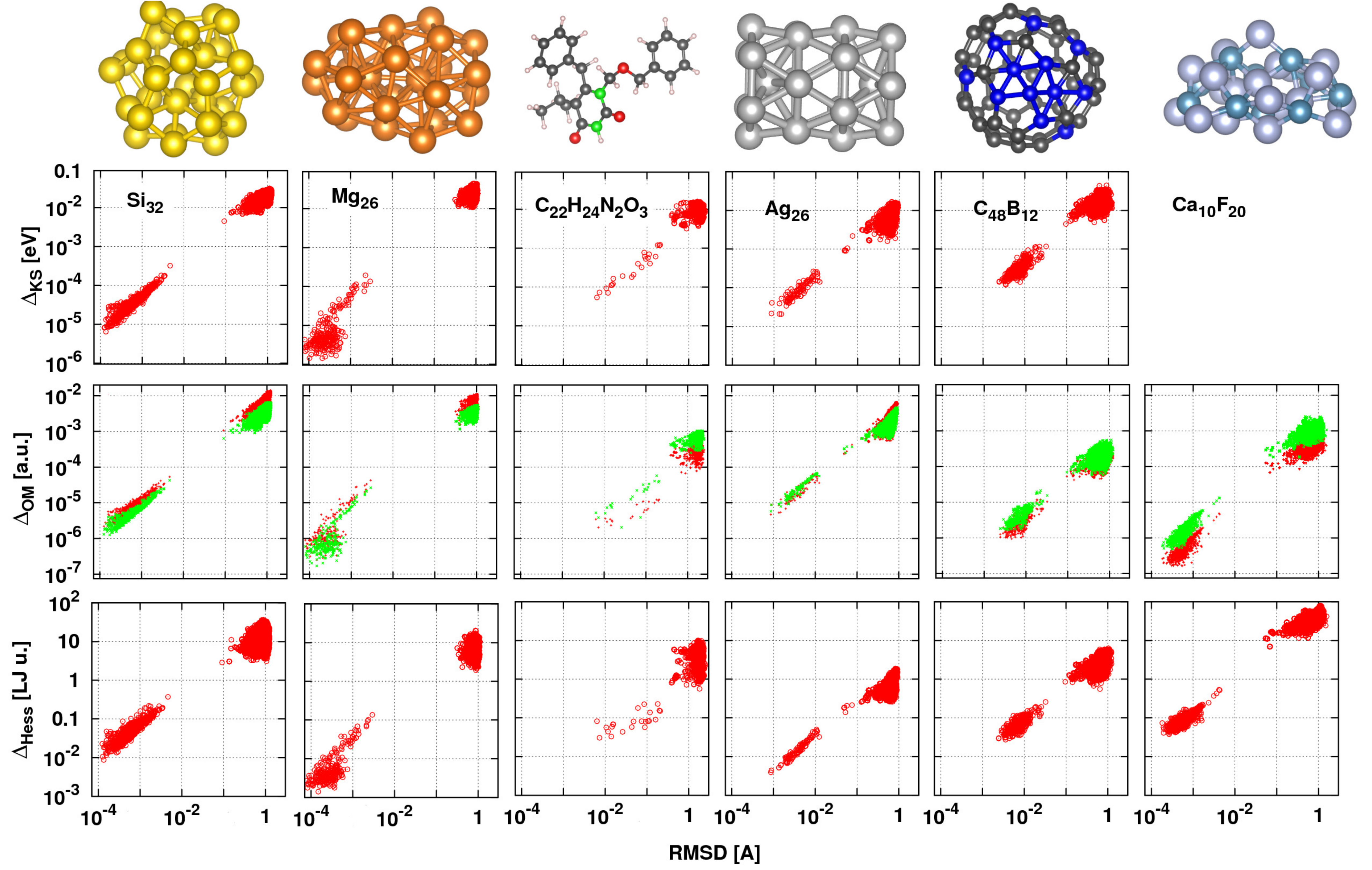}
    \caption{Comparison of distances between local minimum-energy configurations of various clusters (rows) constructed based on the sorted eigenvalues of the Kohn-Sham Hamiltonian matrix (first row), the overlap matrix (second row), and the Lennard-Jones Hessian matrix, and plotted against a permutation-invariant RMSD. For the overlap matrix, results are shown for matrices based only on $s$-type orbitals (red) and both $s$ and $p$ orbitals (green). Details of the different systems and the fingerprint construction are discussed in Ref.~\citenum{sade+13jcp}.   
    Reproduced with permission from Ref.~\citenum{sade+13jcp}. Copyright 2013 American Institute of Physics.
    }
    \label{fig:goedecker-fingerprints}
\end{figure*} 

Understanding the way a representation converts the Cartesian coordinates of atoms into features is necessary to make sense of any subsequent analysis, because any explicit or implicit assumption made in the structure-feature map will be reflected in the unsupervised analyses based on those features\cite{ceri19jcp}.
An example of this is given in  Figure~\ref{fig:shibuta}, that shows the effect of using rotationally variant or invariant features (respectively, $\rep<\nlm||\rho_i>$ and $\rep<\nnl*||\frho_i^2>$) to analyze a simulation of undercooled iron\cite{shib+16am}.
Atoms are colored according to a two-dimensional projection  describing the associated environments, in this case obtained using a kernel principal component analysis\cite{scho+98nc} built on the feature vectors $\bfeat(A_i)$.
Using orientation-dependent features makes it possible to distinguish more clearly the presence of multiple grains, and would be useful, for instance, to investigate the texture of the nanocrystalline sample, much like one would do with an electron backscattering diffraction analysis.
Using invariant features highlights that all nanocrystals have the same structure, and makes it possible to recognize the disordered environments at the grain boundaries.
This kind of analysis can also be used to elucidate the properties of different representations, investigating the effect of different choices on the unsupervised analysis of a well-understood system to better appreciate the relation between structure and features.

In this Section we summarize recent developments, and identify clear insights, related to the use of representations to determine the similarity between structures, to perform clustering and dimensionality reductions analyses, and to build models that go beyond the injective structure-property map that we have used this far.

\subsection{Features, distances, kernels}

Before delving into the use of structural representations to visualize and classify atomic configurations, let us recall the link between feature vectors $\bfeat(A_i)$, that are associated to structures or environments, and distances or kernels, that express the relationship between two of these entities.
For example, given a feature vector $\bfeat$, it is possible to define a distance using e.g. a Euclidean metric, $\dst(A_i, A'_{i'})^2 = \left\| \bfeat(A_i)-\bfeat(A'_{i'})\right\|^2$, and use as a kernel the scalar product $\krn(A_i, A'_{i'})=\bfeat(A_i)\cdot\bfeat(A'_{i'})$, or a non-linear function, e.g. an exponential of a squared distance $\krn(A_i, A'_{i'})=\exp -\gamma \dst(A_i, A'_{i'})^2$.

The opposite is also true: for a given set of configurations $M$, and any (negative definite) distance or (positive definite) kernel\cite{cutu10arxiv} it is possible to construct a set of features that generate the kernel by taking their scalar product -- a practical implementation of the concept of reproducing kernel Hilbert space that underlies kernel methods.
One only needs to construct the kernel matrix $K_{ij}=\krn(M_i, M_j)$, and find its eigenvalues and eigenvectors $\mbf{K}\mbf{u}^{(j)} =\lambda_j\mbf{u}^{(j)}$. It is easy to see that the scalar product between the reproducing features
\begin{equation}
\label{eq:rkhs}
\phi^K_j(A) = \sum_{i\in M} \krn(A,M_i) u_i^{(j)}/\sqrt{\lambda_j}
\end{equation}
computed for two members of the reference dataset yields exactly the value of the kernel function between the two configurations.\cite{scho+98nc}
It is also possible to define a kernel-induced distance 
\begin{equation}
\dst(A,A')^2 = \krn(A,A)+\krn(A,A')-2\krn(A,A').
\end{equation}
Even though different techniques may be formulated more naturally in terms of features, distances or kernels, it is always possible to translate -- at least approximately -- one description into another. 

\subsection{Measuring structural similarity}
\label{sub:measuring-similarity}

Most unsupervised learning algorithms rely on the definition of a metric to tell apart structures depending on their similarity.
A metric that is capable of identifying identical structures is extremely useful in all the applications that aim at automating the search of materials or molecules with desirable properties\cite{ogan-glas06jcp,amsl-goed10jcp,pick-need11jpcm,curt+18jctc,ogan+19nrm}.
This is not an entirely trivial task: in molecular searches, a mismatch in the simple ordering of atomic indices can lead to the failure of metrics based on the alignment of conformers, such as the root mean square distance (RMSD), and the exact calculation of a permutation invariant version would involve combinatorially increasing computational effort.\cite{sade+13jcp}.
In the case of condensed phases, one needs to deal with the problem that the same periodic structure can be described by different choices of unit cell size and orientation.
The requirements for a metric to compare atomic structures are similar to those discussed in Section~\ref{sec:history-wishes}, and have been discussed in great detail in Ref.~\citenum{sade+13jcp}:
a good metric needs to be invariant to rotations, translations, and permutations\cite{ferr+15jcp}, and still be capable of telling distinct structures apart\cite{zhu+16jcp}. The comparison between the resolving power of different metrics has been often determined using distance-distance correlation maps\cite{sade+13jcp,de+16pccp,pozd+20prl,pars+20mlst}, such as those shown in Figure~\ref{fig:goedecker-fingerprints}, that compare the distance between pairs of structures in a reference dataset, as computed by two metrics.
In the most extreme case, one observes pairs structures that are identical based on a metric, and distinct based on another -- indicating the presence of a manifold of degenerate structures that are distinct, but cannot be told apart by one of the distances\cite{pozd+20prl}.

\begin{figure}
    \centering
    \includegraphics[width=1.0\linewidth]{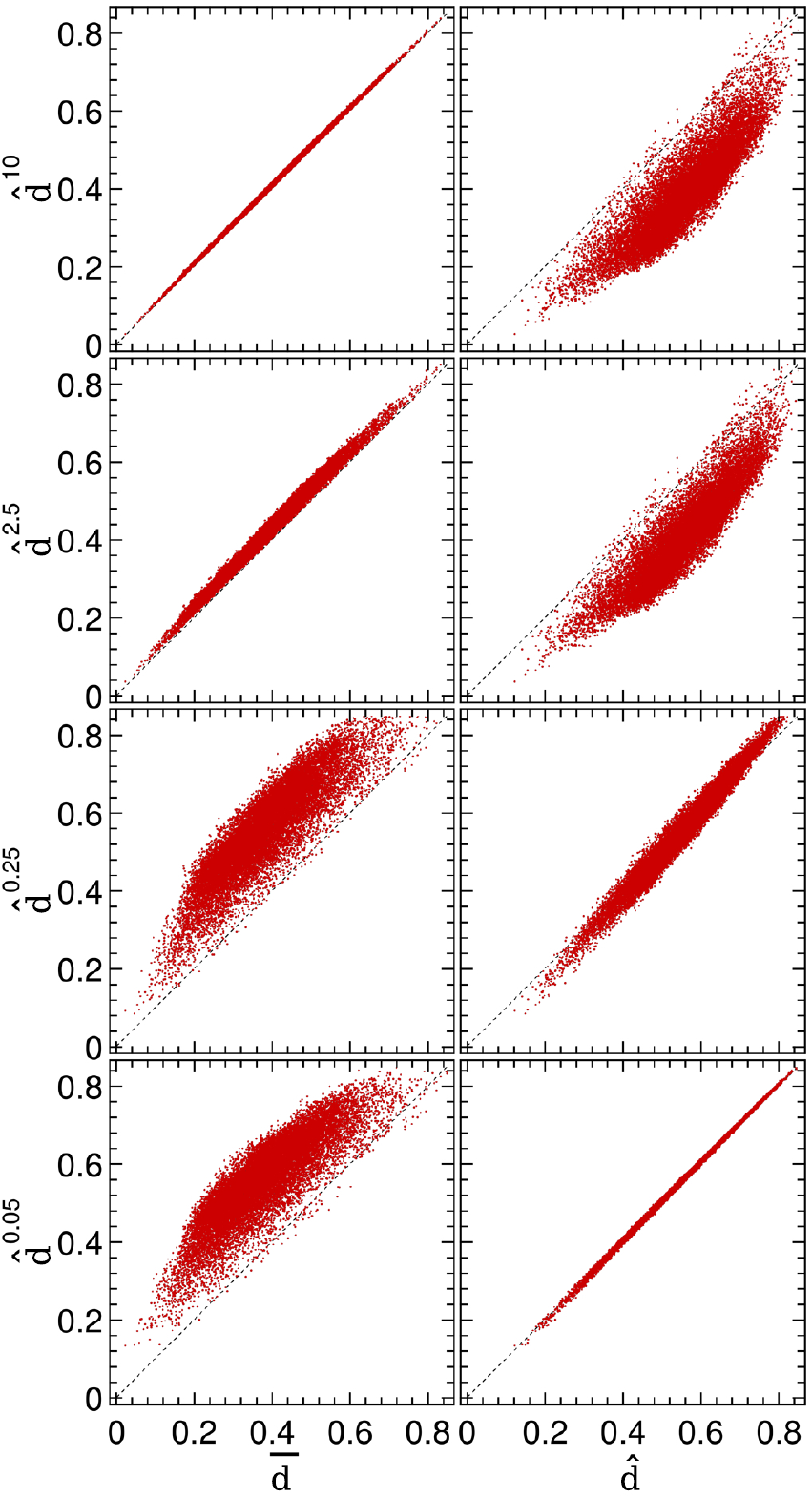}
    \caption{Distance-distance correlation plots comparing the average environment distance~\eqref{eq:d-average} to the best-match~\eqref{eq:d-bestmatch} and REMatch~\eqref{eq:d-rematch} distances, with different values of the entropy regularization parameter $\gamma$.
    The reference structures are taken from the QM7b dataset of small organic molecules\cite{mont+13njp}, and the environments are described by SOAP features $\rep<\ennl||\frho_i^2>$. Reproduced with permission from Ref.~\citenum{de+16pccp}. Copyright 2016 PCCP Owner Societies.
    }
    \label{fig:qm7-rematch}
\end{figure}

\begin{figure*}
    \centering
\includegraphics[width=1.0\linewidth]{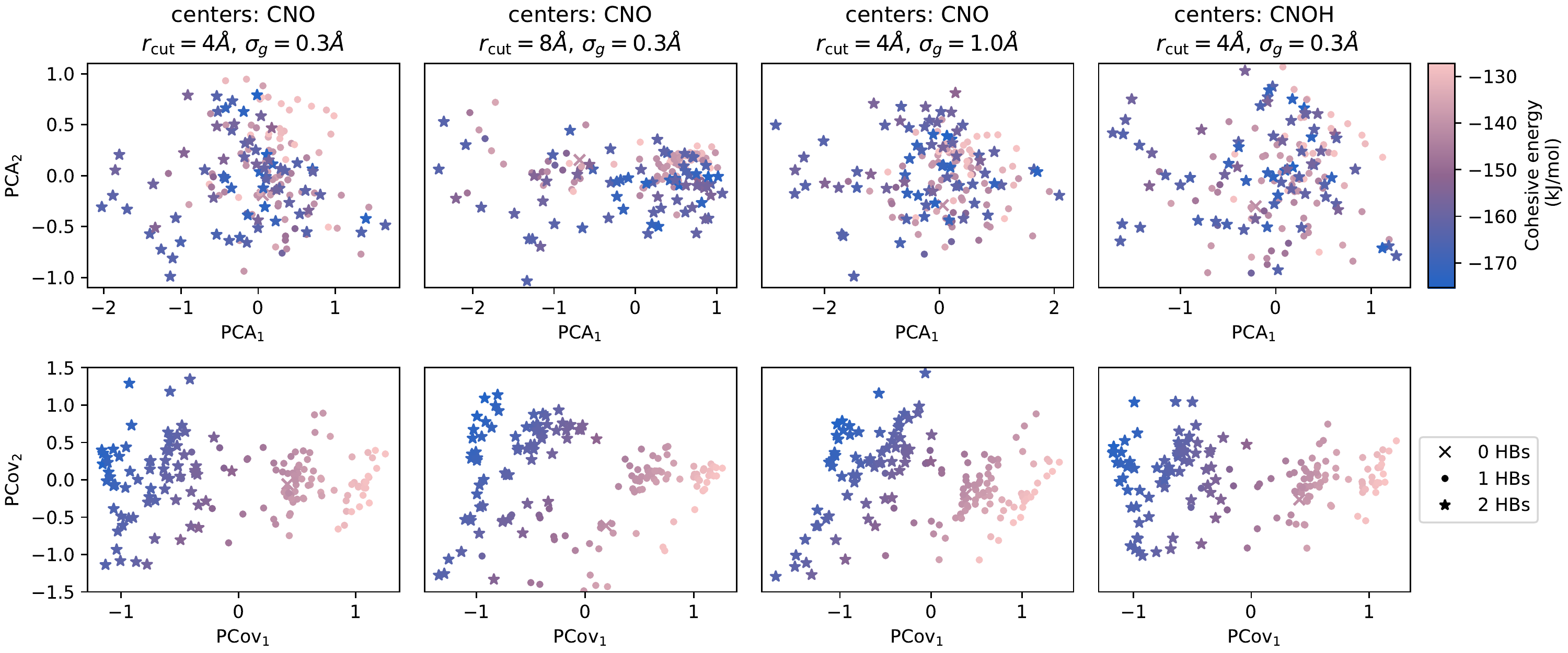}
\caption{Each map describes a set of 156 low-energy polymorphs of 21 different isomers of azaphenacene. The configurations are the same subset of the structures from Ref.~\citenum{yang+18cm} that was used in Ref.~\citenum{helf+20mlst}.
Each point corresponds to a structure, color-coded based on its lattice energy, and with a symbol that indicates the number of hydrogen bonds per molecule, identified with a self-consistent definition\cite{gasp-ceri14jcp}.
The top row reports the first two principal components from a principal component analysis of SOAP $\rep|\frho_i^2>$ structures. The bottom row shows maps obtained using KPCovR\cite{helf+20mlst}. Each column is computed using different SOAP hyperparameters, as indicated in the plot titles.
}
    \label{fig:aza-pcov}
\end{figure*}

An important aspect when defining a metric for structural comparison is the fact one is often interested in measuring the dissimilarity between entire structures, $\dst(A,A')$. Most of the representations we discussed this far are designed to compare atom-centered environments, and therefore yield $\dst(A_i,A'_{i'})$. As a practical example, we define $\dst$ as the Euclidean distance between the feature vectors,
\begin{equation}
\dst^2(A_i,A'_{i'})\equiv \left\| \bfeat(A_i)-\bfeat(A'_{i'})\right\|^2 \label{eq:dist-rep}.
\end{equation}
Different ways of combining atom-centered representations to obtain a structure-level comparison are discussed and benchmarked in Ref.~\citenum{de+16pccp}, using a construction based on the definition of global \emph{kernels}. Here we present the same strategies, but express them directly in terms of distances. The two formulations are equivalent when using the kernel-induced distance.

The simplest global distance can be defined as a mean over all environment pairs,
\begin{equation}
\bar{\dst}^2(A,A') = \frac{1}{N_A N_{A'}}\sum_{i\in A, i'\in A'} \dst^2(A_i,A'_{i'}).
\label{eq:d-average}
\end{equation}
Using the abstract notation  $\rep|A_i>$ rather than $\bfeat(A_i)$ to highlight the connection with the definition of the global representation $\rep|A; \frho^{2}>$ as the sum of environmental $\rep|A; \frho_i>$ (see Section~\ref{sub:translational-sym}) it is easy to see that 
\begin{multline}
\bar{\dst}^2(A,A') = \frac{1}{N_A N_{A'}}\sum_{i\in A, i'\in A'}\|\rep|A'_{i'}>-\rep|A_i>\|^2 = \\
\left[  \sum_{i\in A} \frac{\rep|A_i>}{N_A} - \sum_{i'\in A'} \frac{\rep|A'_{i'}>}{N_{A'}} \right]^2 \equiv \|\rep|\overline{A'}>-\rep|\overline{A}>\|^2,
\end{multline}
i.e. that the average environment distance $\bar{\dst}^2(A,A')$ can be computed by taking the Euclidean distance between the mean of the environment's features in the two structures.
This construction is very natural, and consistent with an additive decomposition of properties in a regression model, but potentially lacks resolving power: two structures with very different environments could end up having a similar value of the average feature vector.

An alternative way to determine a global metric involves finding the best match between the environments of the two structures, defining
\begin{equation}
\!\!\!\hat{\dst}^2(A,A') =\!\! \operatorname*{argmin}_{\bP\in\mathbb{U}^{N_A\times N_{A'}} }\sum_{i\in A, i'\in A'} \!\!\!\dst^2(A_i,A'_{i'}) P_{ii'}~\label{eq:d-bestmatch}
\end{equation}
where $\mathbb{U}^{N_A\times N_{A'}}$ is the set of $N_A\times N_{A'}$ doubly-stochastic matrices, i.e. matrices with positive entries such that sums of rows and columns all equal $1/N_A$ and $1/N_{A'}$ respectively.
When $N_A=N_{A'}$, the optimal $\bP$ contains only zeros and $1/N_A$, and the problem can be construed as a linear assignment problem, and solved in $O(N_A^3)$ time using the Hungarian algorithm\cite{kuhn55nrlq}.
Much like the case of the use of sorted interatomic  distances as a structural representation (Section~\ref{sub:smoothness}), the process of matching entries in the environment distance matrix introduces discontinuities in the derivatives of the distance metric.
One can solve this problem, obtaining at the same time a scheme with a cost that scales as $O(N_A^2)$ and that can be applied to the comparison of structures of different sizes, by introducing an entropy regularization in Eq.~\eqref{eq:d-bestmatch}
\begin{equation}
\hat{\dst}^\gamma(A,A')^2 =\!\!\!\! \operatorname*{argmin}_{\bP\in\mathbb{U}^{N_A\times N_{A'}} }\!\sum_{i\in A, i'\in A'}\!\!\!\!\!\! P_{ii'}(\dst^2(A_i,A'_{i'}) +\gamma \ln P_{ii'}), ~\label{eq:d-rematch}
\end{equation}
controlled by the magnitude of the parameter $\gamma$.
This approach was introduced in Ref.~\citenum{cutu13nips} for the general problem of solving optimal transport problems and of evaluating the Wasserstein distance between probability distributions, and was first applied in Ref.~\citenum{de+16pccp} to atomistic problems in terms of regularized entropy match (REMatch) kernels. By introducing a non-additive combination of the environments, REMatch kernels and the associated distances offer an increased resolving power compared to the plain average distance~\eqref{eq:d-average}, as demonstrated in Figure~\ref{fig:qm7-rematch}. The figure also shows that Eq.~\eqref{eq:d-rematch} interpolates between the average and the best-match metrics, to which it tends respectively for $\gamma\rightarrow\infty$ and $\gamma\rightarrow 0$.

\subsection{Representations for unsupervised learning}

As stressed in the introduction of this Section, in performing cluster analysis or dimensionality reduction, the choice of featurization is not a neutral one, but introduces a bias that will be visible in the end result of the analysis.\cite{ceri19jcp}.
While sometimes this bias is desirable, such as in Fig.~\ref{fig:shibuta} in which a judicious choice of features makes it possible to emphasize, or ignore, the orientation of grains in a polycrystalline sample, one should resist the temptation to fine-tune parameters that do not have an obvious meaning to obtain a result that reflects a preconceived interpretation of the data.
The top row of Fig.~\ref{fig:aza-pcov} shows how different choices of the hyperparameters of the SOAP powerspectrum (cutoff radius $\rcut$, density smearing $\sigmaa$, and the types of atoms that are used as environment centers) change unpredictably the distribution of the points on the 2D map obtained by principal components analysis of a dataset that consists in different polymorphs of a family of molecular materials\cite{yang+18cm}.
In the first panel, in particular, one can recognize a degree of correlation between the position of the points, and intuitive structural and energetic properties, such as the number of H-bonds, and the lattice energy.
The correlation is however far from perfect, and with other reasonable choices of hyperparameters it disappears almost completely.

One possible approach to make unsupervised models less dependent on the details of the underlying featurization is to combine them with an element of supervised learning.
This includes, for instance, combining or contrasting density-based clustering with (kernel) support vector machines classification\cite{helf+19fmb}.
Even more explicitly, one can combine a variance-maximization scheme analogous to PCA with the regression of a target property, as in principal covariates regression (PCovR)\cite{dejo-kier92cils}.
In PCovR one minimizes a loss built as a mixture of a PCA and a linear regression loss, weighted by a mixing parameter $\alpha$
\begin{multline}
\ell = \sum_i \alpha \left\|\bFeat - \bFeat \bP_{\Xi T}\bP_{T\Xi }\right\|^2 \\+ (1-\alpha) \left\|\mbf{Y} - \bFeat \bP_{\Xi T}\bP_{TY}\right\|^2
\label{eq:pcovr-loss}.
\end{multline}
The matrix $\bP_{\Xi T}$ projects from the feature space to a low-dimensional latent space, $\bP_{T\Xi}$ reconstructs an approximation of the full-dimensional feature vector based on its latent-space embedding, and $\bP_{TY}$ regresses the property matrix $\mbf{Y}$ using the latent-space coordinates as inputs.
By explicitly looking for a latent-space projection that allows to regress linearly a target property, one forces the dimensionality reduction to identify a subspace of the chosen features that correlates well with one or more quantities of interest.
The lower row of Fig.~\ref{fig:aza-pcov} is obtained using a recent kernel extension of this method (KPCovR\cite{helf+20mlst}) attempting simultaneously to maximise the spread of data and the kernel regression of the lattice energy, giving equal weight to the two components ($\alpha=0.5$).
Not only points on the resulting map correlate very well with the target: one observes that also structural parameters such as the H-bond counts are now clearly separated between different regions, and the appearance of further groups of well-clustered structures that correspond to similar isomers of azaphenacene\cite{helf+20mlst}.
What is perhaps more important,  introducing an explicit supervised learning target leads to maps that are more consistent across different choices of hyperparameters.
Thus, (K)PCovR reduces the arbitrariness of the description, and mitigates the risk of implicitly introducing an unknown bias by deliberate or accidental tuning of the hyperparameters of the representation.

\begin{figure}[tbhp]
    \centering
    \includegraphics[width=1.0\linewidth]{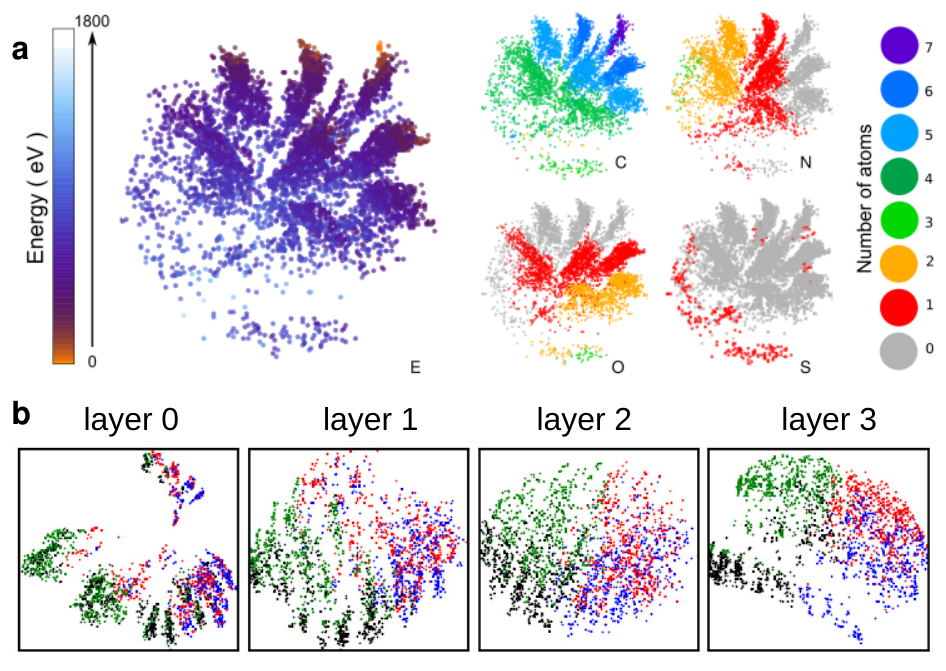}
\caption{
(a) A sketch-map\cite{ceri+11pnas} representation of the QM7 molecular dataset\cite{mont+13njp} based on a SOAP kernel distance. Each point corresponds to one molecule. Left: points are colored according to the atomization energy; right: points are colored according to composition. Adapted with permission from Ref.~\citenum{de+16pccp}. Copyright 2016 PCCP Owner Societies.
(b) Principal component analysis (PCA) on the multiple layers of a deep NN learning simultaneously the 14 properties of the QM7 molecular dataset, using Coulomb matrix features as representation.
Each point (molecule) is colored according to the rule: $E$ and HOMO (highest occupied molecular orbital energy)
large $\rightarrow$ red; $E$ large and HOMO small $\rightarrow$ blue; $E$ small and HOMO large $\rightarrow$
green; $E$ and HOMO small $\rightarrow$ black. The NN extracts, layer after layer, a representation of the chemical space that better captures the multiple properties of the molecule. 
Reproduced from Ref.~\citenum{mont+13njp}. Copyright 2013 American Chemical Society.
}
    \label{fig:montavon-qm7}
\end{figure}

\begin{figure}[tbhp]
    \centering
    \includegraphics[width=1.0\linewidth]{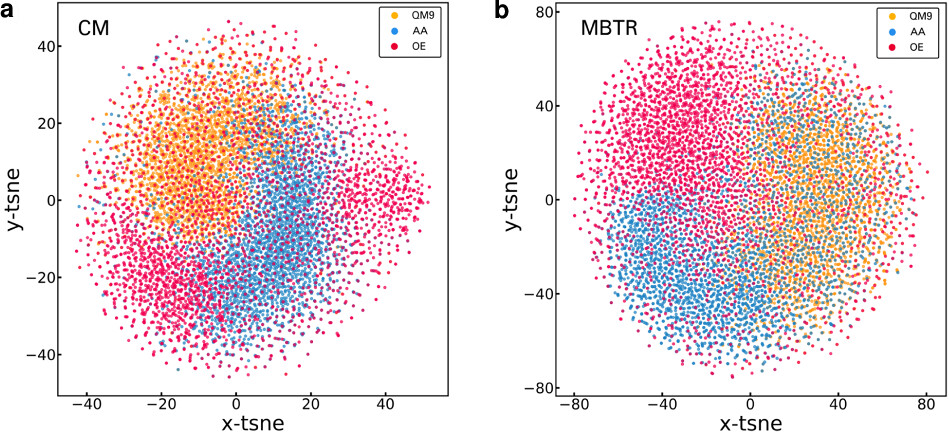}
    \caption{2D maps obtained applying the t-SNE dimensionality reduction algorithm\cite{maat-hint08jmlr} to three different molecular datasets -- the systematic enumeration of 9-non-H-atoms molecules in QM9\cite{rama+14sd}, the conformers of aminoacids in the Berlin aminoacid dataset AA\cite{ropo+16sd} and the large molecules extracted from the Cambridge structural dataset of the OE dataset\cite{scho+16jpcl} . Panel (a) uses a Coulomb matrix representation, panel (b) uses the MBTR features\cite{huo-rupp17arxiv}. 
    Reproduced with permission from Ref.~\citenum{stuk+19jcp}. Copyright 2019 American Institute of Physics.
    }
    \label{fig:datasets-maps}
\end{figure}

\subsection{Analyzing representations and datasets}

The unsupervised  analysis of a dataset helps building an intuitive understanding of complicated structure-property relations for a material or a class of materials.
Given the ``black box'' nature of many machine-learning models (and the fact that even the rigorously-defined density correlation features we focus on in this review have a high-dimensional nature and non-trivial relationship to the actual atomic structure) low-dimensional projections of the feature space can also be useful to gain a better understanding of the structure of feature space.
For example, Fig.~\ref{fig:montavon-qm7}a tells us less about the QM7 dataset\cite{mont+13njp} (that contains small organic molecules containing C, H, N, O, S, Cl) than about the SOAP features that underlie the representation: the unsupervised analysis shows that the chemical composition is the most clear-cut differentiating characteristic when looking at this dataset through SOAP lenses.
Fig.~\ref{fig:montavon-qm7}b visualizes the same QM7 data using a different representation, based on the Coulomb matrix, and shows how successive layers of a neural network transform these features into non-linear combinations that correlate very well with the target properties. Thus, this visualization helps understand how a highly-nonlinear function transforms a description of the system into combinations that can be more easily  used for regression, and diagnose the inner workings of the deep neural network.

A final ``introspective'' application of this kind of analysis involves examining the structure of a dataset -- not as a way to learn about the atomistic configurations it contains, but about its makeup, or the relationship with other datasets.
An example is given in Fig.~\ref{fig:datasets-maps}, showing the comparison between the chemical space covered by three databases of organic molecules, with QM9 and AA being mostly disjoint, and the more diverse OE molecules encompassing both the other sets. Other examples of this kind of analysis are discussed in Section~\ref{sec:applications}.

\begin{figure}
    \centering
    \includegraphics[width=1.0\linewidth]{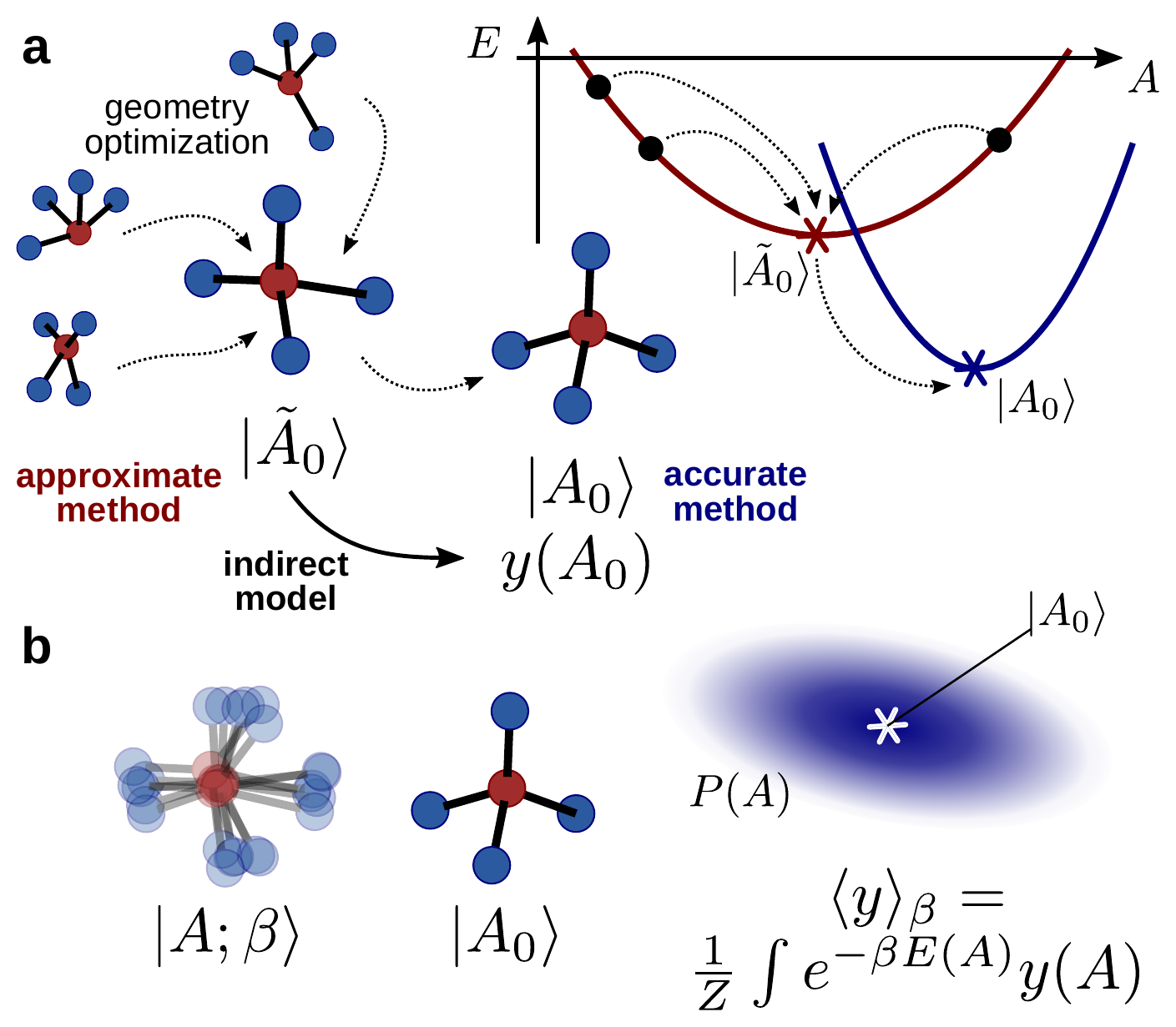}
\caption{A schematic overview of the process of using atomic structure representations to predict properties that are not directly associated with the starting structure. (a) prediction of the properties of the minimum-energy configuration of a structure; the problem can be made well-posed by using a cheap approximate method to optimize the structure, and taking the representation of this approximate structure as the input to regress accurate energy and geometry. (b) prediction of a property that is associated with a thermodynamic average; the minimum energy structure can be taken as a proxy for the ensemble, but a more formally precise ``ensemble representation'' is also possible. }
    \label{fig:indirect}
\end{figure}

\subsection{Indirect structure-property relationships}
\label{sub:indirect-struc-prop-rel}
The one-to-one mapping between an atomic structure and its representation is one of the key requirements to achieve accurate ``surrogate quantum models'' of atomic-scale properties.  However, it can also be a limitation whenever one wants to describe properties that are not strictly associated with the specific configuration at hand.
For example, consider the databases of molecular properties (e.g. the QM9 dataset\cite{rama+14sd}) that have been extensively used as a benchmark, and have been a powerful driving force behind the development of the representations we describe here. The typical benchmark involves taking a structure whose geometry has been optimized at the DFT level and use it to predict the DFT energy -- an exercise that is manifestly of little practical utility.
A more useful approach, instead, would be using a non-optimized structure to predict the properties of the nearest local configurational optimum. As shown in Fig.~\ref{fig:indirect}a, this is conceptually problematic, because we are now trying to achieve a many-to-one mapping.
A possible solution is to map each distorted geometry to an idealized one, or to use a lower level of theory to determine an unique structure $\tilde{A}_0$. Thus, the many-to-one mapping is realized by the local optimization procedure, and the corresponding representation $\rep|\tilde{A}_0>$ can be used to uniquely identify the entire  basin of attraction of the local minimum.
Only for the training structures, this geometry is optimized further at a higher level of theory, obtaining the structure $A_0$ for which properties are meant to be computed.
When the model is fitted, the relationship between $\tilde{A}_0$ and its high-quality counterpart is learned implicitly.
This kind of ``indirect'' model has been used, for instance, in Ref.~\citenum{bart+17sa}, where structures optimized at the semiempirical PM7\cite{stew13jmm} level were used to predict CCSD energetics computed for a DFT-optimized version of the same compound.
While the error was almost twice as large as a model using directly $\rep|A_0>$ as input, chemical accuracy could be reached when discarding \emph{from the training set} structures for which the DFT-optimized structure was too different from the PM7-optimized geometry.

A similar conceptual problem arises when one wants to build models for properties that are associated with a thermodynamic state rather than a precise structure, such as a melting point or solubility of a material.
The problem is very well understood in the context of cheminformatics, where molecular-graph descriptors can be thought as representing the entire set of molecular conformers. In the technique known as 4D-QSAR, ``ensembles'' of conformers are used to build fingerprints that encompass explicitly the structural variability of each compound\cite{andr+10m}. 
These two approaches can also be applied while using the kind of representations discussed in the present review. Typically, and particularly if the ensemble consists in relatively small fluctuations around equilibrium, one might take a representative structure (e.g. the minimum energy configuration) and use its $\rep|\tilde{A}_0>$ as a proxy of the thermodynamic state (Fig.~\ref{fig:indirect}b).
The case in which the target property can be estimated as an ensemble average can be formulated very elegantly in the case of a linear model.  Consider for instance the mean of a property $y$ over the Boltzmann distribution at inverse temperature $\beta$, $P(A)= e^{-\beta E(A)}/Z$,
\begin{equation}
\langle y\rangle_\beta \equiv \frac{1}{Z}\int \D{A} e^{-\beta E(A)} y(A)
\end{equation}
where $Z=\int \D{A} e^{-\beta E(A)}$ is the canonical partition function.
Exploiting the linear nature of the representation one can define an ``ensemble ket''
\begin{equation}
\rep|A; \beta> \equiv\frac{1}{Z} \int \D{A} e^{-\beta E(A)} \rep|A>.
\end{equation}
With this definition, one could use a linear model for $y(A)$ with weights $\rep<q||y>$ and see that
\begin{equation}
\langle y\rangle_\beta\approx \sum_q \rep<y||q>\rep<q||A; \beta>,
\end{equation}
which is convenient because it allows using properties of configurations and of ensembles on the same footings -- and possibly combining them in a single training exercise. The same approach can also be applied in a kernel setting, computing the ensemble average of the reproducing kernel Hilbert space vector associated with the structures.

\section{Efficiency and effectiveness} \label{sec:efficiency}

We have discussed in Section~\ref{sec:history-wishes} how most of the existing choices of representations share profound similarities, and shown, in Section~\ref{sec:symmetry-fields}, that many alternative schemes can be formally related to each other by means of a linear transformation, smoothening or a limit operation.
However, this is not to say that in practical applications they are entirely equivalent.
The computational cost of evaluating them, and their performance in classifying structures, and in regressing their properties, is determined by the choice of basis functions. Even for formally equivalent representations, the condition number of the linear transformation between them and their corresponding bases have significant impact on the numerical behavior of the computed coefficients and the quantities derived from these coefficients.

\begin{figure*}
    \centering
    \includegraphics[width=1.0\linewidth]{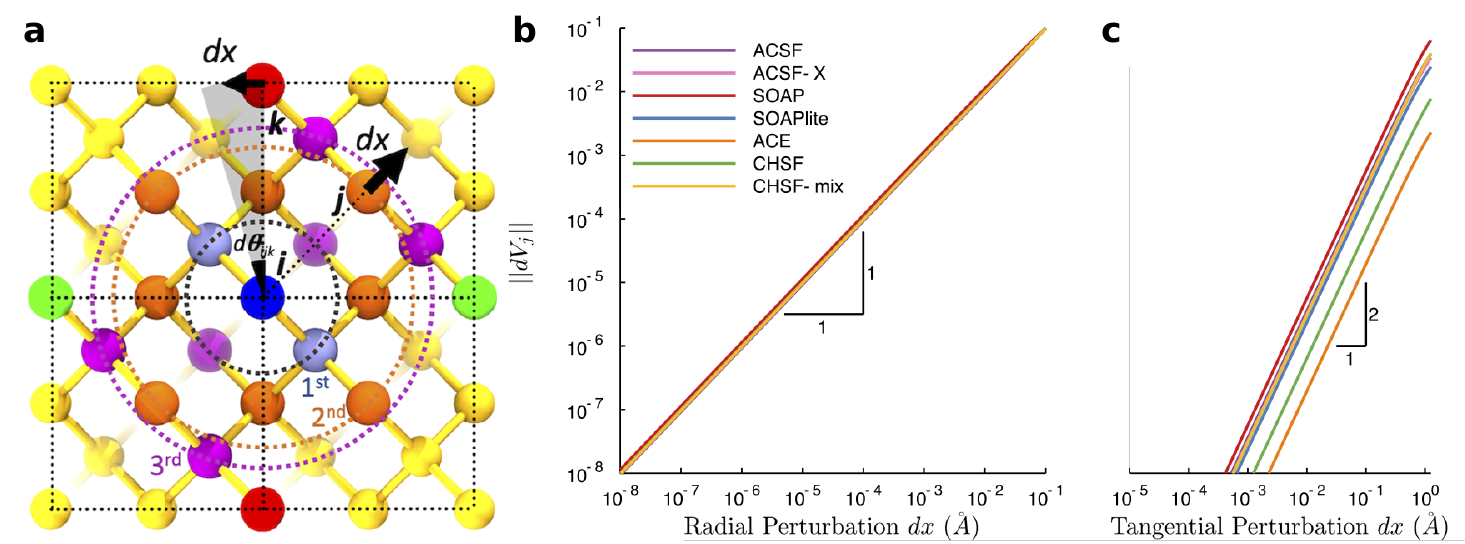}
    \caption{(a) a $4\times4\times4$ Si cell, that is taken as the reference structure for  radial and tangential perturbation of the neighbors of the central atom. (b,c)
    Norm of difference of atomic descriptors on atom i as a neighboring atom j is perturbed from its reference position. (b) A radial perturbation yields a linear change in the features;   (c) a tangential
perturbations in a high symmetry direction for the first shell yields a quadratic change in the features. 
Reproduced with permission from Ref.~\citenum{onat+20jcp}. Copyright 2020 American Institute of Physics.
}
    \label{fig:sensitivity}
\end{figure*}

\subsection{Comparison of features}\label{sub:feat-compare}

A preliminary question when comparing alternative choices of features for the description of atomic structures and/or environments is that of establishing an objective way of assessing their relative merits.
The performance when used in the regression of useful atomic-scale properties is an obvious criterion, but such a comparison is intimately intertwined with the target property and the regression algorithm.\cite{fabe+17jctc,nguy+18jcp,zuo+20jpcl}.
Very recent efforts have attempted to characterize different representations in terms of their information content -- for instance through the eigenvalue spectrum of the covariance or kernel matrix associated with a dataset, the decrease in accuracy when reducing the number of features\cite{onat+20jcp}, or the sensitivity of the features to atomic displacements.

This latter approach can be realized by directly comparing the separation in feature space against finite displacements of the atoms\cite{onat+20jcp}, or through an analysis of the Jacobian $J_{jk}=\partial\rep<k||A_i>/\partial \br_j$\cite{pars+20mlst}.
The sensitivity of the features to small changes of the atomic positions indicates their usability and performance in regression of classification tasks. \citet{onat+20jcp} analysed the effect of random perturbations in crystalline environments, finding that, for features based on atomic density correlations, displacements of atoms in the environment usually cause a linear response. One notable deviation from this trend are perturbations along some high-symmetry directions in atomic environments carved from perfect crystals, where the response to displacements is second-order, implying that the representations cannot capture these types of deformations (Fig.~\ref{fig:sensitivity}).
However, as discussed in reference~\citenum{onat+20jcp} the types of symmetric deformations applied in the study correspond to reflection operations.
Due to the body-correlation order considered, features are invariant to mirror symmetry, and so the observed loss of sensitivity is not unexpected.
Analizing the response of the features to perturbations in terms of the Jacobian, as in Ref.~\citenum{pars+20mlst}, has the advantage of characterizing fully the sensitivity at a given point. The Jacobian should have six zero principal values, corresponding to rigid rotations and translations of the environment. Additional zeros could be associated with the presence of a continuous manifold of degenerate structures. In some cases, as demonstrated by the finite-displacement deformation in Fig.~\ref{fig:sensitivity}b, high-symmetry configurations can result in directions with zero gradient that have no adverse effect on the accuracy of a model built on the density correlation features.

Another comparison between different bases is to analyse the landscape defined by the similarity or distance between environments, $\dst(A_i,A'_{i'})$ where the environment $A_i$ is kept fixed.
The distance between the atom-centered environments $A_i$ and $A'_{i'}$ can be defined as
the Eucledian distance between feature vectors, Eq.~\eqref{eq:dist-rep}.
Written as a function of the Cartesian coordinates of $A'_{i'}$, $\dst(A_i,A'_{i'})$ is a scalar field which will have a global minimum manifold where the field is exactly zero, corresponding to equivalent environments $A_i$ and $A'_{i'}$  that are related by symmetry operations. 
Whether there are other manifolds at exactly $\dst(A_i,A'_{i'})=0$, corresponding to the same features resulting from symmetrically nonequivalent environments is related to the question of completeness (Section~\ref{sub:geo-completeness}). In practical applications, the shape of the global minimum manifold has also implications for the numerical evaluation. In particular, one could examine how different $A_i$ and $A'_{i'}$ may be for $\dst(A_i,A'_{i'}) < \epsilon$ where $\epsilon$ is a small number. Using a random search approach, the numerical sensitivity of the feature landscape has been analysed in Ref~\citenum{bart+13prb}. Reference structures $A_i$ were perturbed and then reconstructed by minimising the distance $\dst(A_i,A'_{i'})$, and the optimised structures compared to the reference ones. For small numbers of neighbors in the reference environment, all the examined representations performed similarly well, but only SOAP was capable of accurately reconstructing the reference environments of more than 12 neighbors. As we have seen in earlier sections, this differences can be attributed to the choice of basis functions other representations use, although it should be noted that SOAP distances and similarities converge in the limit of a complete basis, therefore the actual form of the basis might affect the convergence, and the computational cost of the representation, but does not impact its resolving power.

\begin{figure}[tbp]
\includegraphics[width=1.0\linewidth]{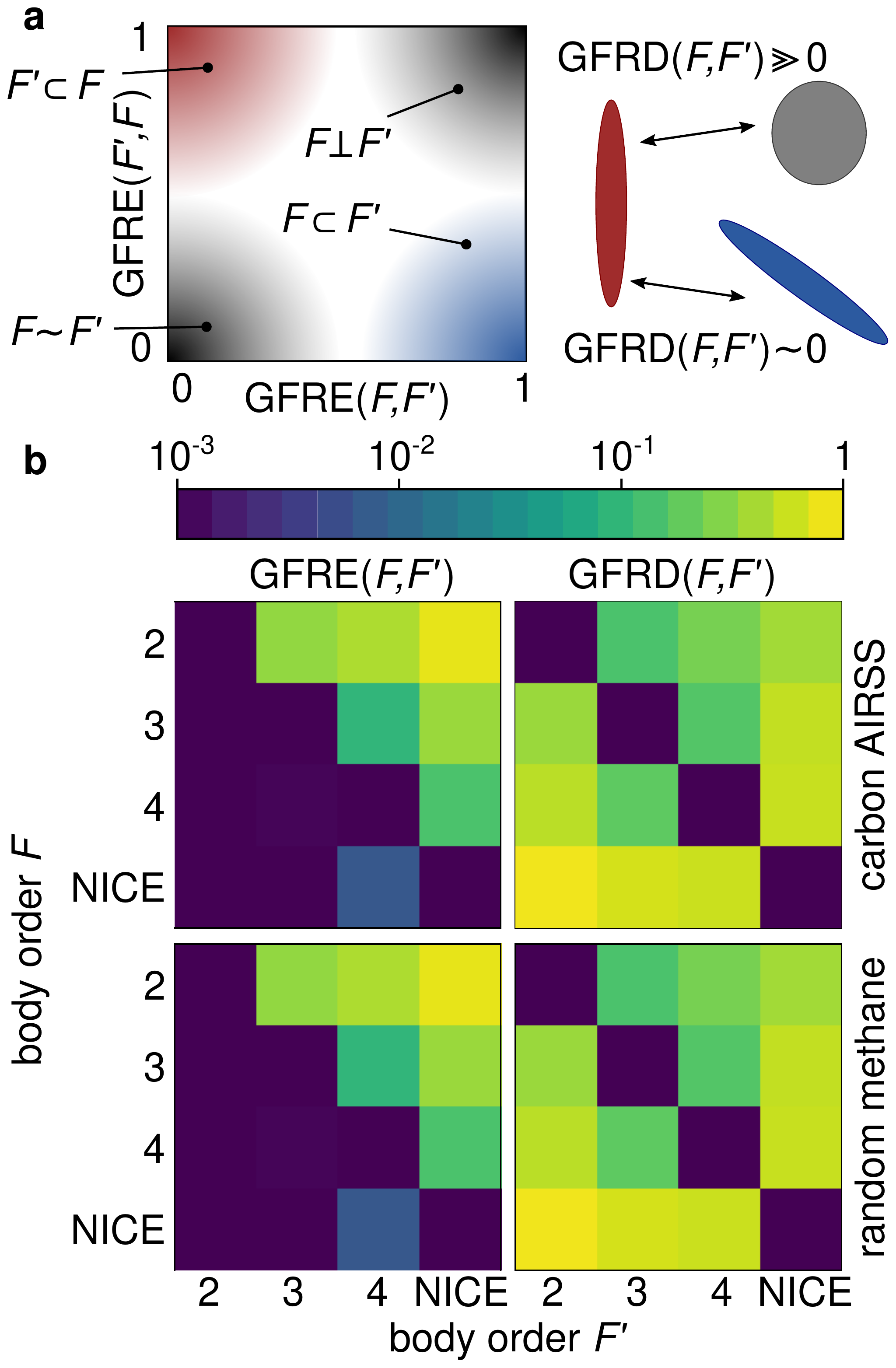}
\caption{
(a) Schematic depiction of the interpretation of the error  (GFRE, Eq.~\eqref{eq:GFRE}) and the distortion (GFRD, Eq.~\eqref{eq:GFRD}) that describe the relationship between two feature spaces.
(b) Comparison between density correlation features of different order, as well as the NICE features\cite{niga+20jcp} up to $\nu=4$, computed in terms of the GFRE and GFRD, for a data set of random \ce{CH4} configurations\cite{matcloud20a} and hypothetical carbon allotropes predicted by AIRSS\cite{pick-need11jpcm,matcloud20b}.
Adapted from Ref.~\citenum{gosc+21mlst}. Copyright 2021 IOP Publishing under Creative Commons Attribution 4.0 International License \url{https://creativecommons.org/licenses/by/4.0/}.
\label{fig:gfrm-bodyorder}%
}
\end{figure}

A more explicit comparison between pairs of representations can be obtained by evaluating the error one incurs when using a set of features, arranged in a feature matrix $\bFeat$ in which each row corresponds to a sample in a reference dataset, to linearly reconstruct a second featurization of the same structures or environments $\bFeat'$, defining a global feature space reconstruction error
\begin{equation}
\text{GFRE}(\bFeat,\bFeat') = \min_{\bP} \sqrt{{\norm{\bFeat'_\text{test} - \bFeat_\text{test} \bP }^2}/n_\text{test}}.
\label{eq:GFRE}
\end{equation}
$\bP$ is a linear regression weight matrix obtained on a training subset of the rows of $\bFeat$ and $\bFeat'$, and both sets of features are assumed to be standardised.\cite{gosc+21mlst}
The GFRE can be extended to also incorporate non-linearity in the mapping, either by a locally-linear approach, or by using a kernelized version. 
Loosely speaking, it measures the relative amount of information encoded by the two feature spaces, and is not symmetric.  $\text{GFRE}(\bFeat,\bFeat')\ll\text{GFRE}(\bFeat',\bFeat)$ indicates that the featurization underlying $\bFeat$ is more informative than that used to build $\bFeat'$, and vice versa.
$\text{GFRE}(\bFeat,\bFeat')\approx\text{GFRE}(\bFeat',\bFeat)\approx 0$ implies that the two featurizations contain similar information (Fig.~\ref{fig:gfrm-bodyorder}a). 
A similar asymmetric measure of similarity between feature spaces can be defined by comparing the resolving power of the corresponding metrics\cite{glielmo2021arxiv}, translating the information that is present in distance-distance correlation plots (Sec.~\ref{sub:measuring-similarity}) into a quantitative measure of information content.

Having $\text{GFRE}(\bFeat,\bFeat')\approx\text{GFRE}(\bFeat',\bFeat)\approx 0$ does not mean that $\bFeat$ and $\bFeat'$ they are equivalent and can be used interchangeably. 
One could emphasize more some structural correlations than others: imagine for instance multiplying by a large constant the entries of one column.
This kind of distortions, which can have a substantial impact on the performance of models built on $\bFeat$ or $\bFeat'$, can be measured by defining a global feature space distortion (GFRD)
\begin{equation}
\text{GFRD}(\bFeat,\bFeat') = \min_{\mbf{Q}\in\mathbb{U}} \sqrt{{\norm{\bFeat_\text{test}\bP - \bFeat_\text{test} \mbf{Q} }^2}/n_\text{test}}.
\label{eq:GFRD}
\end{equation}
$\mbf{P}$ is the same projection matrix that enters the definition of the GFRE (so that $\bFeat\bP\approx \bFeat'$), and $\mbf{Q}$ is the unitary transformation that best aligns $\bFeat$ and the best linear approximation of $\bFeat'$.

\begin{figure*}[tbhp]
\centering
\includegraphics[width=0.9\linewidth]{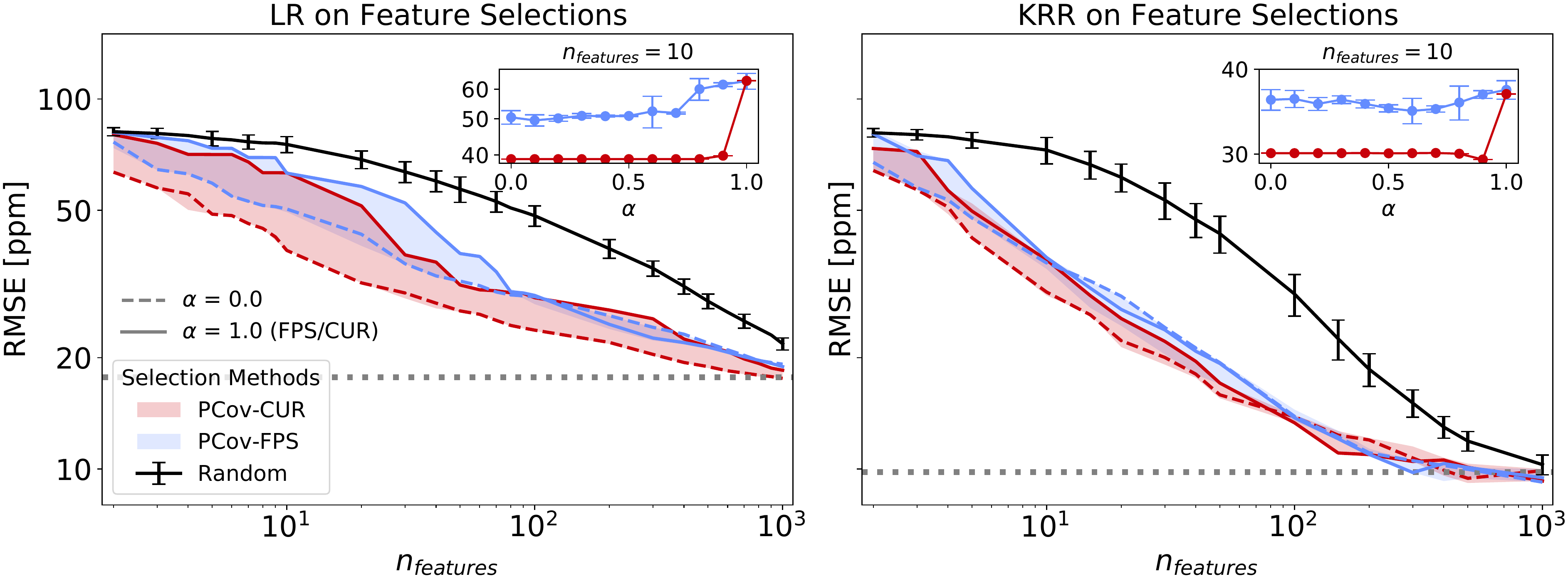}
\caption{Test-set error in the prediction of a linear regression (left) and kernel ridge regression (right) model of the nuclear chemical shieldings of atoms in a set of molecular materials, as a function of the number of features used in the model. Features are selected using the FPS and CUR methods (full lines) from a set of 2520 SOAP features. The shaded areas indicate the range of values obtained varying the mixing parameter $\alpha$ in a principal covariate-augmented version of the methods. Adapted from Ref.~\citenum{cers+21mlst}. Copyright 2021 IOP Publishing under Creative Commons Attribution 4.0 International License \url{https://creativecommons.org/licenses/by/4.0/}.
}
    \label{fig:pcov-cur}
\end{figure*}

If both GFRE and GFRD are zero, then the linearly independent components of $\bFeat$ and $\bFeat'$ are related by a unitary transformation, which implies that distances and scalar products between feature vectors are equal in $\bFeat$ and in $\bFeat'$.
Figure~\ref{fig:gfrm-bodyorder} demonstrates the use of these measures to compare $\rep|\frho_i^\nu>$ features of different body order. The asymmetry is very clear, with higher-order features containing more information than their lower-order counterparts.
Note that -- in view of the linear nature of the mapping -- this is not entirely obvious: formally, $\nu=1$ features are \emph{not} linearly dependent on higher-$\nu$ features, and so these observations reflect the specific nature of the atom-density field whose correlations are being represented, and the nature of the structures in the benchmark datasets.
The figure also includes invariants built with the N-body iterative contraction of equivariants (NICE) framework, that are designed to capture most of the information up to high body orders. The truncation of the expansion, that is necessary to keep the evaluation of $\nu=4$ order features affordable, leads to a small residual GFRE when reconstructing the full $\nu=3$ features.
The GFRD is rather large between all featurizations, indicating that -- even though higher-order features contain sufficient information to describe lower-order correlations -- they weight the information differently, which is why it is often beneficial to treat different orders of correlation separately in the construction of interatomic potentials.\cite{bart-csan15ijqc,deri-csan17prb,glie+18prb,veit+19jctc}

\subsection{Feature selection}
\label{sub:feature-selection}
Numerical feature vectors $\bfeat(A_i)$ are the result of a basis set expansion of the abstract atom-centered representations, which are, for practical purposes, truncated.
A concrete discretization of the symmetrized $\nu$-correlations is obtained by choosing a finite subset from the set of all possible features,
\begin{equation}
\bq \subset \bq_{\rm total} := \big\{
            (n_\alpha l_\alpha)_{\alpha = 1}^\nu,
            \quad \nu \in \mathbb{N} \big\}.
\end{equation}
(A choice of $(n_\alpha,l_\alpha)_{\alpha}$ naturally induces a choice of $m_\alpha$ and symmetrized features.)
The role of the discretization $\bq$ is very different for linear and nonlinear models and therefore warrants a brief comment:
For nonlinear models we typically only require {\em geometric completeness} (see Section~\ref{sec:incompleteness}), which means that the feature set can be chosen to be {\em minimal} but in a way so that all possible configurations, or at least all configurations of interest (e.g. from a training set) can be distinguished in a stable and smooth way. While it is an open problem to characterize precisely what this entails, we generally expect that relatively small feature sets on the order of hundreds for single-species scenarios could be sufficient.

On the other hand, converging a linear model requires eventually letting the discretisation $\bq$ converge to the full feature set $\bq_{\rm total}$, which in practice leads to a much larger set $\bq$ and in particular higher correlation-orders $\nu$ to achieve a desired accuracy, e.g. on the order $O(10'000)$ features for single-species models. The additional cost in training and evaluating the features is of course offset by the fact there is no additional cost in evaluating the nonlinear models. Due to the large feature sets the selection of effective subset of $\bq$ may be even more important in the linear setting. In particular it will be crucial to {\it a priori} choose {\em sparse} subsets of $\bq_{\rm total}$ rather than tensor-product sets due to the combinatorial explosion of the number of features with high $\nu$ (curse of dimensionality). For example, a total-degree $D$ discretisation,
\begin{equation} \label{eq:totaldegree}
\begin{aligned}
     \bq(\nu^{\rm max}, D)  =
    \big\{
        (n_\alpha l_\alpha m_\alpha)_{\alpha = 1}^\nu \,:\,\, \nu \leq \nu^{\rm max} &    \\
        \sum_\alpha n_\alpha + l_\alpha \leq D   \,&
    \big\}
\end{aligned}
\end{equation}
was used by \citet{Bachmayr2019}, while closely related a priori sparsifications were used by \citet{shap16mms,braa-bowm09irpc} in all cases demonstrating accuracy/performance competitive with or outperforming nonlinear models.

\paragraph*{Data-driven selections}
When using high-body order features, some of the components can be related by non-trivial linear dependencies, that can be enumerated numerically\cite{Bachmayr2019,niga+20jcp}. The construction does not ensure that there is no other linear dependence that is specific to a given dataset, meaning that feature vectors could potentially be compressed  even further without noticeable deterioration in the quality of the representation.
The benefits of the compression are clear: if only a few components need to be evaluated, significant efficiency gains may be realised both in computational effort and storage requirements.

Thus, the objective of feature selection or truncation is to find a subset of features that retain the information content of the original, untruncated representation.
This is to be contrasted with dimensionality reduction techniques, that apply a linear transformation on the full feature vector to generate a lower dimensional representation. These only reduce the computational cost of operations that are applied on the reduced feature vectors: the whole feature vector must be evaluated first, before being able to determine its projections.

A simple example of a feature selection strategy is the farthest point sampling (FPS) technique\cite{elda+97ieee}. One chooses an initial column $\bfeatc_{c_0}$ (indexed by $c_0$) of the feature matrix, and then iterates selecting the columns that maximize the Haussdorf distance to the previously selected columns
\begin{equation}
c_{m+1} = \operatorname{argmax}_j \left\{ \min_{i \in \mbf{c}_m}\left\|\bfeatc_i - \bfeatc_j\right\|\right\},
\end{equation}
effectively identifying the indices $\mbf{c}$ of the features that have the most diverse values across the data set.
FPS has also been used in a similar manner, but on the rows of $\bFeat$ in order to select a representative set of data points.\cite{ceri+13jctc,bart+17sa,de+16pccp}
The CUR matrix decomposition\cite{maho-drin09pnas}, instead, generates a low-rank approximation of the feature matrix $\bFeat$, in the form
\begin{equation}
\bFeat \approx   \mathbf{C} \mathbf{U} \mathbf{R}.
\end{equation}
Unlike singular value decomposition, CUR uses the actual columns ($\mathbf{C}$) and rows ($\mathbf{R}$) of $\bFeat$. To make the selection, a \emph{leverage score} is associated with each feature $c$
\begin{equation}
    \pi_c = \frac{1}{k}\sum_{i=1}^{k} (\mathbf{v}_{i})_c^2
    \textrm{,}\label{eq:cur-leverage}
\end{equation}
based on the right singular vectors $\mathbf{v}_i$ of the singular value decomposition of $\bFeat$. $k$ is usually taken to be the approximate rank of $\bFeat$. Features may be selected in a probabilistic procedure or simply based on their score.
Imbalzano \textit{et al}.\cite{imba+18jcp} argued that the scores associated with feature vector components which are linearly dependent are close, therefore the selection can easily result in a redundant set.
Instead, in Ref.~\citenum{imba+18jcp} a greedy algorithm based on the CUR decomposition was suggested, where features were selected iteratively.
The feature with the highest score is selected, and the columns of $\bFeat$ are orthogonalised relative to the column corresponding to the selected feature.
The scores are updated in each step, so the linear dependence of already selected features are removed. This iterative scheme often performs better when using a very small value of $k$ in constructing the $\pi_c$, Eq.~\eqref{eq:cur-leverage}.

Fig.~\ref{fig:pcov-cur} shows that a data-driven selection of the most relevant/diverse features makes it possible to achieve models with an accuracy that approaches that of the full model while reducing the number of components by a factor of about 3 (for linear regression) or 10 (for KRR). Particularly for intermediate sizes of the selection, the improvement in accuracy with respect to a random selection can be dramatic.
Both FPS and CUR methods can be improved further by incorporating information on the properties associated with the structures,\cite{cers+21mlst} as in Eq.~\eqref{eq:pcovr-loss}. Including a supervised component by setting $\alpha<1$ in the feature selection usually leads to more performing models, as shown  in Fig.~\ref{fig:pcov-cur}.
Feature selection methods can be applied to any flavor of density correlation features. Imbalzano \textit{et al}.\cite{imba+18jcp} used a reference data set on liquid water\cite{mora+10pnas} and a large set of systematically generated ACSFs. Evaluating the RMSE of the predicted energies and forces revealed that automatic selections performed by a CUR or FPS approach may achieve similar performance to features selected based on chemical intuition and heuristics, while keeping approximately the selection size.
A dramatic reduction in numbers of features is also possible for the SOAP power spectrum, and a data-driven selection of the most important components has quietly become commonplace to accelerate SOAP-based ML models\cite{wilk+19pnas,enge+19pccp,benm+20prb}.
A more systematic investigation of the effectiveness of feature selection for many commonly used atomic descriptors has been recently reported by \citet{onat+20jcp}, who analysed how accurately the original feature vector can be reconstructed from the reduced set, as well as the performance on a practical regression task.

\begin{figure}[tbhp]
\includegraphics[width=1.0\linewidth]{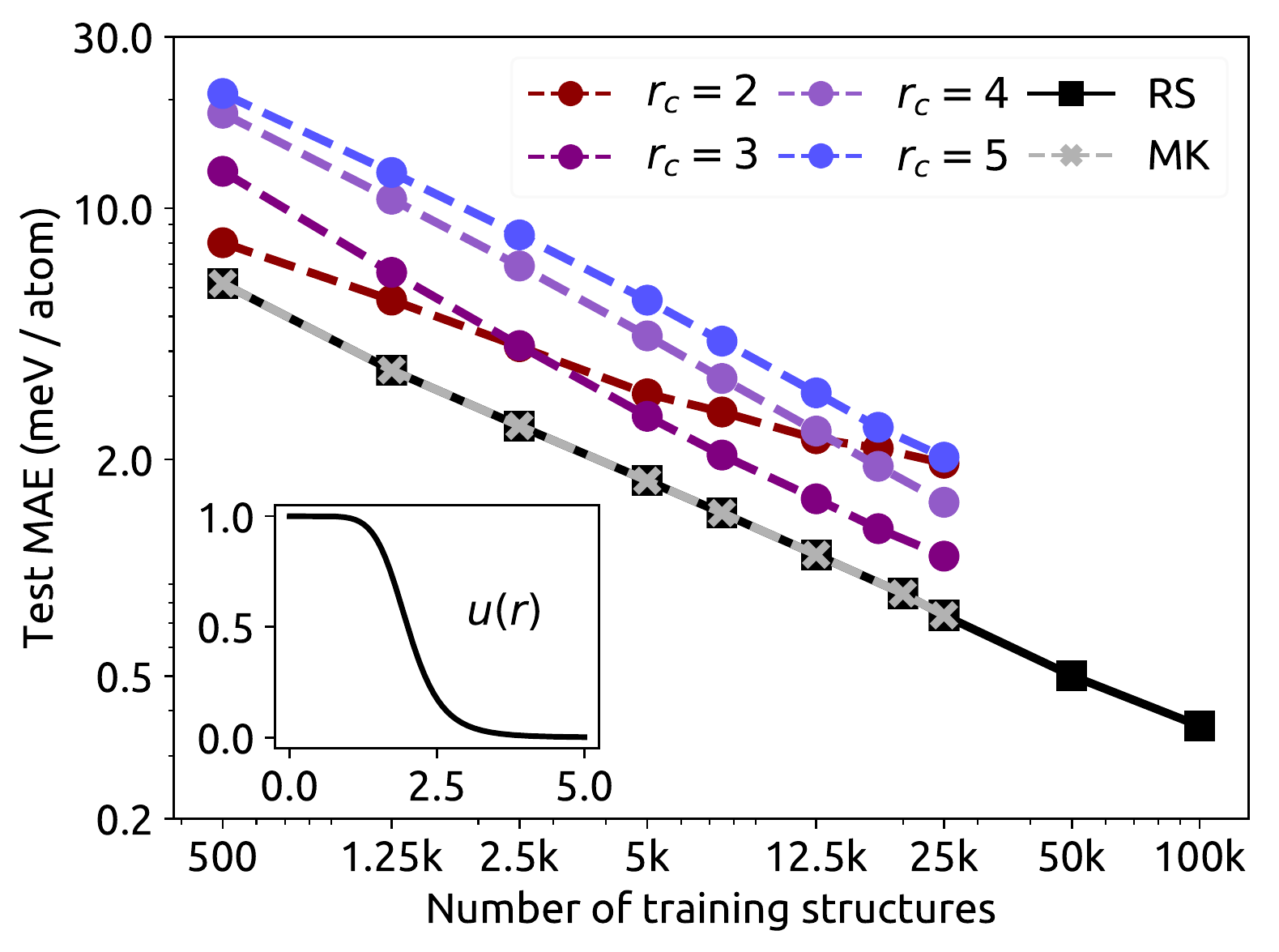}
\caption{\label{fig:lc-rcut}
Learning curves for the atomization energy of molecules in the QM9 data set\cite{rama+14sd}. Four of the lines show the
MAE on the test set for kernel regression models based on SOAP ($\rep|\frho_i^2>$) features with different cutoff radii (dashed lines graduating from red to blue). The other lines show the MAE on the test set for the optimal radially-scaled (RS) and
multiple-kernel (MK) SOAP models (black and grey lines respectively). In
every model, the features were constructed with very converged hyperparameters, $\nmax = 12$ and $\lmax=9$. The inset
shows the radial-scaling function $u(r)$ from $r = 0$\AA{} to $r = 5$\AA{} with the
parameters that were found to minimize the ten-fold cross validation MAE
on the optimization set through a grid search, $r_0 = 2$\AA{} and $m = 7$. The multiple-kernel model combines the  $\rcut= 2, 3, 4$ and RS kernels in the ratio 100'000 : 1 : 2 : 10'000, and the learning curve agrees with the RS result to within graphical accuracy.
Error bars are omitted because they are as small
as the data point markers. Note that errors are expressed on a per-atom
basis. Error per molecule expressed in kcal/mol can be obtained
approximately by multiplying the scale by 0.4147, that is computed based
on the average size of a molecule in the QM9 database.
Reproduced with permission from Ref.~\citenum{will+18pccp}. Copyright 2018 PCCP Owner Societies.
}
\end{figure}

\begin{figure}[tbhp]
\includegraphics[width=0.95\linewidth]{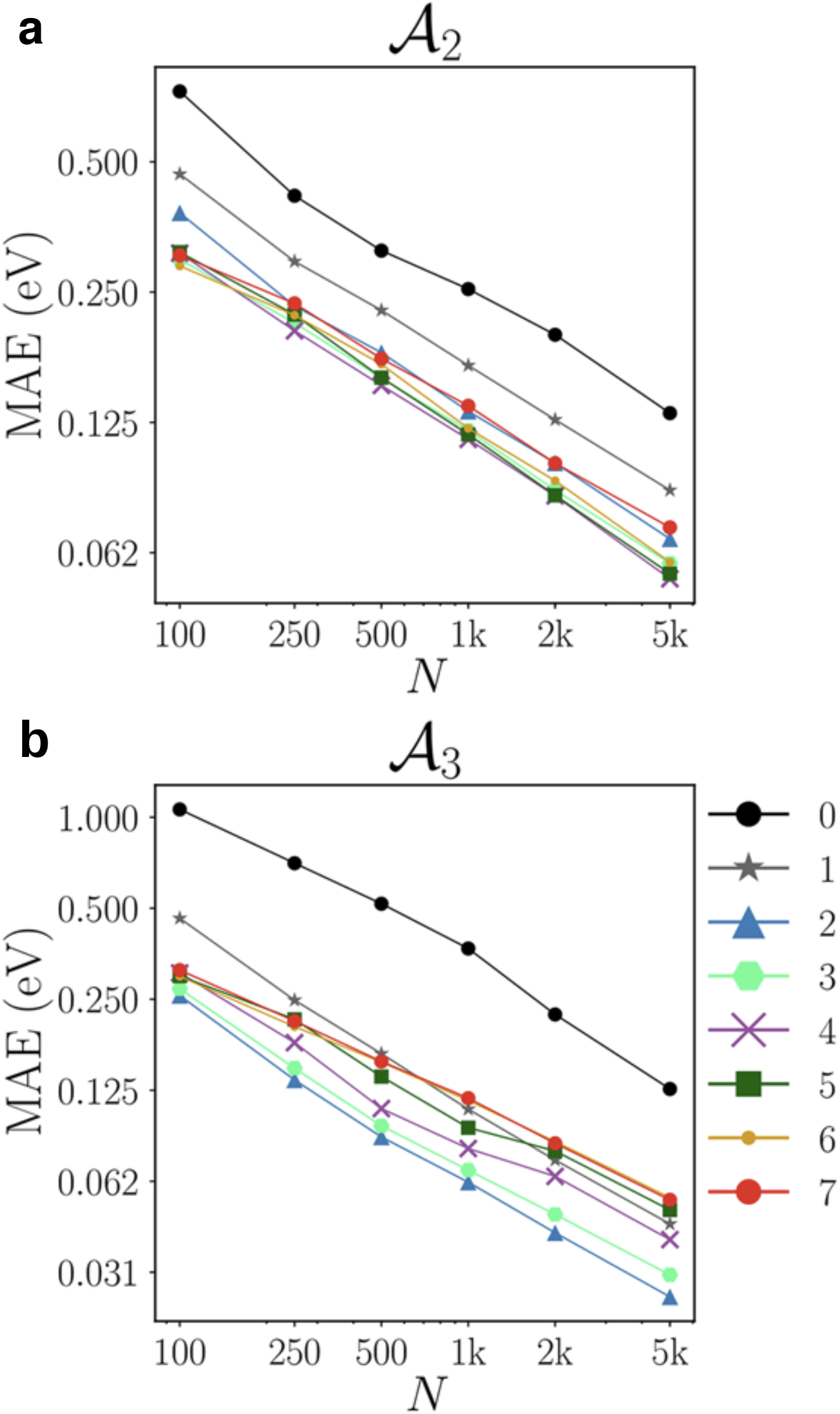}
\caption{\label{fig:lc-fchl}
Optimization of the exponents in scaling power laws. a) Out-of-sample
MAE for atomization/formation energy predictions as a function of training
set size on the QM9 dataset. Learning curves are generated using KRR with
a 2-body FCHL representation. The legends indicate the exponent $n_2$ used in the scaling power law, $\xi_2(d)$. Leftb) Out-of-sample MAE for atomization/formation energy predictions as a function of training set size on the QM9 dataset. Learning curves are generated using KRR with a 3-body FCHL representation. The legends indicate the exponent $n_3$ used in the scaling power law, $\xi_3(d)$. In order to compare results to Fig.~\ref{fig:lc-rcut}, the ordinates must be divided by 18. Adapted from Ref.~\citenum{fabe+18jcp}.
}
\end{figure}

\subsection{Feature optimization}
\label{sub:feature-optimization}

As discussed in \cref{sub:non-linear-model}, non-linear models optimize the description of their inputs by generating new features that are best correlated with the target property, or that are adapted to the structure of the dataset.
For instance, taking products of 2-body features results in an effective representation that incorporates some, but not all, features of body order 3, 4\dots
In some cases it is possible to find an expression for the effective representation associated with a kernel model\cite{bart+13prb,glie+18prb,chri+20jcp}, while other cases (most notably deep neural network models) put less focus on the interpretability of the intermediate features, and act largely as data-driven `black boxes'.
Alternatively, feature optimization can be performed explicitly on the representations presented in \cref{sub:lm-rep}.
Such optimization could take the form of the choice of basis functions.
In the Behler-Parrinello framework, it is customary to select a small number of atom-centered symmetry functions based on experience and heuristics\cite{behl11jcp}.
An optimization of the hyperparameters by gradient descent has also been proposed\cite{gao+19jcp} to obtain more accurate models based on atom-centered symmetry functions.

When considering systematically-convergent implementations of density-correlation features, the optimization of the basis set is less crucial, although one may want to reduce the size of the basis for the sake of computational efficiency, as discussed in Section~\ref{sub:feature-selection}.
That is not to say that the details of the practical implementation of the features does not change the behavior of a model built upon them. Optimizing  hyperparameters such as cutoff radius, density smearing, basis set cutoff, affects how naturally the features correlate with the target property, which is one of the factors determining how quickly a regression model becomes capable of performing accurate predictions\cite{huan-vonl16jcp}.
For example, the smearing of the atom density, or the truncation of the basis set, should reflect the natural scale over which the target properties vary.
Similarly, the size of the local environment determined by $\rcut$ relates to the typical decay length of interactions, as mentioned in \cref{sub:locality}, but it also changes the effective dimensionality of feature space, which affects the accuracy of the model in a non-trivial way.
Consider the learning curves shown in Fig.~\ref{fig:lc-rcut}, that report on the prediction accuracy, as a function of the train set size, for a kernel ridge regression model of molecular atomization energies, based on SOAP features that differ by the value of $\rcut$.
A very large cutoff $\rcut=5$\AA{} does not yield the best performance, despite providing information on a wider range of distances. In fact, one observes the need to balance the complexity of the model and the available data: a very short-range $\rcut=2$\AA{} yields the most effective description in the data-poor regime, but the accuracy of the corresponding model saturates due to lack of information on non-covalent interactions.
Combining multiple representations in a ``multi-kernel'' model (which is effectively equivalent to concatenating multiple feature vectors, each scaled separately) yields consistently better performances\cite{bart+17sa,paru+18ncomm}.
The weighting of different components -- that can be optimized by cross-validation -- indicates the relative importance of correlations on various length-scales. The fact that large-$\rcut$ features carry low weight in the optimal combination suggests that an improvement of performance can be obtained by calibrating the distance-dependent contributions of neighbors to the environment description.
This can be achieved by introducing a radial scaling function (indicated as $u(r)$ in Ref.~\citenum{will+18pccp}, as $f(r,r_j,\rcut)$ in Ref.\cite{caro19prb} and as $\xi_\nu(d)$ in Ref.~\citenum{fabe+18jcp}) that downweights the contributions of atoms in the far field.
As shown in Fig.~\ref{fig:lc-fchl} for the FCHL representation\cite{fabe+18jcp}, the choice of the form of this scaling can change the accuracy of the model by more than a factor of 2.
A similar effect is seen in Fig.~\ref{fig:lc-rcut} for the case of SOAP features.
It is also worth noting that the cutoff function usually adopted in Behler-Parrinello-like frameworks decays rapidly well before reaching $\rcut$ -- suggesting that a similar optimization is implicitly at play.\cite{behl11pccp}
Optimization of a radial scaling function has become commonplace, and most recent applications based on the SOAP power spectrum rely on it to achieve consistently optimal performance in both the data-poor and data-rich regime.

\begin{figure}[tbhp]
\includegraphics[width=1.0\linewidth]{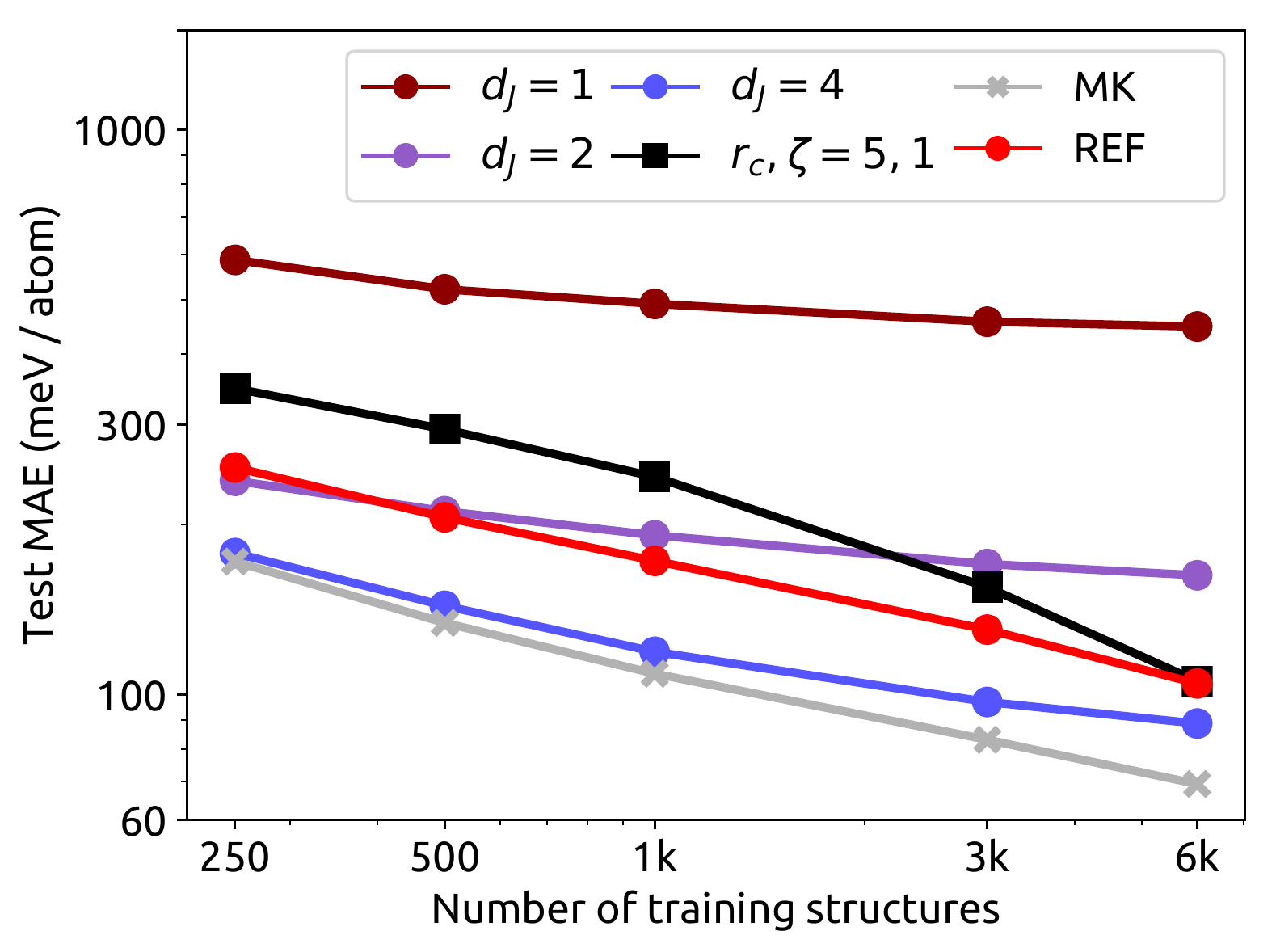}
\caption{\label{fig:lc-elpa}
Learning curves for a model of the cohesive energy of a database of elpasolite structures, each containing a random selection of four elements chosen among 39 main group elements.\cite{fabe+16prl} The standard SOAP
curve is shown in black, the best curve from Ref.~\citenum{fabe+16prl} is shown in bright red (REF) and the curves obtained with an alchemical model with reduced dimensionality $d_J$ are shown in dark red ($d_J$ = 1), purple ($d_J = 2$) and blue ($d_J$ = 4).
The multiple-kernel model (shown in grey) combines three standard SOAP kernels with different cutoff and one alchemically optimized kernel with $d_J=4$. Reproduced with permission from Ref.~\citenum{will+18pccp}. Copyright 2018 PCCP Owner Societies. }
\end{figure}

\begin{figure}[tbhp]
\includegraphics[width=1.0\linewidth]{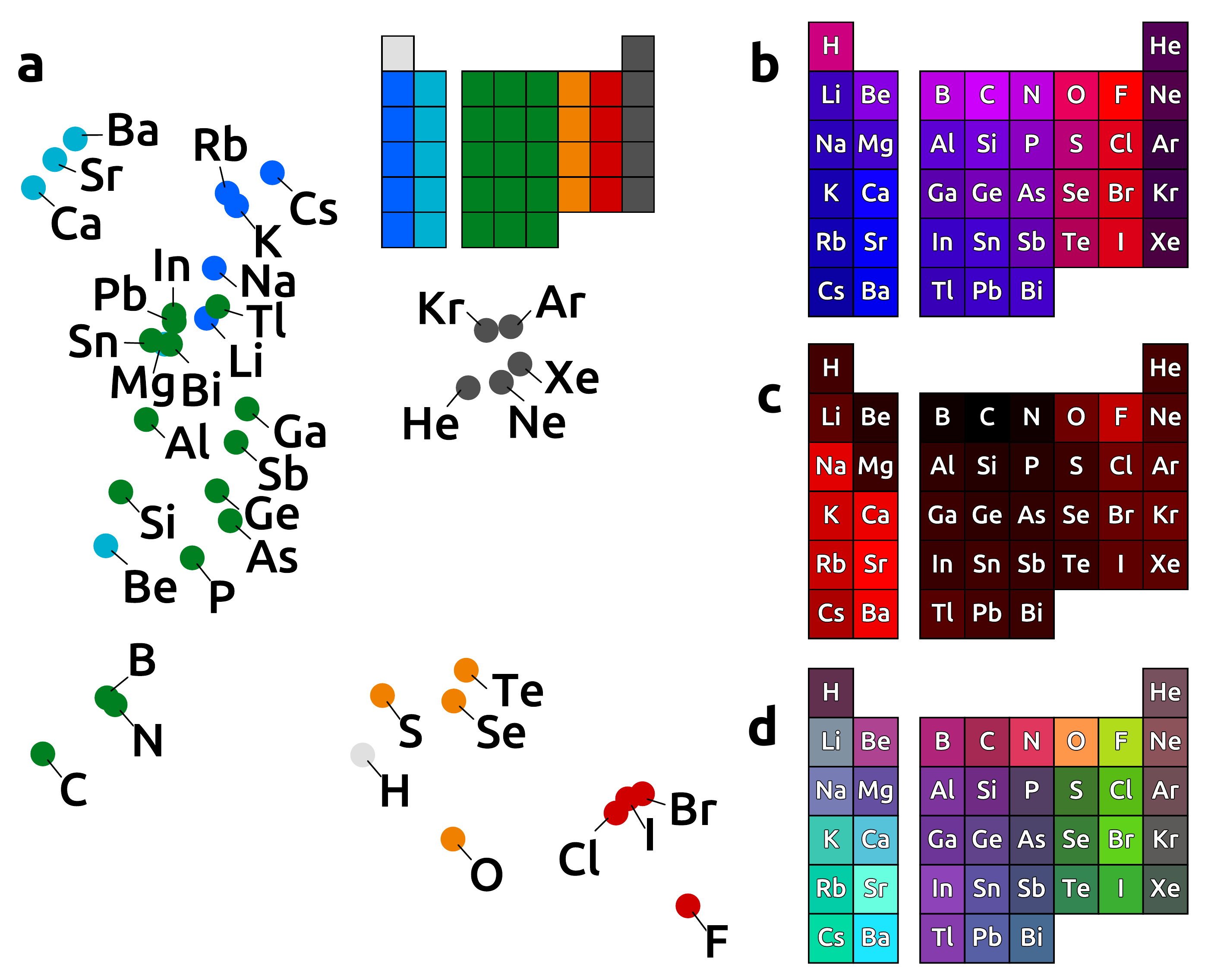}
\caption{\label{fig:periodic-elpa}
Data-driven representations of the chemical space. (a) A 2D map of
the elements contained in the elpasolite data set,\cite{fabe+16prl} with the coordinates corresponding to $\rep<1||a>$ and $\rep<2||a>$  for the case with $d_J = 2$ (see also Fig.~\ref{fig:lc-elpa}). Points are colored
according to the group. (b) A periodic table colored according to the
coordinates in the 2D chemical space. $\rep<1||a>$ corresponds to the red channel
and $\rep<2||a>$ to the blue channel. (c) A periodic table colored according to $\rep<1||a>$
(red channel) for a 1D chemical space. (d) A periodic table colored according
to 4D chemical coordinates ($\rep<1||a>$: red channel, $\rep<2||a>$: green channel, $\rep<3||a>$: blue
channel, $\rep<4||a>$: hatches opacity)
Reproduced with permission from Ref.~\citenum{will+18pccp}. Copyright 2018 PCCP Owner Societies. }
\end{figure}

Rather than optimizing the correlations between \emph{geometric} features and the target properties, one can attempt to build features that incorporate a notion of chemical similarity between different elements.
The idea was introduced in terms of an alchemical similarity kernel in Ref.~\citenum{de+16pccp}, that is also a core component of the FCHL framework\cite{fabe+18jcp}, but has been implemented in different forms in the context of atom-centered symmetry functions\cite{artr+17prb,gast+18jcp,rost+18jcp} and of generic atom-density correlation features\cite{will+18pccp}.
In general terms, the idea is to achieve a reduction of the dimensionality of the chemical space, writing formally a (linear) projection of the elemental features
\newcommand{\etilde}[0]{\tilde{\e}}
\begin{equation} \label{eq:alchemy}
    \rep<\etilde|=\sum_a \rep<\etilde||a>\rep<a|,
\end{equation}
where the coefficients $\rep<\etilde||a>$ enact the projection between the elemental and the ``alchemical'' basis.
The reduction in the dimensionality of the feature space can be substantial: for powerspectrum ($\nu=2$) features, the number of components scales quadratically with the number of species, and so even just halving the dimension of the chemical space reduces the number of powerspectrum features by 75\%{}:
\begin{multline}
\rep<\etilde_1 n_1; \etilde_2 n_2; l||\frho_i^2>=\\
\sum_{a_1a_2}\rep<\etilde_1||a_1>\rep<\etilde_2||a_2>
\rep<\ennl{}||\frho_i^2>.
\label{eq:alchemy-soap}
\end{multline}

Figure~\ref{fig:lc-elpa} demonstrates how reducing the dimensionality of chemical space helps achieving a transferable, accurate model with a small number of training structures.
By comparing learning curves with different degrees of compression, one sees that there is a similar data/complexity interplay as observed for radial correlations.
A low-dimensional alchemical space is beneficial in the data-poor regime, as it allows the model to make an educated guess about the interactions of pairs of elements that are not represented in the training set.
Learning curves with low chemical complexity, however, saturate in the limit of large training set, because they generate features that are not sufficiently flexible, and cannot describe the differences between elements.

The optimization of both geometrical and compositional components of the density-correlation features can be construed as a linear transformation of the kets (or, when seen in terms of the linear kernels built on such features, as the action of a Hermitian operator\cite{will+19jcp}).
The requirement that such transformations do not affect the symmetry properties of the features restricts form they can take -- for instance, they cannot mix different $l$ or $m$ dependent channels.
These observations imply that (1) \emph{linear} feature optimizations do not change the nature of the representations, and can be applied equally well to any implementation of $\rep|\frho_i^\nu>$ features, (2) as long as the linear transformation is full rank, there is no loss of information, which means that the observed change in performance is linked to the details of the regression scheme, such as regularization in linear or kernel models.

As a final remark, let us mention that a critical analysis of a feature-optimization effort often reveals insights into the physical-chemical properties of the system being studied and the target properties.
For instance comparing models of the energy using different $\rcut$ can be used to infer relationships between the length and energy scales\cite{bart+17sa}, and the inspection of the chemical mapping coefficients in Eq.~\eqref{eq:alchemy} can be used to construct a data-driven periodic table of the elements (see Fig.~\ref{fig:periodic-elpa}).
The use of interpretable, physics-inspired features can also be used to provide intuitive chemical insights by the construction of \emph{knock-out models}\footnote{A knockout mice is a genetically-modified mouse in which one or more genes have been inactivated. The effect on the development of the animal can be used to understand the role played by the affected gene(s).} in which for instance correlations are restricted to 2-bodies, the cutoff reduced to first or second neighbors.
The impact of these artificial restrictions on the features information content, and therefore on the asymptotic performance of the model, indicates how important 3 or higher-body order interactions are, or how much long-range effects are relevant to determine the value of the target property.\cite{helf+19jcp}

\begin{figure*}
\begin{minipage}{0.75\linewidth}
\includegraphics[width=1.0\linewidth]{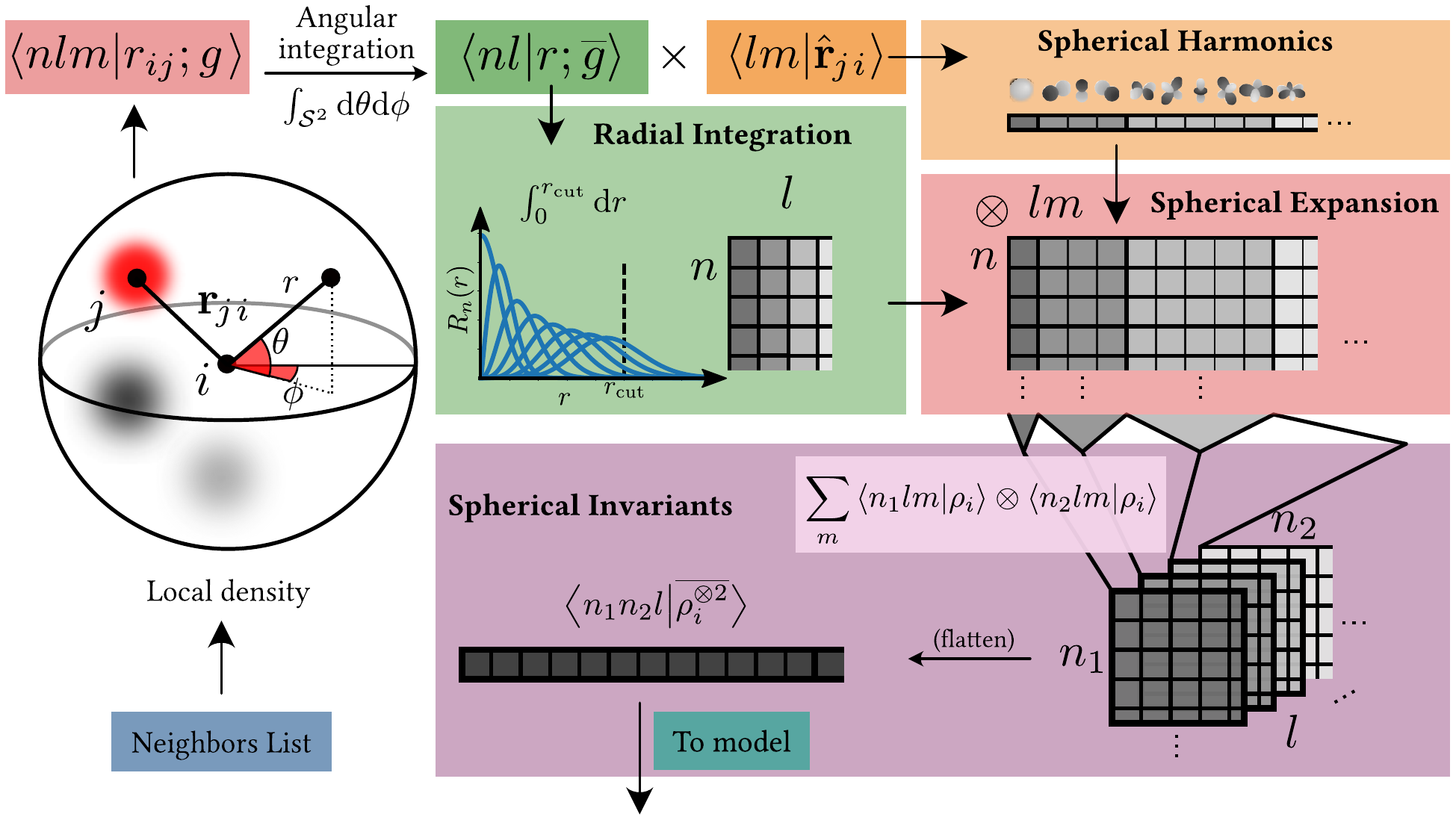}
\end{minipage}\hfill
\begin{minipage}{0.22\linewidth}
\caption{Schematic overview of the process
of expanding the density in a radial and
angular basis set, and recombining those
to form spherical invariants (or covariants). 
Reproduced with permission from Ref.~\citenum{musi+21jcp}. Copyright 2021 American Institute of Physics.
\label{fig:expansion-schematics}}
\vfill\vspace{32mm}\ 
\end{minipage}
\end{figure*}

\subsection{Efficient implementation}\label{sub:implementation}

Despite encoding similar information content, the differences in formulation of competing structural representations may lead to large variations in implementation and performance.
A first fundamental divide is between evaluation of features by summing over clusters of $\nu$ neighbors and computing $\nu$ tensor products of atomic densities (see \cref{sub:density-trick}).
Consider the case of evaluating ``SOAP-like'' ACSF by cluster sum (cost $\nmax^2 \lmax \nneigh^2$) and by density expansion (cost $\nneigh \nmax \lmax^2$ for the density, and $\nmax^2 \lmax^2$ for the SOAP evaluation).
Despite the adverse scaling of ACSF computed as a sum over clusters of neighbors, these representations can be implemented efficiently\cite{zuo+20jpcl,nguy+18jcp,Kamath2018} by relying on a careful selection of the features (discussed in \cref{sub:feature-selection}), reuse of parts of the computations, parallelism and GPU acceleration.\cite{sing+19jctc,Gao2020,Lu2020}
In fact, when computing a linear model which is explicitly equivalent to a $(\nu+1)$-body order potential, the low-order terms can be more efficiently evaluated as a sum over neighbors\cite{glie+18prb,pozdnyakov2019arxiv}

In line with the general focus of this review, we concentrate in particular on the efficient implementation of atom-density representations.
As we shall see, roughly the same considerations apply to both those representations that are usually built on a \emph{smooth} atom density,\cite{will+18pccp,will+19jcp,niga+20jcp} that generalize the construction of the SOAP powerspectrum and bispectrum,\cite{bart+13prb} and those that are usually computed in a way that corresponds to a $\delta$-like density, such as ACE\cite{drau19prb,drau20prb} and MTP.\cite{shap16mms,novi+21mlst}
Indeed, both families of representations rely on three steps: (i) expansion of the local atom density on a suitable basis, e.g. \cref{eq:density-nlm}, (ii) computation of $\nu$ tensor products of the expansion, and then (iii) contraction over the correlations to obtain equivariant features (Fig.~\ref{fig:expansion-schematics}).
While these three steps have been implemented in different ways, their efficient implementation relies on similar considerations.

\paragraph*{Atomic density expansion}
\Cref{eq:body-order-x} provides the blueprints for a broad class of $(\nu+1)$-body atom-density representation.
Practical implementations differ by the type of localized function used to construct the local atom density (see \cref{eq:rhoi-ket}), and by the radial and angular basis used for its expansion.
As discussed in \cref{sub:lm-rep}, spherical harmonics are a natural angular basis, but other choices are possible.
For instance, the MTP representation projects the atomic density onto a tensor product of direction vectors leading to the covariant \textit{moment tensor}\cite{mick+13jcp,shap16mms}
\begin{equation}
   \mathbf{M}_n^{\otimes \nu}(\rho_i) =  \sum_{j\in A_i} P_{n,\nu}(r_{ji})\,\,\underbrace{\brhat_{ji}\otimes\brhat_{ji}\ldots \otimes\brhat_{ji}}_{\nu \text{ times}}
\end{equation}
where $P_{\mu,\nu}$ is a radial function. \emph{Invariant} components can be obtained by combining and contracting products of the elements of these tensors.
The tensor product basis is directly related to spherical harmonics, as shown in appendix B.2 of Ref.~\citenum{Bachmayr2019}.
The performance of the MTP representation,\cite{zuo+20jpcl}  which relies on an efficient recursive evaluation of the basis functions\cite{shap16mms}, are a testament to the effectiveness of this basis choice.

As shown in Section~\ref{sub:lm-rep}, the  choice of an angular basis of spherical harmonics simplifies greatly the evaluation of \cref{eq:body-order-x}, that can be written in terms of contractions of density coefficients $\rep<\nlm||\frho_i>$ (cf.   \cref{eq:density-nlm}).
If the environment-centered density is written in terms of a sum of density functions $g(\bx-\br_{ji})\equiv\rep<\bx||\br_{ji}; g>$, peaked at the neighbors positions, the expansion coefficients can be written as the accumulation
\begin{equation}
\rep<\nlm||\frho_i> =
\sum_{j\in i} \fcut(r_{ji})\rep<\nlm||\br_{ji}; g>
\end{equation}
of terms that correspond to an expansion over a basis of radial functions $\rep<x||nl>$ and spherical harmonics  $\rep<\bxhat||lm>$ of contributions coming from Gaussians centered on each neighbor
\begin{equation}
\rep<\nlm||\br_{ji}; g> =
\int\D{\bx}
\rep<nl||x>\rep<\lm||\bxhat>
\rep<\bx||\br_{ji}; g>.\label{eq:g-nlm-r-integral}
\end{equation}
In the $g\rightarrow\delta$ limit, the contribution from the $j$-th neighbor amounts simply to a product of the radial and angular functions evaluated at $\br_{ji}$,
\begin{equation}
\rep<\nlm||\br_{ji}; \delta> =
\rep<nl||r_{ji}>\rep<\lm||\brhat_{ji}>.
\end{equation}
Similar to the 1D case discussed in Sec.~\ref{sub:smearing}, \emph{for a given choice of radial basis} the smearing of the density can be achieved by a mollification of the basis:
\begin{multline}
\rep<\nlm||\br_{ji}; g> =
\\\int\D{\bx}
\rep<nl||x>\rep<\lm||\bxhat> 
\int\D{\bx'} g(\bx - \bx')
\rep<\bx'||\br_{ji}; \delta > \\=
\int\D{\bx'}\rep<\bx'||\br_{ji}; \delta >  \int\D{\bx}
\rep<nl||x>\rep<\lm||\bxhat> g(\bx - \bx') \\
=\int\D{\bx'}\rep<\bx'||\br_{ji}; \delta >  \rep<\lm||\bxhat'> \rep<nl; g||x'> \\ \equiv \rep<\nlm; g||\br_{ji}; \delta>,
\label{eq:g-nlm-mollified}
\end{multline}
where we use $\rep<nl; g|$ to indicate the radial term that results from the Gaussian convolution. 
Each of these terms can be computed very efficiently, exploiting in particular the fact that all orders of the spherical harmonics and their derivatives can be computed using recursion relations.\cite{Limpanuparb2014,Bachmayr2019,drau19prb}

It might appear that using a smooth atom density $g$ complicates substantially the evaluation of \cref{eq:g-nlm-r-integral}. However when $g$ is a spherical Gaussian with standard deviation $\sigma$, the integral over $\D{\bxhat}$ can be computed analytically\cite{kauf-baum89jpb}
\begin{equation}
\int\D{\bxhat} \rep<\lm||\bxhat>
\rep<x\bxhat||\br_{ji}; g> =
\rep<x; l; ||r_{ji}; g>\rep<\lm||\brhat_{ji}>.
\label{eq:g-integral-angular}
\end{equation}
where the radial integral reads
\begin{equation}
\rep<x; l;||r; g> =
4\pi e^{-r^2/2\sigma^2}\!\! x^2 e^{- x^2/2\sigma^2}\mathsf{i}_{l}\!\left( x r/\sigma^2\right),
\end{equation}
so one gets
\begin{equation}
\rep<\nlm||\br_{ji}; g>=
\rep<\lm||\brhat_{ji}>
\int\D{x}
\rep<nl||x>
\rep<l; x|| r_{ji}; g>.
\end{equation}
The radial part of the integral
\begin{equation}
\int\D{x}
\rep<nl||x>
\rep<l; x||r_{ji}; g> = \rep<nl||r_{ji}; g>
\label{eq:g-integral-radial}
\end{equation}
can be computed numerically for any form of the radial basis resulting in $\nmax\lmax n_{grid}$ evaluations of special functions.
For instance, the original implementation of the SOAP representation uses a numerically orthogonalized, equispaced Gaussian basis.\cite{bart+13prb}
Alternatively, this integral might also be performed analytically by using Gaussian type orbitals (GTO) as the radial basis\cite{gris+18prl,Himanen2020}, $\rep<x||nl; \text{GTO}>$. This choice makes it possible to compute the coefficients of the smeared density as easily as for the $g\rightarrow \delta$ case
\begin{equation}
\rep<\nlm;\text{GTO} ||\br_{ji}; g> =
\rep<\lm||\brhat_{ji}>
\rep<nl;\text{GTO} ||r_{ji};g>,
\end{equation}
where the only overhead comes from having to compute $\order{\nmax\lmax}$ terms for the radial part and its orthonormalization.
This is asymptotically cheaper than combining radial and angular terms -- which requires $\order{\nmax\lmax^2}$ multiplications per neighbor -- but can be substantial in practical cases, because the analytical integrals in \cref{eq:g-integral-angular,eq:g-integral-radial} yield non-standard special functions.

\newcommand{\ghat}[0]{\hat{g}}
To reduce this overhead, one can choose a form of the atomic density that is symmetric about $\br_i$ instead of $\br_{ji}$\cite{caro19prb}
\begin{equation}
\rep<\bx||\br_{ji};\ghat>=\exp[-\frac{(x-r_{ji})^2}{2\sigma^2_{r}}-\frac{r_{ji}^2}{\sigma^2_{\perp}} (1 - \brhat_{ji} \cdot \hat{\bx} )].
\end{equation}
Together with a choice of radial functions that do not depend explicitly on $l$, this  allows factorizing the radial integral~\eqref{eq:g-integral-radial} as
\begin{equation}
   \int\D{x}
\rep<n||x>
\rep<x; l||r_{ji}; \ghat> = \rep<n||r_{ji};\ghat> \rep<l||r_{ji};\ghat>.
\label{eq:g-integral-radial-caro}
\end{equation}
Coupled with the polynomial basis proposed in Ref.~\citenum{bart+13prb}, these expansion coefficients can be computed efficiently using recurrence relations in the radial and angular coefficients.
More in general, the cost of evaluating the radial integrals $\rep<\nl|| r; g>$ can be made negligible by using splines to approximate the value of the special functions resulting from the integrals, or the numerical integration of basis functions for which there is no analytical expression.
Another aspect that does not affect the asymptotic scaling of the expansion, but can significantly influence the prefactor, involves the evaluation of spherical harmonics.\cite{Limpanuparb2014,drau20prb} Several well-established techniques can be used to speed up the calculation of $Y^m_l$, including the use of real-valued spherical harmonics, the use of recurrence relations, and the use of formulations that are entirely written in terms of the Cartesian components of $\brhat_{ji}$.

\begin{figure}
    \centering
\includegraphics[width=1.0\linewidth]{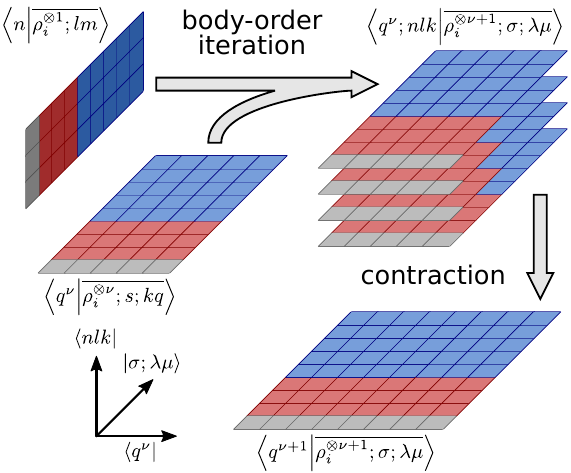}
\caption{A schematic representation of the NICE framework. A hierarchy of $N$-body equivariant features is built by iterative combination with the atom density coefficients, and the exponential increase in feature space size is kept at bay by successive contractions. 
Reproduced with permission from Ref.~\citenum{niga+20jcp}. Copyright 2020 American Institute of Physics.}
    \label{fig:nice-scheme}
\end{figure}

\paragraph*{Symmetrized $n$-body correlations}

The density coefficients $\rep<\enlm||\frho_i>$ are then combined to compute invariant (or covariant) features.
Formally, the evaluation of the symmetry-adapted features -- both those built using only local $\rep|\frho_i>$ features, and the multi-scale features that combine $\rep|\frho_i>$ and $\rep|V_i>$ -- involves a tensor product of $\nu$ sets of density coefficients to yield density correlations in the uncoupled basis $\rep<(\enlm_i)_{i=1}^{\nu}|$, and then a contraction along the $m_i$ indices, that generates the equivariant features expressed in the coupled basis $\rep<(\enlk_i)_{i=1}^{\nu}|$.
A technical difficulty one has to keep in mind when implementing the calculation of equivariant features is that the angular $(l,m)$ indices have an irregular memory layout, with $-l\le m \le l$.
Depending on the hardware architecture, it might be beneficial to store the coefficients in a regular $(\lmax+1)\times(2\lmax+1)$ array, padded with zeros.

A more substantial challenge associated with the increase of the body order is that both the number of linearly independent features and the cost of evaluating each of them based on a naive contraction of the tensor products of density coefficients (e.g. based on the expressions in Ref.\cite{drau19prb}) increase exponentially with $\nu$.
Even though the exponential scaling is related to the expansion parameters $\lmax$ and $\nmax$, and not on the number of neighbors, as it would be the case for the calculation of the features as a sum over clusters of $\nu$ atoms (see Sec.~\ref{sub:density-trick} and~\ref{sub:linear-models}), it makes the enumeration of a complete linear basis prohibitively expensive.
The recurrence relations\cite{niga+20jcp} of~\cref{eq:rho-recursion} (or the equivalent ones for the invariant features proposed in Ref.\cite{Bachmayr2019}) make it possible to evaluate individual equivariant features with a cost that scales only linearly with $\nu$.
To beat completely the exponential scaling, these recursive expressions should be combined with feature selection schemes such as those discussed in Section~\ref{sub:feature-selection}.
For example, the $n$-body iterative contraction of equivariant (NICE) features incorporates a selection/contraction step at each level of the iteration.
For each equivariant component $\rep<\q'||\rrhoislm{\nu}>$, one determines (e.g. by principal component analysis, or just by dropping some components) a set of coefficients $U^{\nu;\sigma\lambda}_{\q'\q}$ that can be used to reduce the dimensionality of the features
\begin{equation}
\rep<\q^{\nu;\sigma\lambda}||\rrhoislm{\nu}> = \sum_{\q'} U^{\nu;\sigma\lambda}_{\q\q'} \rep<\q'||\rrhoislm{\nu}>.
\label{eq:nice-contraction}
\end{equation}
Given that this operation only mixes features with the same equivariant behavior, it is then possible to perform an iteration equivalent to~\cref{eq:rho-recursion} to increase the body order further
\begin{multline}
\rep<\q||\rrhoislm{(\nu+1)}>\equiv
\rep<\q^{\nu;\tau k}; n l k||\rrhoislm{(\nu+1)}> =\\ \delta_{\sigma(\tau(-1)^{l+k+\lambda})}
\sum_{m}  \cg{l m}{k (\mu -m)}{\lambda\mu} \\ \times\rep<n||\frho[lm]_i^1>
\rep<\q^{\nu;\tau\lambda}||\frho[\tau;k(\mu-m)]_i^\nu>.
\label{eq:nice-contract-recursion}
\end{multline}
Note that in the first line we use the loose definition of the indices in the bra-ket notation (Section~\ref{sub:bra-ket}): the $\nu+1$ term can be indexed explicitly, with a notation that recalls the lower-order terms that are combined to obtain it; once it is computed, the granularity of the indexing becomes irrelevant, and a flat index can be used to streamline the notation.
With this combination of expansion and contraction only the components that contribute significantly to the description of the structural diversity of the dataset, or to the prediction of the target properties, are retained to evaluate higher-order correlations.

An alternative perspective for developing efficient implementations is to represent invariant or equivariant properties $\y(A_i)$ in terms of the {\em unsymmetrized} correlations,
\[
\begin{aligned}
    \y(A_i) &\approx
    \sum_{\nlm_1\ldots \nlm_{\nu}}
    \rep<\y||\nlm_1\ldots n_\nu l_\nu m_\nu>
    \\
    & \qquad \qquad \times \rep<\nlm_1\ldots \nlm_\nu||\rho_i^{\otimes \nu}>,
\end{aligned}
\]
with the desired symmetries imposed through constraints on the coefficients
$\rep<\y||n_1l_1m_1\cdots n_\nu l_\nu m_\nu>$. While this perspective imposes additional complexity on regression schemes it is convenient for fast {\em evaluation} of a fitted model (with coefficients now ensuring the correct symmetries) since the coupling coefficients need not be stored or evaluated anymore. An efficient evaluation now requires a recursion for the unsymmetrized correlations
\[
    \rep<\nlm_1;\ldots \nlm_\nu||\rho_i^{\otimes\nu}>
    = \prod_{\alpha = 1}^\nu \rep<n_\alpha l_\alpha m_\alpha || \rho_i >,
\]
which is relatively straightforward to construct,\cite{Bachmayr2019}
the key challenge being to retain only the $(n_\alpha,l_\alpha,m_\alpha)_\alpha$ features that give rise to non-zero coefficients. %

\begin{figure}[tbp]
\includegraphics[width=0.99\linewidth]{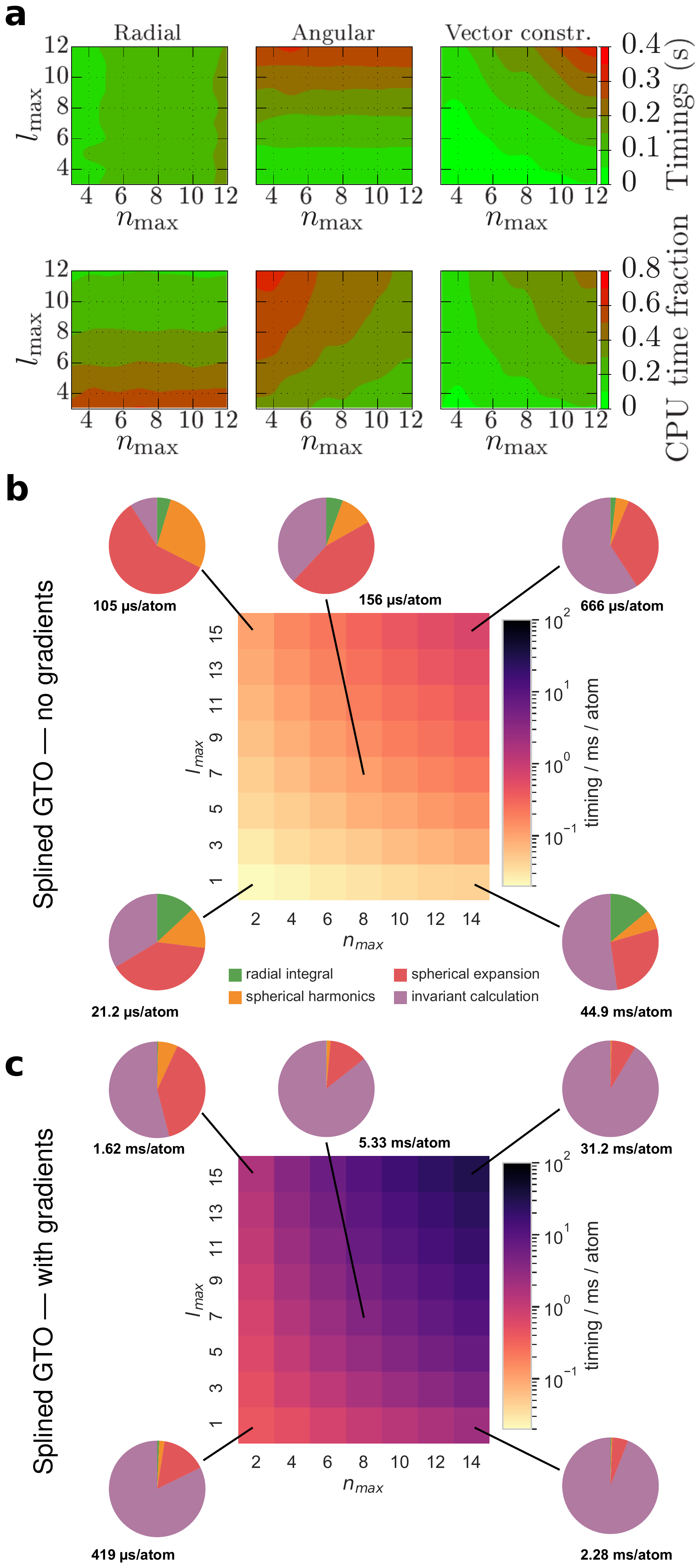}
\caption{
(a) Single-core timings for the evaluation of 
radial expansion, angular expansion, and SOAP vector construction for an atomic structure containing 10'000 randomly-placed atoms, using the implementation discussed in Ref.~\citenum{caro19prb} and as a function of $(\nmax,\lmax)$. 
Reproduced with permission from Ref.~\citenum{caro19prb}. Copyright 2019 American Physical Society.
(b) Single-core timings for the evaluation of SOAP features for a dataset of molecular crystals\cite{musi+19jctc}, using the implementation in librascal\cite{LIBRASCAL}, as a function of $(\nmax,\lmax)$; to compare with panel (a), consider that the presence of 4 distinct chemical elements corresponds roughly to a fourfold increase of $\nmax$. The breakdown of the total timing in the different steps of the calculation is shown for a few representative sizes of the expansion.
(c) As in (b), including also the calculations of the gradients of the features with respect to atomic positions. 
Reproduced with permission from Ref.~\citenum{musi+21jcp}. Copyright 2021 American Institute of Physics.
\label{fig:soap-timings}
}
\end{figure}

\begin{figure*}[tbp]
\begin{minipage}{0.64\linewidth}
\includegraphics[width=1.0\linewidth]{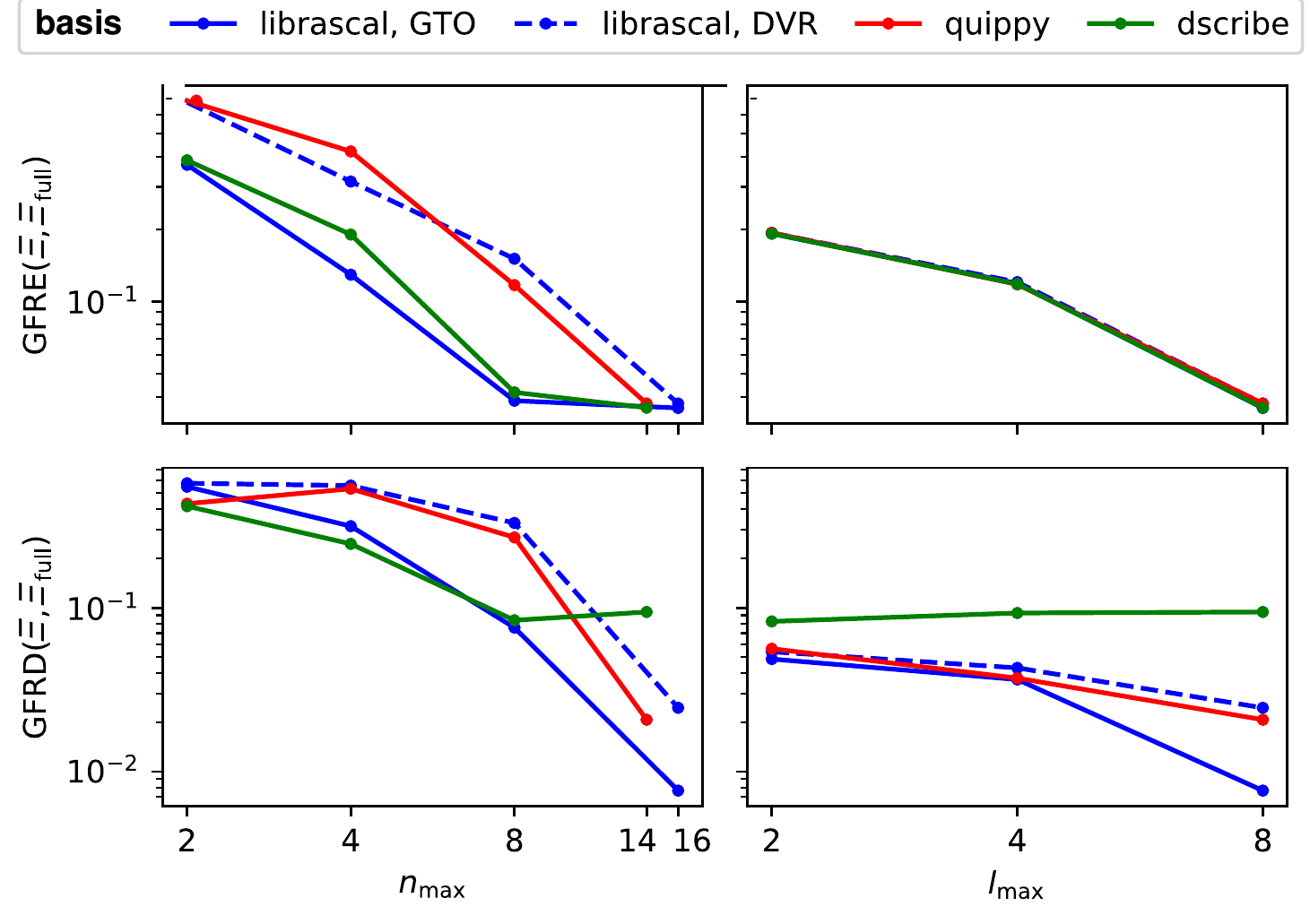}
\end{minipage}
\
\begin{minipage}{0.34\linewidth}
\caption{The panels demonstrate the convergence of SOAP features as computed for 10'000 random \ce{CH4} configurations\cite{matcloud20a} using the radial bases implemented in different codes, and measured in terms of the error one incurs when linearly predicting fully-converged features $\bFeat_\text{full}$ ($\nmax=24,\lmax=12$, computed with the GTO implementation in librascal) using features with $\lmax=8$ and growing values of $\nmax$ (left) and with $\nmax=14$ (16 for librascal) and growing values of $\lmax$ (right).
Top panels show the linear reconstruction error (GFRE) which measures the amount of information that cannot be linearly decoded from the coarser features. Bottom panels show the reconstruction distortion (GFRD) which measures the additional error one makes when limiting the reconstruction to an orthogonal transformation.\cite{gosc+21mlst}
}\label{fig:feature-reconstruction}
\end{minipage}
\end{figure*}

\subsection{Packages to evaluate atom-density representations}

To provide a practical example of the use of software to compute representations, we compare three packages, namely quippy\cite{QUIP}, dscribe\cite{Himanen2020} and librascal\cite{LIBRASCAL,musi+21jcp}, that are open source, and can be easily used in a Python code.
We do not discuss the internals of the implementations, but show code snippets that can be readily used to evaluate descriptors of atomic structures, primarily focusing on the SOAP powerspectrum. All examples use the Atomic Simulation Environment\cite{hjor+17jpcm}, and atomic structures are assumed to be stored in the variable \verb+structures+, an instance of ASE's \verb+Atoms+ object. In all of these implementations, the descriptor vectors are returned as \verb+numpy.array+ objects, from which kernel values may be obtained by computing the dot products between descriptor vectors. 
We also do not discuss the computational efficiency of the different codes, which is still the subject of very active development. Fig.~\ref{fig:soap-timings} provides some representative timings from librascal, and from a recent implementation of SOAP that uses non-Gaussian atomic densities\cite{caro19prb}. 
The wildly different breakdown of the computational effort as a function of the basis set size, and the large overhead associated with the evaluation of the gradients of the features highlight some of the implementation challenges.

The quippy python package is based on the QUIP suite with the GAP extension, which provides the \verb+descriptors+ module. QUIP must be downloaded and built using a Fortran compiler before quippy, which uses f90wrap to access the compiled functions in QUIP via python interfaces. In the GAP implementation, Gaussian radial basis functions are used, placed at equal intervals, and orthogonalized.
\begin{lstlisting}
from quippy.descriptors import Descriptor
soap = Descriptor("soap cutoff=3.5 cutoff_transition_width=0.0 atom_sigma=0.3 n_max=8 l_max=6")
# returns a dictionary containing the features and connectivity information
features = soap.calc(structures)
# features["data"] is a numpy array with the shape (\n_environments,n_features)
\end{lstlisting}
The \verb+Descriptor+ object is initialised using a string containing the kernel parameters in a \verb+key=value+ format, with some keys being mandatory.

The \verb+dscribe+ package\cite{Himanen2020} implements multiple descriptors, including SOAP, MBTR\cite{huo-rupp17arxiv} and ACSF. A python interface is used to interact with calculator functions written in C/C++, ensuring efficient evaluation. The main difference between the SOAP implementation of \verb+quippy+ and \verb+dscribe+ are the choice of radial basis functions, which are spherical primitive Gaussian Type Orbitals (GTOs), orthogonalised using the method suggested by L\"owdin\cite{lowd50jcp}. Alternatively, cubic or higher order polynomials may also be chosen.
In analogy with the definition of GTOs used in quantum chemistry, the radial basis has an explicit dependence on $l$.
\begin{lstlisting}
from dscribe.descriptors import SOAP
soap = SOAP(
    rcut=3.5,
    nmax=8,
    lmax=6,
    sigma=0.3,
    species=["H","C"]
)
# returns a (n_environments,n_features) numpy array
X = soap.create(structures)
\end{lstlisting}
The the python object providing the descriptor is constructed from the class \verb+SOAP+ and specifying the parameters in the initialisation arguments.

The package librascal also provides a variety of descriptors, but chiefly focuses on the calculation of density-based representations, including SOAP and the $\nu=1$ and $\nu=3$ correlations. The back-end, written in C++, can be accessed from python interfaces.
Exploiting the spirit of the general construction of $\rep|\frho_i^\nu>$ features, librascal implements two kinds of radial functions, namely a family of GTO-like radial functions\cite{gris+19book} as well as a discrete variable representation (DVR) basis, corresponding to a real-space evaluation of the symmetrized density using a Gauss-Legendre quadrature rule.
\begin{lstlisting}
from rascal.representations import SphericalInvariants
hypers = {
    'soap_type': 'PowerSpectrum',
    'interaction_cutoff': 3.5,
    'radial_basis': 'GTO', # or 'DVR'
    'max_radial': 8,
    'max_angular': 6,
    'gaussian_sigma_constant': 0.3,
    'gaussian_sigma_type': 'Constant',
    'cutoff_smooth_width': 0.0,
    'normalize' : False
}
soap = SphericalInvariants(**hypers)
# returns a (n_environments,n_features) numpy array
X = soap.transform(structures).get_features(soap)
\end{lstlisting}
The \verb+SphericalInvariants+ object uses the \verb+transform+ method to compute SOAP features, that are stored internally in a sparse format, in which each dense block corresponds to a $(\e,\e')$ pair of elemental densities. These features can be used to compute scalar-product kernels between two environments, or cast to a dense array through the \verb+get_features+ method.

These three packages all compute ``SOAP'' features, but differ in the choice of basis functions. Much as with electronic structure codes, that often yield  results that differ significantly despite performing nominally the same type of calculations,\cite{leja+16science} one cannot expect to be able to combine the features computed by one package with the regression weights computed by another.
It is however important to assess whether the features are equivalent in a less stringent sense, e.g. whether they contain analogous information, and whether they converge to the same limit when the expansion parameters $(\nmax, \lmax)$ are increased.
Figure~\ref{fig:feature-reconstruction} demonstrates the convergence of the GFRE and GFRD (see Section~\ref{sub:feat-compare} and Ref.~\citenum{gosc+21mlst}) between small-$(\nmax, \lmax)$ features and a highly converged $\bFeat_\text{full}$ featurization. In all cases we consider GFRE$(\bFeat_\text{full},\bFeat)$ is at least one order of magnitude smaller than GFRE$(\bFeat,\bFeat_\text{full})$.
One sees that, reassuringly, in all cases the feature reconstruction errors converge towards zero. For $\nmax=16$ all choices of radial bases are essentially converged, and the residual error is due to the convergence of the angular channels.
Since all implementations use equivalent spherical harmonics expansions, the convergence with the angular cutoff $\lmax$ is nearly identical.
The convergence rate of the radial bases, however, is not the same. The GTO bases in librascal and dscribe have similar amounts of information (although they are not fully equivalent, as they are parameterized differently), and converge faster than the bases used in quippy and the librascal DVR implementation.
The GFRD also converges to zero for most implementations -- meaning that in the complete basis set limit the corresponding features become equivalent. The implementation in dscribe is an exception, with a GFRD saturating at approximately 0.1, suggesting that implementation details lead to persistent differences in the weighting of different kinds of correlations even when $(\nmax, \lmax)$ increase beyond the values that are typically used in practice.

\section{Applications and current trends}\label{sec:applications}

In this Section we report some representative applications that highlight different aspects of the representations discussed in this Review -- demonstrating how an understanding of the nature and properties of the structure/features mapping can be used to construct efficient and insightful machine-learning models.

\begin{figure}[hbpt]
    \centering
\includegraphics[width=1.0\linewidth]{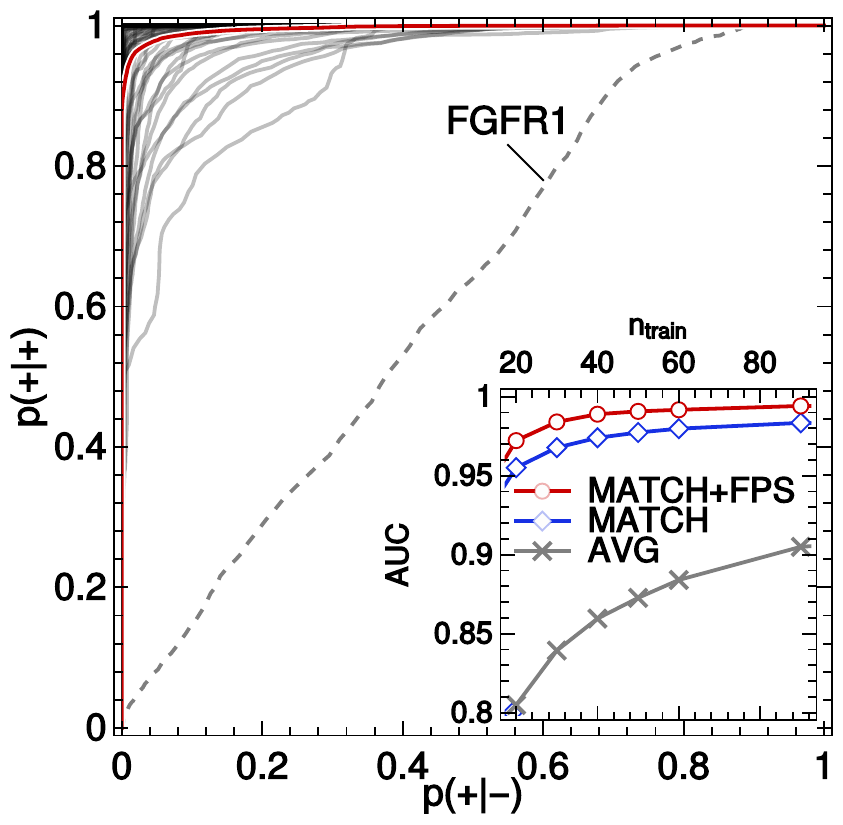}
\caption{ROCs of binary classifiers based on a SOAP kernel, applied to the prediction of the binding behavior of ligands and
decoys taken from the DUD-E\cite{Mysinger2012}, trained on 60 examples. Each ROC corresponds to one specific protein receptor, and plots the fraction of true positives $p(+|+)$ against the fraction of false negatives $p(+|-)$. The red curve is the average over the individual ROCs. The
dashed line corresponds to receptor FGFR1, which contains inconsistent data in the version of the DUD-E at the time of the original publication\cite{bart+17sa}. Inset: AUC performance measure as a function of the number of ligands used in the training, for the ``best match''-SOAP kernel (MATCH) and average molecular SOAP kernel (AVG).  
Reprinted with permission from Ref.~\citenum{bart+17sa}.  © The Authors, some rights reserved; exclusive licensee AAAS. Distributed under a \href{http://creativecommons.org/licenses/by-nc/4.0/}{Creative Commons Attribution License 4.0 (CC BY-NC)}.
}
    \label{fig:dude}
\end{figure}

\subsection{Best match kernels for ligand binding}

Contrary to the problem of predicting interatomic potentials, or other extensive properties, the affinity between a protein and a small drug-like molecule does not fit well into the mold of an additive property model.
The structure of the ligand must allow for the active portion of the molecule to fit in the binding pocket of the target protein, and the nature of the chemical groups in this ``warhead'' portion are more important to determine the strength of the interaction than peripheral portions of the molecule.
Figure~\ref{fig:dude} shows the accuracy of a classifier based on SOAP features, that aims to distinguish active components from decoys for a given target protein. The targets and the ligands, as well as their ``ground truth'' binding behavior are taken from the database of useful decoys, enhanced (DUD-E)\cite{Mysinger2012}. The performance of the classifier is represented in terms of the receiver operating characteristic (ROC) curves (the ROC curve of a perfect classifier would run along the left and top margins of the plot, while a classifier that is as good as random would run along the diagonal), and their area under the curve (AUC) (the AUC is the integral of the ROC, and roughly corresponds to the fraction of molecules that are classified correctly).
The AUC plot, in the inset of Fig.~\ref{fig:dude} shows that a model based on an average metric -- that describes each molecule as the average of its environments, Eq.~\eqref{eq:d-average} -- performs rather poorly, which is unsurprising given the highly non-additive nature of the binding affinity.
Using a ``best-match'' kernel (equivalent to the distance in Eq.~\ref{eq:d-bestmatch}, and implemented in practice as the small-$\gamma$ limit of the REMatch kernel\cite{de+16pccp}) improves dramatically the accuracy of the classifier, bringing the AUC to well above 0.95. A judicious choice of the training structures, based on farthest point sampling, accelerates even further the convergence of the classifier with train set size.
This application provides an example of how local representations can be combined in a non-additive way, resulting in a dramatic improvement of the machine-learning performance for a problem in which non-additive behavior is to be expected.

\begin{figure}[bt]
\centering
\includegraphics[width=\linewidth]{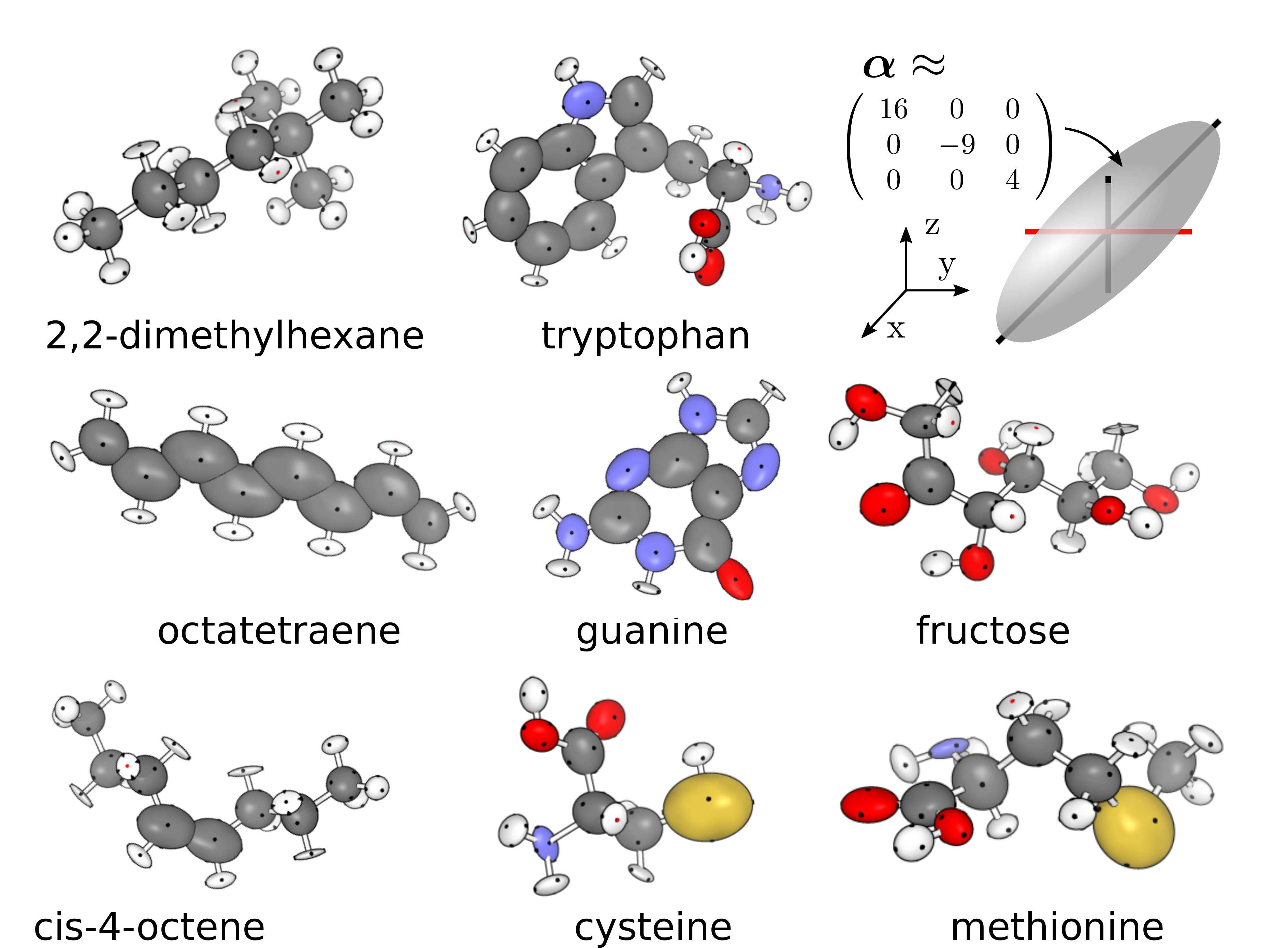}
\caption{Predicted atomic contributions to the total CCSD polarizability tensor for a selection of the showcase dataset as reported in Refs.~\citenum{wilk+19pnas,yang+19sd}. The ellipsoids are aligned along the principal axes of $\boldsymbol{\alpha}_i$, and their extent is proportional to the square root of the corresponding eigenvalue. The principal axes are shown, and are colored based on whether the corresponding eigenvalues are positive (black) or negative (red). 
Reproduced with permission from Ref.~\citenum{wilk+19pnas}.
Copyright 2019 National Academy of Sciences.
}
\label{fig:anisotropies_showcase}
\end{figure}

\subsection{Tensorial features and polarizability}

Some of the early examples of machine-learning models leveraging covariant features focused on the prediction of dielectric response functions, such as the dipole moment $\boldsymbol{\mu}$ (and the equivalent bulk quantity, polarization), polarizability $\boldsymbol{\alpha}$ (and the closely-related electronic dielectric constant) as well as higher-order terms, such as the first hyperpolarizability $\boldsymbol{\beta}$.
We discuss the case of the static  dipole polarizability $\boldsymbol{\alpha}$ as a representative case that highlights many of the current ideas and applications.
In its Cartesian form, $\boldsymbol{\alpha}$ is a symmetric tensor, fully determined by six components $(\alpha_{xx},\alpha_{yy},\alpha_{zz},\alpha_{xy},\alpha_{xz},\alpha_{yz})$.
In order to build a machine-learning model based on equivariant density correlation features, it is more convenient to apply a unitary transformation that casts it into its irreducible spherical components (ISCs). The spherically symmetric term, $\alpha^{(0)}_0$, corresponds to the trace of the tensor, while the 5 anisotropic components, $\alpha^{(2)}_{\{-2,-1,0,+1,+2\}}$ transform collectively as $\lambda=2$ spherical harmonics, and can be computed using recursive relationships that are explicitly reported in Ref.~\citenum{stone1975}. A clear advantage of this construction is that, unlike the components of the Cartesian tensor, the two ISCs of $\boldsymbol{\alpha}$ can be independently represented by the equivariant density-based features corresponding to $\lambda=0$ and $\lambda=2$, relying on a linear prediction model similar to the one reported in Eq.~\eqref{eq:tenspred}.\cite{gris+18prl,wilk+19pnas,gris+19book}

\begin{figure}[tb]
\centering
\includegraphics[width=0.95\linewidth]{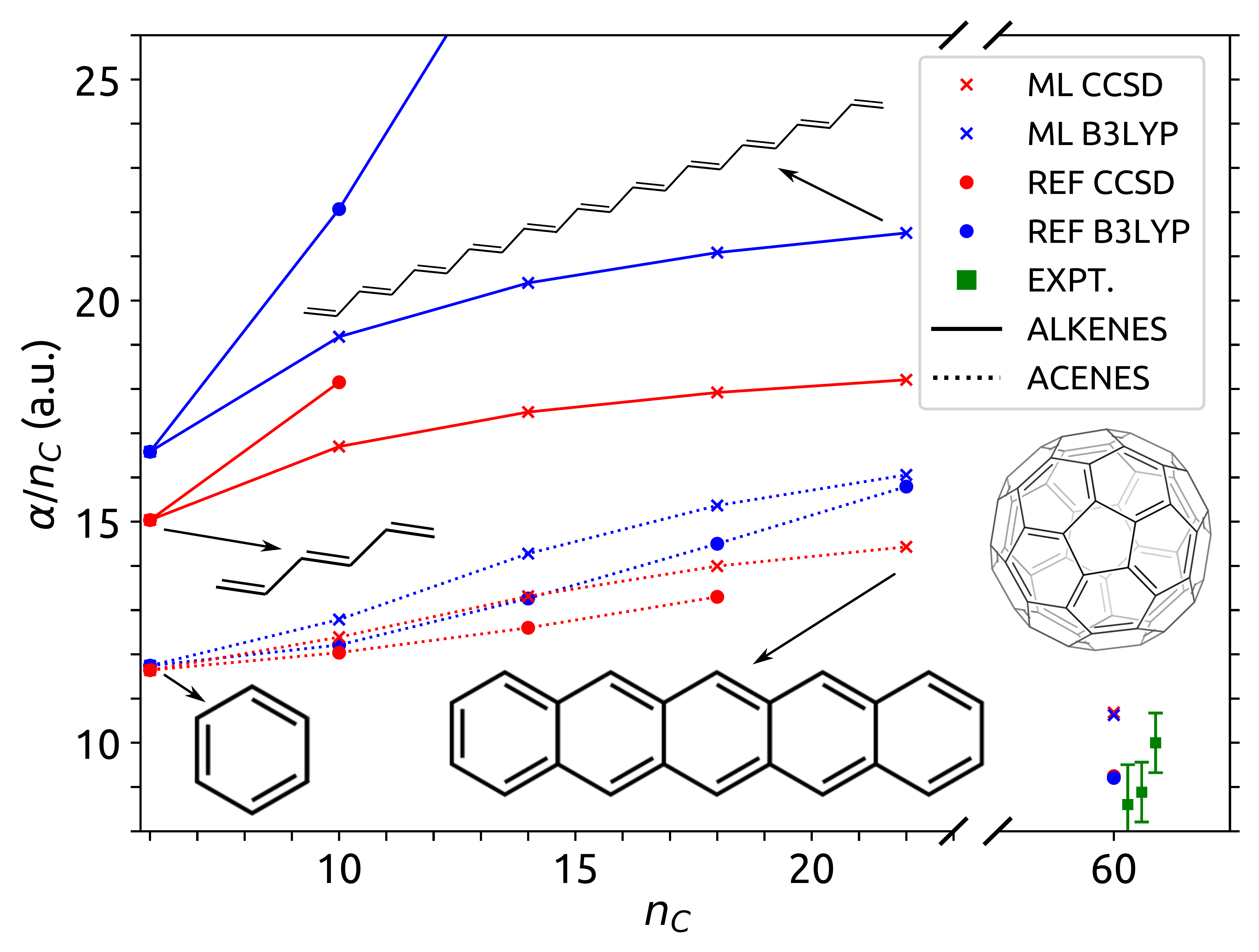}
\caption{
Polarizability per carbon atom ($\alpha/n_C$) vs. number of carbons ($n_C$) for the series of s-\textit{trans} alkenes (from \ce{C6H8} to \ce{C22H24}, full line) and acenes (from benzene to pentacene, dotted line), as well as fullerene (\ce{C60}). The green squares (and error bars) indicate the experimental measurements for \ce{C60}\cite{FullereneReview}. Results are provided from DFT (blue) and CCSD (red) calculations, as well as the corresponding AlphaML models. 
Reproduced with permission from Ref.~\citenum{wilk+19pnas}.
Copyright 2019 National Academy of Sciences.
}\label{fig:conjugated}
\end{figure}

The inherent locality of the model means that the tensor prediction can be broken down in the sum of individual atomic contributions $\boldsymbol{\alpha}=\sum_i\boldsymbol{\alpha}_i$.
These local components can be combined to make predictions on larger, and more complex molecules than those included in the training set. This transferability was exploited in the AlphaML model\cite{wilk+19pnas,alphaml} to fit against coupled-clusters (CCSD) reference values, computed on small organic molecules from the QM7b dataset\cite{mont+13njp,yang+19sd}, and predict on 52 larger ``showcase'' molecules that are at the limit of what is computable with state-of-the-art quantum chemistry methods. On these molecules, the error of AlphaML against the CCSD reference (0.24 a.u./atom) was less than half the discrepancy between CCSD and DFT (0.57 a.u./atom).

An additive model also provides predictions for the local contributions to $\boldsymbol{\alpha}$, which are represented in Fig.~\ref{fig:anisotropies_showcase}, in terms of ellipsoids aligned along the principal axes of $\boldsymbol{\alpha}(A_i)$.
Even though these components do not have to be physically meaningful -- given that the only training target is given by \emph{total} polarizabilities -- the local $\boldsymbol{\alpha}(A_i)$ reflect some chemical insights, e.g. the model predicts large components when centering the representation on the highly-polarizable sulfur atoms, as well as along the directions where the molecules are highly polarizable.
Highly conjugated molecules are also interesting  because they exhibit a non-additive behavior of the polarizability, due to the vanishing HOMO-LUMO gap. Due to the spatial nearsightedness of the representation, the model breaks down when asked to predict the polarizability of large polyenes and polyacenes based on the information learned on simpler and smaller molecular units. This is well represented in Fig.~\ref{fig:conjugated}, where the prediction of $\boldsymbol{\alpha}$ is tested for conjugated carbon-based molecules of increasing size, including fullerene\cite{wilk+19pnas}.

\begin{figure}[tbp]
  \centering
  \includegraphics[width=\linewidth]{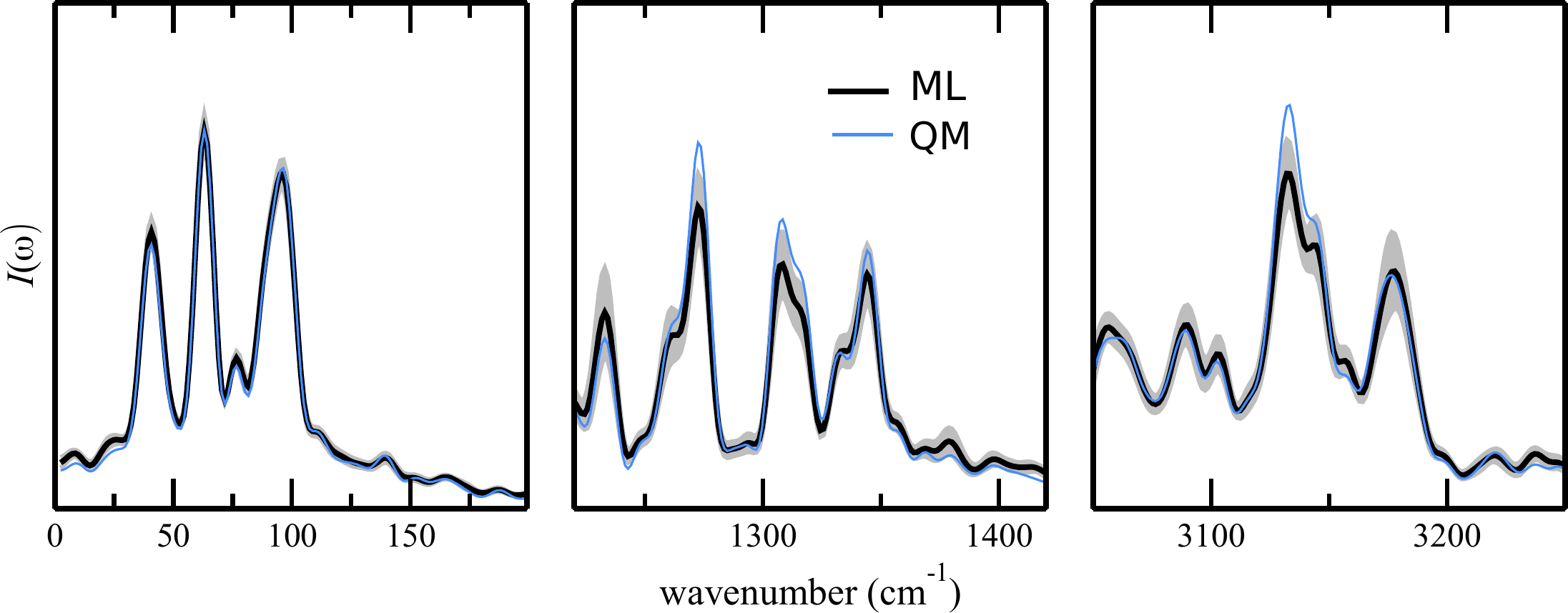}
  \caption{(\emph{black line}) Raman spectrum prediction of paracetamol form-I averaged over 16 different training models. Each training model is obtained by a random subselection of 2000 configurations over a total of 2500. (\emph{shaded area}) Standard deviation of the predicted spectra over the 16 models, calibrated with a likelihood maximization procedure described in Ref.~\citenum{musi+19jctc}. (\emph{blue line}) Reference \emph{ab initio} Raman spectrum. 
  Adapted from Ref.~\citenum{raim+19njp}. Copyright 2019 IOP Publishing under Creative Commons Attribution 4.0 International License \url{https://creativecommons.org/licenses/by/4.0/}.
  }
  \label{fig:raman}
\end{figure}

The prediction of $\boldsymbol{\alpha}$ using equivariant features can also be extended to the condensed phase and provides a crucial ingredient to compute Raman spectra. An example of this is reported in Ref.~\citenum{raim+19njp}, where the polarizability of crystal polymorphs of paracetamol are predicted along a full molecular dynamics trajectory, thus allowing for the calculation of the Raman intensity in terms of the polarizability correlation spectrum.
As shown in Fig.~\ref{fig:raman}, given the local nature of the polarizability response in this kind of systems, accurate Raman intensities and lineshapes can be predicted for the entire range of frequencies.
The low cost associated with computing dielectric response functions by ML models using symmetry-adapted features makes it possible to routinely evaluate condensed-phases infrared and Raman spectra including also a description quantum mechanical nature of the nuclei\cite{kapi+20jcp} -- a task that until very recently required enormous computational effort\cite{mars-mark17jpcl}.

\begin{figure}[tbp]
    \centering
  \includegraphics[width=\linewidth]{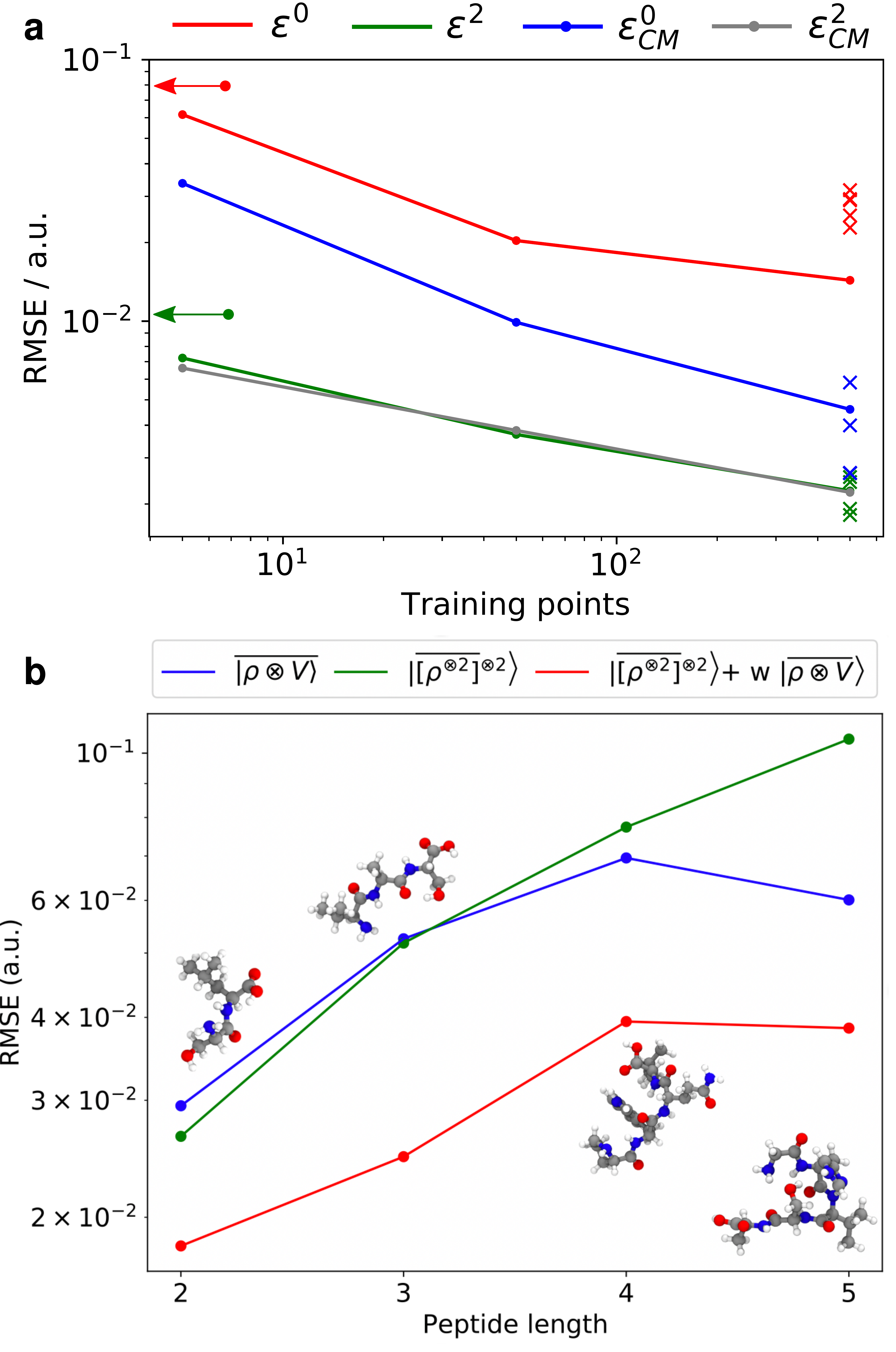}
\caption{(a)
Learning curves of the $\lambda=0$ and $\lambda=2$ components of the dielectric response tensor $\boldsymbol{\epsilon}_\infty$ of water, through direct learning (red and
green lines, respectively) and indirect learning going through the Clausius-Mossotti relation (blue and gray).
The testing data set consists of 500 independent configurations.
Arrows indicate the intrinsic standard deviation of the testing
samples. Crosses show the predictions for 5 hexagonal ice
structures using the ML model trained on liquid water.
Adapted with permission from Ref.~\citenum{gris+18prl}.
Copyright 2018 American Physical Society
(b) Absolute RMSE in learning the $\lambda=0$ spherical tensor of polarizability of polypeptides as a function of the peptide length. The model was trained on 27428 single amino acids and 370 dipeptides. The error was computed on 30 dipeptides, 20 tripeptides, 16 tetrapeptides and 10 pentapeptides respectively. The curves correspond to a LODE model (blue) a squared-kernel SOAP model (green) and a hybrid model mixing the two kernels (red). Adapted with permission from Ref.~\citenum{gris+21cs}. Copyright 2020 Royal Society of Chemistry.
}
    \label{fig:nonlocal-alpha}
\end{figure}

\subsection{Long-range and non-local responses}

The clear breakdown of a ML model based on local features that is apparent in Fig.~\ref{fig:conjugated} is representative of a general limitation of density-based features. There are essentially two approaches one can take to tackle the issue of the non-locality of the structure-property relations, both of which are illustrated in Figure~\ref{fig:nonlocal-alpha}.
One approach is to learn a proxy of the target property, which has a more localized nature, and which can then be easily manipulated to obtain the end result. The top panel of Fig.~\ref{fig:nonlocal-alpha}, adapted from Ref.~\citenum{gris+18prl}, is an example of this approach. The electronic dielectric response $\boldsymbol{\epsilon}_\infty$ of bulk water is affected by a collective, macroscopic electrostatic effect that is captured, in the continuum limit, by well-known expressions such as the Clausius-Mossotti relation, $\boldsymbol{\alpha}=V(\boldsymbol{\epsilon}-1)/(\boldsymbol{\epsilon}+2)$, that links $\boldsymbol{\epsilon}_\infty$ to an effective molecular polarizability.  This effective $\boldsymbol{\alpha}$ is more readily learnable by a local model, leading to better accuracy and transferability in predicting  $\boldsymbol{\epsilon}_\infty$.

A different approach is needed when there is no obvious transformation of the target property to a more local version, as is the case for the polarizability of conjugated hydrocarbons. In these cases, one needs a model that is able to describe arbitrary non-local correlations.
Long-range representations such as  multiscale LODE features (Section~\ref{sub:long-range} and~\ref{sub:ms-model}) are particularly attractive, in that they combine a long-range character (coming from the potential field) with an additive decomposition that provides the transferability needed to extend the prediction to systems of increasing size.
This is demonstrated in Fig.~\ref{fig:nonlocal-alpha}, where the multiscale LODE model is tested for predicting the isotropic component of the polarizability of a series of polypeptides of increasing length.
While the prediction at small peptides lengths share a similar accuracy as that obtained using a pure density-based representation, the inclusion of the potential field greatly decreases the prediction error when considering longer molecular chains.
A large, overall improvement of the prediction accuracy is observed when adopting an optimized, weighted  combination between local and LODE features.
These result suggests that the inclusion of long-range features within the regression model provides a better description of the intermediate-range interactions, and that by adjusting the relative importance of local and delocalized terms the model can be trained only on small molecules, and extrapolate reliably across systems of increasing size.
Very similar findings were reported on the transferability of models of molecular dipoles\cite{veit+20jcp} -- where however the splitting between local and long-range physics was achieved by combining different regression models rather than by different choices of features.

\begin{figure}[bp]
 \includegraphics[width=0.9\linewidth]{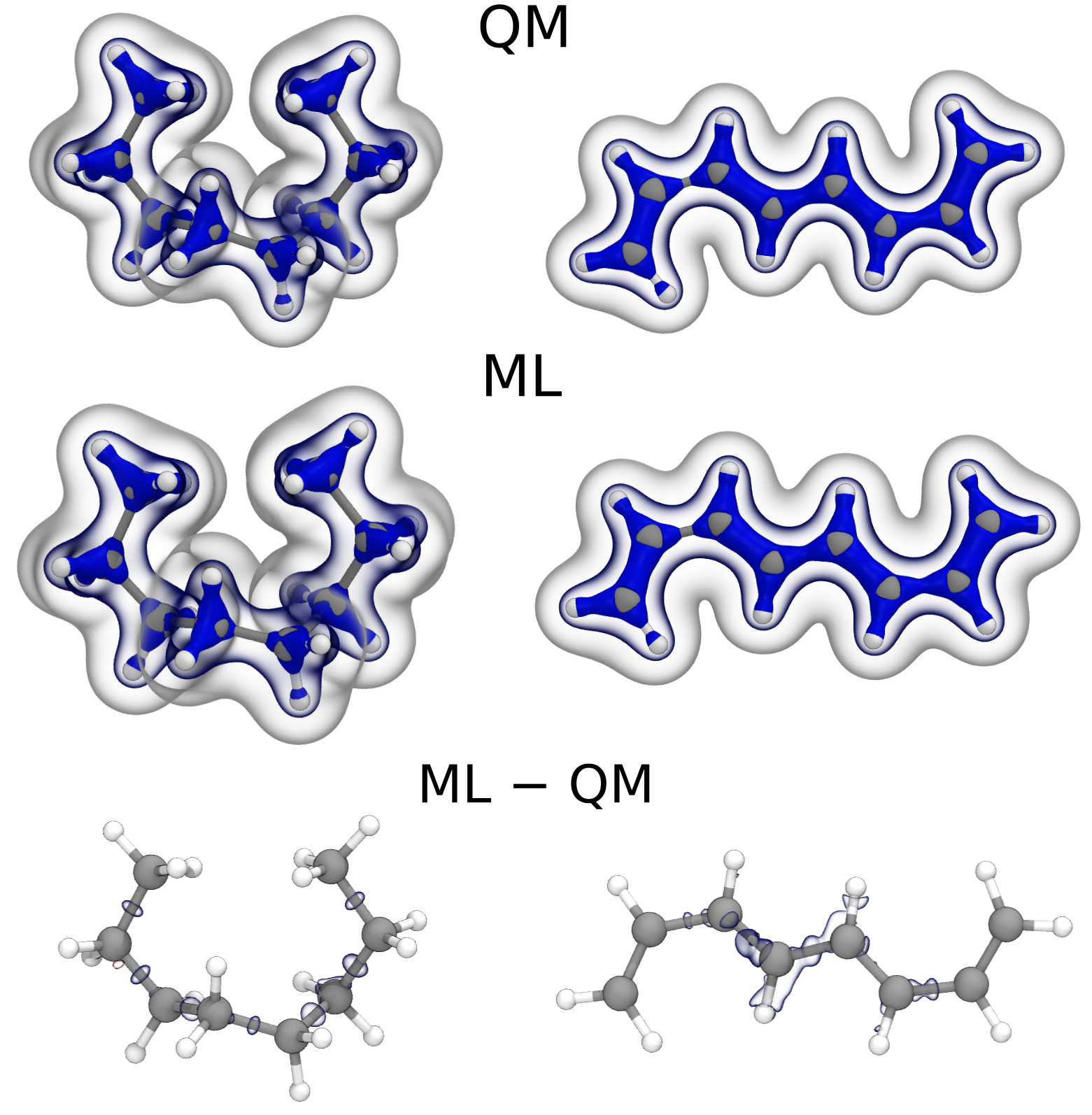}
\caption{Extrapolation results for the valence electron density of one octane (\emph{left}) and one octatetraene (\emph{right}) conformer, using a model trained on butadiene and butane. (\emph{top}) DFT/PBE density isosurface at 0.25, 0.1, 0.01 Bohr$^{-3}$, (\emph{middle}) machine-learning prediction isosurface at 0.25, 0.1, 0.01 Bohr$^{-3}$, (\emph{bottom}) machine-learning error, red and blue isosurfaces refer to $\pm$ 0.005 Bohr$^{-3}$ respectively. 
Reproduced from Ref.~\citenum{gris+19acscs}. Copyright 2018 American Chemical Society.
\label{fig:extrapolation}}
\end{figure}

\newcommand{\trho}{\tilde{\rho}}
\newcommand{\tc}{\tilde{c}}
\subsection{Electronic charge densities}

Another relevant scenario where the data-driven prediction of a quantum property benefits from a representation that relies on the use of local and equivariant features is the electron density $\trho(\br)$ of an atomic structure.\footnote{We use $\trho$ and $\tc$ to refer to electron density and its expansion coefficients, to distinguish them from the similar terms used for the atom density.}
The density is a scalar field, and has been modeled with some success by predicting its value at a specific point by an invariant representation \emph{centered on that point}\cite{alre+18cst,chan+19npjcm}.
Given that (particularly in the case of an all-electron calculation) atomic nuclei are a natural vantage point to decompose the overall electron density, one may want instead to model $\trho(\br)$ as a sum of atom-centered contributions, $\trho(\br)=\sum_i \trho(A_i; \br)$.
These atom-centered terms can then be conveniently decomposed as a sum of local functions, at the price of adopting a multi-centered non-orthogonal basis for the expansion i.e.,
\begin{equation}
\trho(A_i,\br) = \sum_{n\lambda\mu} \tc_{n\lambda\mu}(A_i) R_n(\left|\br-\br_i\right|) Y^\lambda_\mu\left(\widehat{\br-\br_i}\right)
\end{equation}
where $R_n$ represent some suitably optimized radial functions (for instance those used in resolution of the identity methods in quantum chemistry\cite{whit73jcp}) and $\tc_{n\lambda\mu}(A_i)$ correspond to the non-orthogonal expansion coefficients, that depend on the arrangement of atoms in the environment $A_i$.
These coefficients must transform in a covariant fashion with a rotation of the environment, and each $\lambda$-component can be independently predicted  using equivariant features of the corresponding order -- for instance with a linear model
\begin{equation}
\tc_{n\lambda\mu}(A_i) \approx \sum_{\q} \rep<\tc_{n\lambda}||\Q> \rep<\Q||A_i; \rrhoislm{\nu}>,
\end{equation}
even though current implementations use a kernel regression scheme\cite{gris+19acscs,fabr+19cs}.
The non-orthogonality of the basis used to represent $\trho(\boldsymbol r)$, implies that the learning phase has now to be performed considering all the different density components at the same time\cite{gris+19acscs,fabr+19cs}.
While this may sound as a computational drawback of the model, it also improves the locality of the coefficients, which underlies its remarkable transferability across vast conformational and chemical spaces, since the electron density can be effectively learned as a collection of local contributions. This is well exemplified in Figs.~\ref{fig:extrapolation}, where the electron density prediction of C(8) hydrocarbons\cite{gris+19acscs} is tested upon having trained the model on much smaller compounds, with only 4 carbon atoms.
This approach has since been applied to more complex systems, such as oligopeptides\cite{fabr+19cs}, and to the prediction of other scalar fields such as the on-top density\cite{fabr+20jcp}.

\begin{figure}[bp]
 \includegraphics[width=1.0\linewidth]{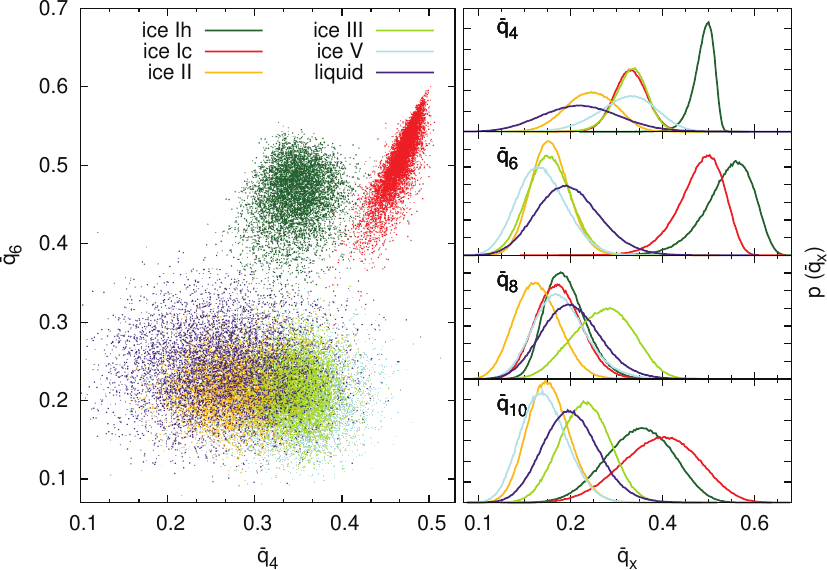}
\caption{Scatter plot and histograms based on Steinhardt order parameters\cite{stei+83prb} $q_n$ computed for simulations of  liquid water and different phases of ice.
Reproduced with permission from Ref.~\citenum{geig-dell13jcp}. Copyright 2013 American Institute of Physics.
\label{fig:geig-dell}}
\end{figure}

\begin{figure*}[tbhp]
\begin{minipage}{0.75\linewidth}
\vspace{0mm}
 \includegraphics[width=1.0\linewidth]{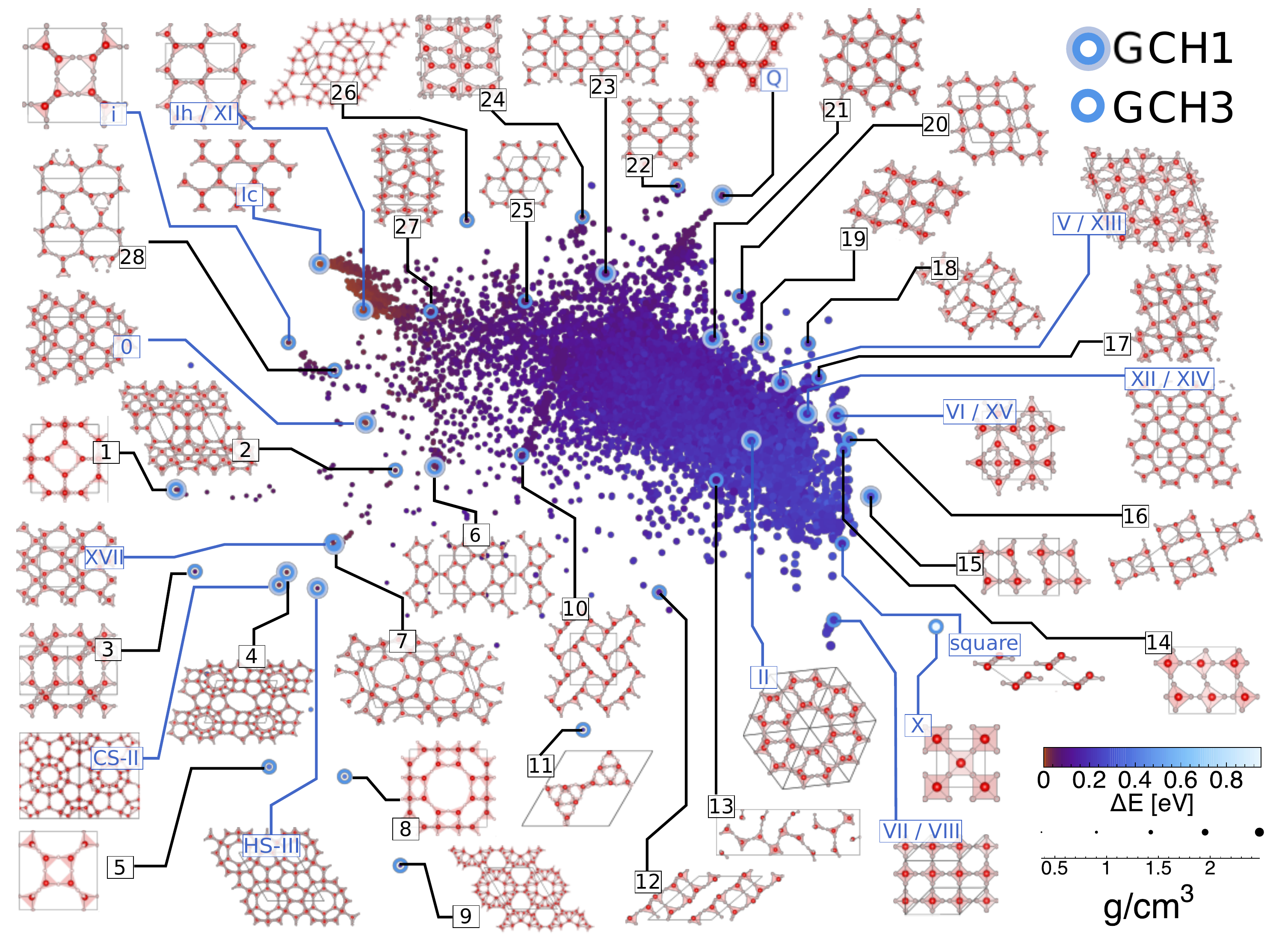}
 \end{minipage}
 \begin{minipage}{0.24\linewidth}
\caption{Sketch map of the structural similarity of 15,869 distinct PBE-DFT geometry-optimised ice structures, as computed by the Euclidean distance in SOAP space. The sketch-map coordinates are obtained by minimizing the error in reproducing this similarity in terms of distances between points on a 2D projection. The density and static lattice energy of each structure are encoded by the size and colour of the respective point, and correlate strongly with the position on the map. Known ice phases are labelled in blue. 34 new candidates are labelled in black and numbered in order of increasing dressed energy relative to a generalized convex hull\cite{anel+18prm} construction. Reproduced from Ref.~\citenum{enge+18ncomm}. Copyright 2018 Springer Nature under Creative Commons Attribution 4.0 International License \url{https://creativecommons.org/licenses/by/4.0/}.
\label{fig:gch-water}}
\end{minipage}
\end{figure*}

\subsection{Structural classification and structural landscapes}

As discussed in Section~\ref{sec:maps}, the choice of a representation to describe atomic structures determines the ``lens'' through which they are interpreted, which in turns has a strong impact on the way unsupervised learning schemes, such as clustering and dimensionality reduction, bring to light recurring patterns, and structure-property relations.
The potential of general-purpose, atom-density correlation features for these tasks has been recognized rather early. Figure~\ref{fig:geig-dell}, adapted from Ref.~\citenum{geig-dell13jcp} shows a classification of snapshots taken from simulations of different phases of water, based on Steinhardt order parameters\cite{stei+83prb}, which are closely related to $\rep|\frho_i^2>$ features, and make it possible to partly differentiate between phases.
In the same study it is shown how a neural network based on atom-centered symmetry functions can be trained to achieve near-perfect classification accuracy.
An even more comprehensive mapping of the phase diagram of water -- in which crystalline and amorphous phases from across the phase diagram, as well as transition pathways between them were considered -- was produced in Ref.~\citenum{piet-mart15jcp}, using permutation-invariant vectors\cite{gall-piet13jcp} as global descriptors for the different configurations.
Abstract structural descriptors are particularly useful when applied to datasets that contain hypothetical structures, generated by a high-throughput procedure\cite{isay+15cm}.
In combination with a dimensionality-reduction scheme\cite{ceri+11pnas}, and with a generalized convex hull construction that attempts to estimate the synthesizability of materials by considering jointly their predicted stability, and the structural similarity to other potential candidates\cite{anel+18prm}, a SOAP representation has been able to rediscover all known (meta)stable ice phases, as well as to propose another 34 structures which might be also stabilizable by pressure, doping, or co-crystallization\cite{enge+18ncomm} (see Fig.~\ref{fig:gch-water}).

\begin{figure}[tbhp]
 \includegraphics[width=1.0\linewidth]{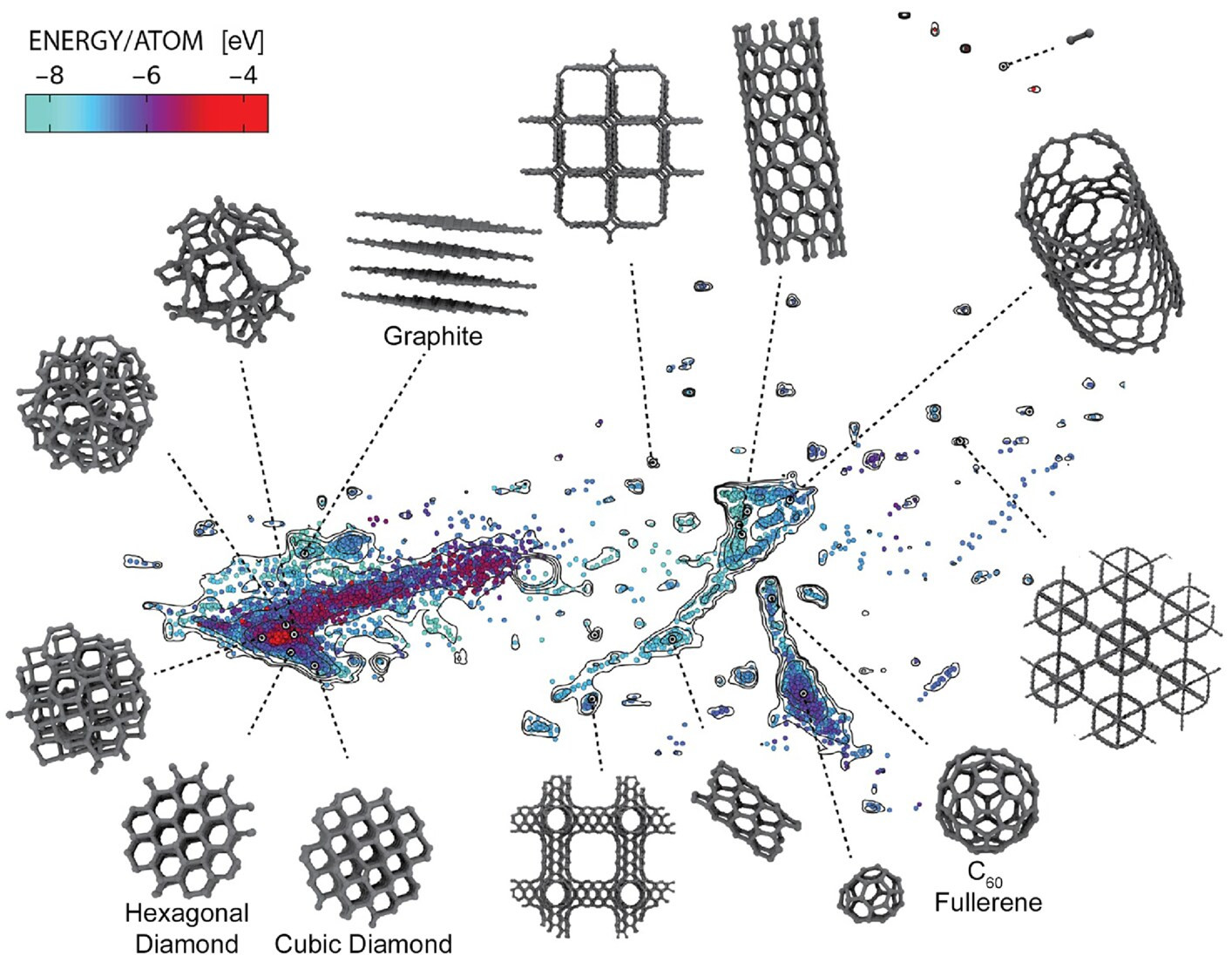}
\caption{Sketch-map describing the makeup of the structures included in the train set of an accurate and transferable potential for carbon.
Selected structures are identified
for graphite, diamond, hexagonal diamond (Lonsdaleite), amorphous carbon and fullerenes. Points are coloured
according to their energy, while contours indicate the density of the database population in a particular region.
Adapted with permission from Ref.~\citenum{rowe+20jcp}. Copyright 2020 American Institute of Physics.
\label{fig:carbon}}
\end{figure}

An incomplete list of applications that use general-purpose features for structural analysis and classification includes: the construction of structure-property maps for small organic molecules\cite{de+16jci,maks+20ijqc}, molecular materials\cite{musi+18cs,yang+18cm,gryn+18jacs}, inorganic perovskites\cite{yang20jmcc}, corrosion inhibitors\cite{wurg+19fm}; the  identification and characterization of defects in solids\cite{shar+18pnas,prie+18am,home+19fm} and self-assembled polymers\cite{gasp+20jpcb,capelli2021chemrxiv}; the classification of secondary-structure patterns in polypeptides\cite{helf+19fmb} and the building blocks of zeolites\cite{helf+19jcp, schw+19nm} and porous materials\cite{nich+20cs};
the classification of different phases in multi-phase materials\cite{diet+17pre,fulf+19jcim}; the characterization of amorphous systems\cite{deri+18cm,zhou+20jmcc,huan+19jmca,caro+20prb,deri+21nature}; the search of stable phases of materials\cite{rein+20pccp}; the determination of the convergence of microsolvation studies of the hydration free energy\cite{basd+20jctc}.

There are also several examples, besides those given in Section~\ref{sec:maps} where low-dimensional maps have been used as a tool to understand the structure of a data set or the nature of a representation.
In Ref.~\citenum{hans+13jctc} a PCA map was used to understand the effect of randomizing the atom ordering on the feature space associated with a Coulomb matrix description of molecules, emphasizing the information loss associated with sorting of the elements -- an alternative route to achieve permutation invariance.
In Ref.~\citenum{de+16pccp}, maps based on different kinds of SOAP kernels provided an understanding of the effect of different approaches to combining environment-level kernels, and of different definitions of an alchemical kernel between chemical elements, on the similarity between molecules as measured by the representation.
In Ref.~\citenum{nguy+18jcp}, maps of a dataset of water oligomers were used to compare the performance of different ML scheme to build 2 and 3-body models of the energy of water clusters.

The use of low-dimensional representations to visualize the structure of a dataset, showing the relationship between different kinds of training structures, identifying regions that are poorly sampled, and determining how new configurations relate to the data the ML model has been fitted, is also gaining traction.\cite{fabe+15ijqc,de+16pccp,bart+17sa,stuk+19jcp,bern+19npjcm,benm+20prb,mons+20nc,deri+20nc}
An example of such map is given in Fig.~\ref{fig:carbon}, showing the diversity of the structures used to train a transferable machine-learning potential for carbon.\cite{rowe+20jcp}
Adopting the same type of representations used for the regression model as the basis of this kind of analysis ensures that the maps describe the same feature space that underlies the fit.

\subsection{3D representations for QSPR and reaction predictions}

Even though the focus of this review is on descriptors of the 3D structure of materials applied to the construction of surrogate models of quantum mechanical properties, there is also growing interest in their application to QSPR tasks.
As we briefly discuss in Section~\ref{sec:history-wishes}, the descriptors that have been traditionally used in cheminformatics are based on a collection of molecular properties, or on molecular graph descriptors that do not depend on the particular  conformation.\cite{Shen2020}
From a conceptual point of view, their coarsness is an advantage, because it is compatible with the definition of thermodynamic properties that are not associated with a single specific configuration, such as solvation and ligand binding free energies.
Nevertheless, there is growing evidence that the use of descriptors incorporating information on the 3D geometries can improve the accuracy of QSPR models, especially  for difficult cases that involve very flexible molecules,\cite{Dastmalchi2012,Jagiello2016} as well as for data analytics approaches for materials informatics\cite{chou+18prm}.
One of the core challenges in these efforts is the determination of the conformer geometries that should be used to evaluate the 3D descriptors, an operation for which several strategies have been explored to enhance the accuracy of QSPR models.\cite{Kuzmin2005,Zankov2020}
As we briefly discuss in \cref{sub:indirect-struc-prop-rel}, one of the most promising research directions involves combining the high fidelity of density based representations with a well-principled construction of ensembles of features.
This is still a very active subject of research, with very encouraging results having been recently demonstrated for the prediction of the solubility of small molecules\cite{wein+21jcp}, the computational screening for antiviral drugs\cite{Axelrod2020}, %
and the prediction of enantioselectivity of organocatalysts\cite{gall+21cs}.

\subsection{Descriptors from electronic-structure theory}
\label{sub:electronic-structure-descriptors}

Another growing trend that is worth a brief mention involves the use of information from electronic-structure calculations in the construction of structural representations. 
The idea has been applied in different forms.  At the simplest level, electronic-structure-based indicators of chemical similarity, obtained for bulk elements, have been used in the construction of elemental similarity kernels,\cite{de+16pccp} to obtain models that are more predictive across chemical space\cite{jinn-asah17jpcl}. 
Alternatively, electronic-structure indicators, such as the local density of states, can be used side-by-side with purely structural representations, yielding a substantial improvement of the accuracy of the model\cite{caro+18cm,aarv+19cm}.

Elements of an electronic structure calculation, such as the charge density\cite{dick-fern19jcp}, the electron density of states\cite{fung+21nc} or the elements of the Fock matrix\cite{welb+18jctc} can be used directly as the basis for a molecular representation. 
This approach requires an electronic structure calculation in order to make predictions for each new structure, which implies a substantial overhead in comparison with methods using as inputs only the atomic positions.
However, the increase in the transferability of the models may well justified the greater computational effort, particularly when using descriptors based on low levels of quantum mechanical theory to predict high-end, accurate molecular properties.\cite{chen+19jcp, qiao+20jcp}

\section*{Conclusions and outlook}

The description of atomic structures in terms of mathematically sound, computationally efficient, and physically-inspired representations has largely driven the extraordinarily successful application of machine-learning schemes to atomic-scale modeling.
Independently-developed representations have undergone a process of convergent evolution to fulfill a concurrent set of requirements, such as symmetry with respect to translations and rotations, smoothness and injectivity -- a clear indication of the importance of these criteria to obtain efficient machine-learning models.
Over the past few years, a more systematic study of the problem of representing atomic structures has clarified the connections between most of the successful representations, and between these and well-established concepts in the statistical physics of liquids ($\nu$-point density correlations) and of alloys (the cluster expansion), as well as with the construction of potential energy surfaces for molecules and the condensed phase.

A formal treatment of symmetries enabled the development of equivariant features that are suitable to build models that automatically obey the same transformation rules as vectors and tensors, making it possible to learn efficiently properties such as dipole moments, polarizability and density fields.
This equivariant formulation can also be used to iteratively increase the body order of a structural representation: An important open question is how to best treat these high-body-order terms, whether by linear models that explicitly include dedicated high-order features, or by non-linear models that generate (some of) them algebraically.
The answer rests both on practical considerations and on the very fundamental, highly non-trivial issue of whether a representation of limited body order provides a complete (injective) description of an atomic structure.  Even though it is possible to build systematically a complete basis to expand in a linear fashion any structure-property relation, it is not clear how to build a \emph{minimal} set of features that guarantees an injective mapping when used as the input of a general \emph{non-linear} model, or how to reduce in an effective manner the size of a complete linear basis. 
A better understanding of the mathematical properties of representations is likely to lead, in the near future, to more robust and better performing implementations, and might also help design better ``deep'' models, by identifying the algebraic manipulations that increase most effectively the expressive power of the features used as inputs.

Another open challenge is how to deal with non-additivity, and with properties that depend on long-range interactions between far-away atoms.
Particularly promising is a long-distance equivariant framework, that can be formulated as as a rather straightforward extension of the same density-correlations scheme that underlies local features, and can be related to a multipole expansion of interactions.
It is yet to be seen whether it can describe more subtle physical phenomena such as quantum delocalization, polarization and charge transfer, and how it compares with more explicitly physically-motivated ``hybrid'' models.
A better control of the multi-scale nature of the interactions, including the use of ``multi-resolution'' features, is likely to be one of the focal point of feature-engineering efforts, which may lead to an incremental -- but nevertheless important -- increase of the accuracy of ML models of matter.
The optimization of features for a specific problem may however impact their general applicability, which is one of the critical advantages of the class of abstract, generic representations we focus on in this review, that can be seen as the point of convergence of molecular potential energy surfaces and condensed-phase potentials.
The quantitative assessment of the mutual information content of alternative descriptors, of their sensitivity to structural deformations, and to the degree to which they correlate with the target properties may serve as a guide to strike a balance between these conflicting goals, and to make better informed choices between alternative frameworks. 

One of the most recent research directions aims at extending even further the reach of the class of descriptors we discuss in this review, by resolving the divide between three dimensional continuous representations and discrete fingerprints, for applications to quantitative structure-property relations. The challenge here is to reconcile the superior resolving power of 3D, atom-density-correlation representations with the fact that traditional cheminformatics tasks aim to predict macroscopic properties, such as solubility or toxicity, that are not associated with an individual configuration, but rather with the ensemble of conformers corresponding to a specific thermodynamic state point.
Another traditional application of cheminformatics is the inverse design of molecules with prescribed (or optimized) properties, and the construction for generative models.
While one could envisage to use 3D representations for this task, a substantial hurdle would be the fact that the map between structure and density-correlation features is not bijective: there are feature vectors that do not correspond to any structure, and even feature vectors that cannot be obtained as a symmetrized correlation of an arbitrary scalar field.
Thus, the unconstrained search for the ``optimal feature vector'' might result in a set of features that do not correspond to an actual structure.
Until this issue is better understood, efforts to use atom-density representations for inverse design should rely on approaches that do not require an inverse feature map.

In the quest for more accurate and efficient machine-learning models of the structure and properties of atomistic systems, physically-motivated concepts have been incorporated into the mathematical representation of atomic configurations, resulting in striking connections with traditional modeling frameworks.
When treading the fine line between data-driven and physics-based approaches, the core question is how to achieve a natural description of well-understood phenomena without giving up the flexibility to model unexpected, complex effects -- and how to build features that can be optimized for a specific application, while still being universally applicable.
A definitive answer to this question is still lacking, but we believe that the general principles that we have summarized in this review may indicate the direction to follow, and provide some guidance to the practitioners who seek to make an informed choice among the ever increasing number of representations for atomic-scale modeling.   

\begin{acknowledgments}
The authors would like to thank Yasushi Shibuta for providing the structures used in Fig.~\ref{fig:shibuta}, and Stefan Goedecker for providing Fig.~\ref{fig:goedecker-fingerprints}, and the many colleagues and friends who discussed with us about this review, and the ideas it summarizes.
FM, MC and AG acknowledge support by the National Center of Competence in Research MARVEL, funded by the Swiss National Science Foundation.
\end{acknowledgments}

\section*{Author biographies}

{\bf F\'elix Musil} studied physics at the EPFL and received his MSc in applied physics in 2015, with a thesis on the modeling of plasma in a fusion reactor. For his PhD he joined in 2016 the group of Prof. Ceriotti at the EPFL to develop and apply methods to investigate structure–property relationships in materials using atomistic modeling and machine learning techniques.

{\bf Andrea Grisafi} studied chemistry at the University of Pisa and Scuola Normale Superiore of Pisa. In 2016, he received his MSc in physical chemistry with a thesis on the statistical mechanics of simple ionic liquids. Since then, he is a PhD student in the group of Prof. Michele Ceriotti at EPFL, where he works on the development of atomic-scale representations that are suitable to incorporate physical symmetries and long-range effects within machine-learning models of
molecular and materials properties.

{\bf Albert P. Bart\'ok} is an Assistant Professor at the University of Warwick. He earned his PhD degree in physics from the University of Cambridge in 2010, his research having been on developing interatomic potentials based on ab inito data using machine learning. He was a Junior Research Fellow at Magdalene College, Cambridge and later a Leverhulme Early Career Fellow. Before taking up his current position, he was a Research Scientist at the Science and Technology Facilities Council. His research focuses on developing theoretical and computational tools to understand atomistic processes.

{\bf Christoph Ortner} is Professor of Mathematics at the University of British Columbia (Canada). After obtaining his doctorate in numerical analysis in 2007 at the University of Oxford (UK) and remaining there as an RCUK fellow, he moved to the University of Warwick in 2011 and to UBC in 2020. His main interests revolve around mathematical and computational aspects of atomistic and multi-scale modeling.

{\bf G\'abor Cs\'anyi} is Professor of Molecular Modelling at the University of Cambridge (UK). He obtained his doctorate in computational physics (2001) from the Massachusetts Institute of Technology (USA), having worked on electronic structure problems. He was in the group of Mike Payne in the Cavendish Laboratory before joining the faculty of the Engineering Laboratory at Cambridge. He is developing algorithms and data driven numerical methods for atomic scale problems in materials science and chemistry.

{\bf Michele Ceriotti} is Associate Professor at the Institute of Materials at the {\'E}cole Polytechnique F{\'e}d{\'e}rale de Lausanne.
He received his Ph.D. in Physics from ETH Z\"urich in 2010, under the supervision of Professor Michele Parrinello. He spent three years in Oxford as a Junior Research Fellow at Merton College, and joined EPFL in 2013, where he leads the laboratory for Computational Science and Modeling. His research interests focus on the development of methods for molecular dynamics and the simulation of complex systems at the atomistic level, as well as their application to problems in chemistry and materials science -- using machine learning both as an engine to drive more accurate and predictive simulations, and as a conceptual tool to investigate the interplay between data-driven and physics-inspired modeling.

\providecommand{\latin}[1]{#1}
\makeatletter
\providecommand{\doi}
  {\begingroup\let\do\@makeother\dospecials
  \catcode`\{=1 \catcode`\}=2 \doi@aux}
\providecommand{\doi@aux}[1]{\endgroup\texttt{#1}}
\makeatother
\providecommand*\mcitethebibliography{\thebibliography}
\csname @ifundefined\endcsname{endmcitethebibliography}
  {\let\endmcitethebibliography\endthebibliography}{}

\clearpage
\begin{figure*}
\includegraphics[width=0.5\linewidth]{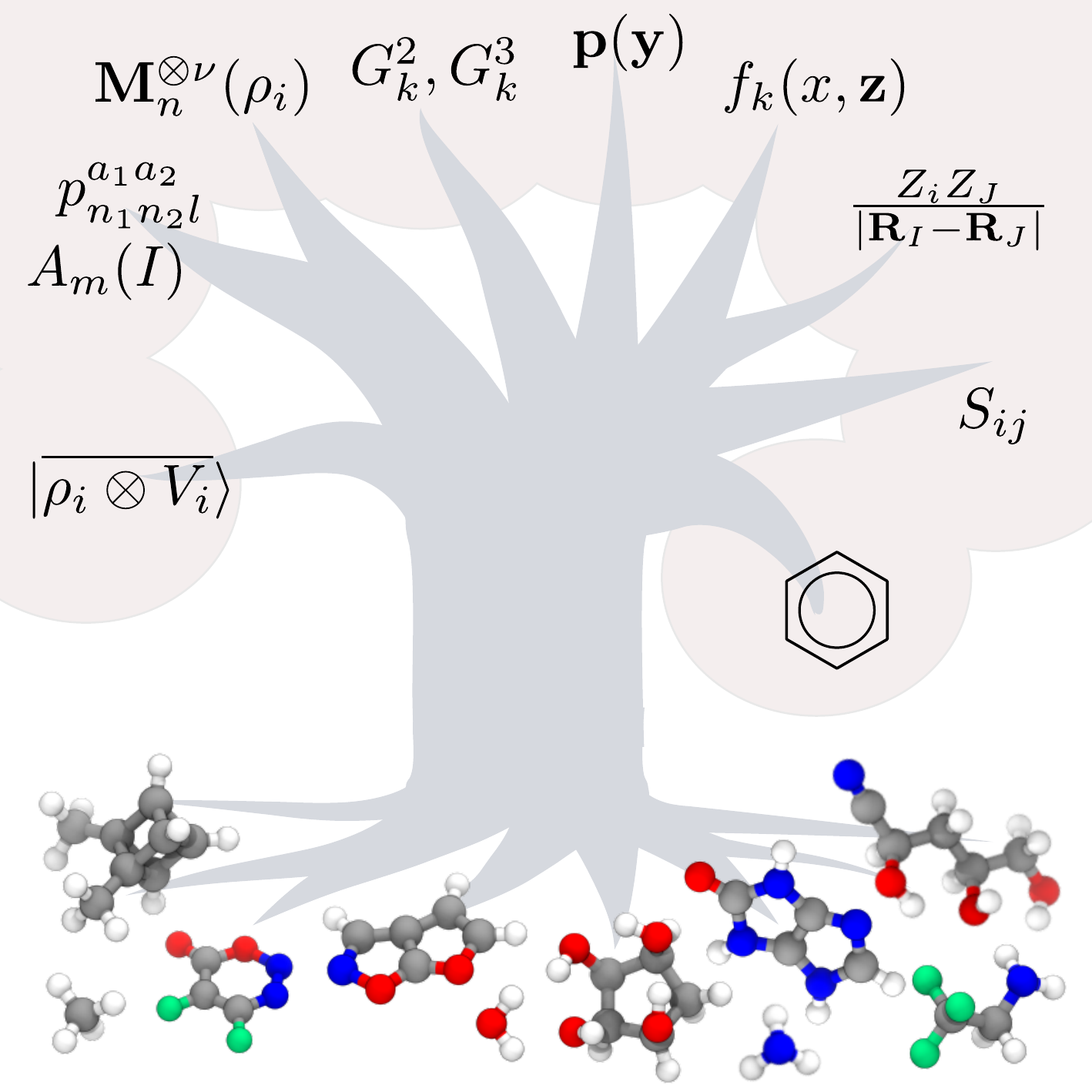}\\
TOC GRAPHICS
\end{figure*}

\end{document}